\documentclass[twocolumn]{aastex63}

\newcommand{\HI}{H{\sc i}}
\newcommand{\kms}{km s$^{-1}$}
\newcommand{\ha}{H$\alpha$}

\newcommand{\mstar}{(1.37$\pm$0.05)$\times$10$^{8}$ M$_{\odot}$}
\newcommand{\mhi}{(8.51$\pm$0.36)$\times$10$^{8}$ M$_{\odot}$}
\newcommand{\mbaryon}{(9.88$\pm$0.36)$\times$10$^{8}$ M$_{\odot}$}
\newcommand{\sfr}{(8.2$\pm$0.4)$\times$10$^{-3}$   M$_{\odot}$ yr$^{-1}$}
\newcommand{\mhalo}{(3.5$\pm$1.2)$\times$10$^{10}$ M$_{\odot}$}
\newcommand{\mhalothreehpd}{1.5$\times$10$^{10}$ M$_{\odot}$}
\newcommand{\concentration}{2.0$\pm$0.36}
\newcommand{\fuzzylower}{$<$ 0.06$\times$10$^{-22}$ eV}
\newcommand{\fuzzyupper}{$>$ 2.7$\times$10$^{-22}$ eV}
\newcommand{\fuzzylowerthreehpd}{$<$ 0.11$\times$10$^{-22}$ eV}
\newcommand{\fuzzyupperthreehpd}{$>$ 3.3$\times$10$^{-22}$ eV}
\newcommand{\sidm}{$<$ 1.63 cm$^{2}$/g}
\newcommand{\wmdm}{$>$ 0.23 keV}
\newcommand{\Vvir}{47 km s$^{-1}$}
\newcommand{\mstarmhalo}{(0.39$\pm$0.13)\%}
\newcommand{\radaccslope}{0.15$\pm$0.11}
\newcommand{\radaccslopesigma}{3.2}
\newcommand{\pooNFW}{1.6}
\newcommand{\pooISO}{6.3}
\newcommand{\pooBu}{4.4}
\newcommand{\innerradius}{0.67 kpc}
\newcommand{\innerslope}{-(0.90$\pm$0.08)}

\makeatletter
\renewcommand\@makefntext[1]{\leftskip=0em\hskip-1em\@makefnmark#1}
\makeatother

\usepackage[super]{nth}
\usepackage{lineno,mathtools}

\shorttitle{A cuspy halo}
\shortauthors{Shi et al.}

\begin{document}

\title{A cuspy dark matter halo}

\correspondingauthor{Yong Shi}
\email{yshipku@gmail.com}

\author[0000-0002-8614-6275]{Yong Shi}
\affil{School of Astronomy and Space Science, Nanjing University, Nanjing 210093, China.}
\affil{Key Laboratory of Modern Astronomy and Astrophysics (Nanjing University), Ministry of Education, Nanjing 210093, China.}

\author{Zhi-Yu Zhang}
\affil{School of Astronomy and Space Science, Nanjing University, Nanjing 210093, China.}
\affil{Key Laboratory of Modern Astronomy and Astrophysics (Nanjing University), Ministry of Education, Nanjing 210093, China.}

\author{Junzhi Wang}
\affil{Shanghai Astronomical Observatory, Chinese Academy of Sciences, 80 Nandan Road, Shanghai 200030, China}

\author{Jianhang Chen}
\affil{School of Astronomy and Space Science, Nanjing University, Nanjing 210093, China.}

\author{Qiusheng Gu}
\affil{School of Astronomy and Space Science, Nanjing University, Nanjing 210093, China.}
\affil{Key Laboratory of Modern Astronomy and Astrophysics (Nanjing University), Ministry of Education, Nanjing 210093, China.}

\author{Xiaoling Yu}
\affil{School of Astronomy and Space Science, Nanjing University, Nanjing 210093, China.}

\author{Songlin Li}
\affil{School of Astronomy and Space Science, Nanjing University, Nanjing 210093, China.}


\begin{abstract}

  The cusp-core  problem is one  of the  main challenges of  the cold
  dark matter paradigm  on small scales: the density of  a dark matter
  halo is  predicted to  rise rapidly toward  the center  as $\rho(r)$
  $\propto$ $r^{\alpha}$ with $\alpha$ between -1 and -1.5, while such a cuspy profile has
  not   been   clearly   observed.     We   have   carried   out   the
  spatially-resolved  mapping   of  gas   dynamics  toward   a  nearby
  ultra-diffuse galaxy (UDG), AGC  242019.  The derived rotation curve
  of dark matter  is well fitted by the cuspy  profile as described by
  the  Navarro-Frenk-White  model,  while  the  cored  profiles
  including both the pseudo-isothermal and Burkert models are excluded. The  halo has
  $\alpha$= \innerslope\, at the innermost radius of \innerradius\,, 
  $M_{\rm halo}$=\mhalo\, and a small concentration of \concentration. AGC
  242019  challenges alternatives of cold dark matter by
  constraining  the  particle   mass  of  fuzzy  dark   matter  to  be
  \fuzzylowerthreehpd\, or \fuzzyupperthreehpd\,, the cross section of self-interacting
  dark matter to  be \sidm, and the particle mass  of warm dark matter
  to be \wmdm, all of which are in tension with other constraints. The modified Newtonian dynamics is also
  inconsistent with a shallow radial acceleration relationship of AGC 242019. For
  the feedback scenario that transforms a  cusp to a core,
  AGC  242019  disagrees with  the  stellar-to-halo-mass-ratio
  dependent  model, but agrees with  the star-formation-threshold
  dependent model.  As a UDG, AGC 242019 is in
  a dwarf-size halo with weak stellar feedback, late formation time,
  a normal baryonic spin and
  low star formation efficiency (SFR/gas).

\end{abstract}




\section{Introduction} \label{sec_intro}

The  cosmological model  of cold  dark matter  and dark  energy, i.e.,
$\Lambda$CDM,  has achieved  tremendous success  in understanding  the
cosmic  structure across  time  on  large scales,  but  this model  is
challenged  by observations  on small  scales such  as the  cusp-core
problem, the  missing dwarf problem, the  too-big-to-fail problem etc.
\citep[for a review, see][]{Weinberg15}.

In cosmological simulations  of cold and collisionless  dark matter, a
dark matter  halo has a density  profile that rises toward  the center
with a power index of -1 to -1.5 \citep{Moore94, Burkert95, Navarro97,
  Moore98,  Ghigna00, Jing02,  Wang20}, referred  as a  cuspy profile.
However, over the past decades,  much shallower or even flat core-like
profiles toward centers have been found  in most, if not all, observed
data of  nearby galaxies  through mapping dynamics  of gas  and stars.
Early studies with HI  interferometric data reveal shallow dark-matter
central  profiles in  individual  galaxies \citep{Carignan89,  Lake90,
  Marc90}.   Studies  with higher  spatial  resolutions  for a  larger
sample of  dwarfs and low-surface-brightness galaxies  further confirm
the  central flatness  of the  rotation  curve, and  derived a  median
dark-matter   density    slope   of   about   -0.2    toward   centers
\citep{deBlok01,  Oh11, Oh15}.   Optical observations  of ionized  gas
such as \ha\, can achieve higher spatial resolutions than the HI data,
and confirmed  the median density  slope of about -0.2  with long-slit
spectra  for  a large  sample  of  dwarf galaxies  \citep{Spekkens05}.
Although the rotation curves from the long-slit spectroscopic data are
sensitive to  the assumed  dynamical center, the  position angle  of the
kinematic major axis etc, further  studies with optical integral field
unit have suggested insignificance of  the above effects and validated
dark-matter     profiles     with     central     shallower     slopes
\citep{KuziodeNaray08, Adams14}.

Among few galaxies whose halos can be described by cuspy profiles, the
one  with the  highest signal-to-noise-ratio  is  DDO 101  that has  a
central   dark-matter    slope   of    -1.02$\pm$0.12   (8.5-$\sigma$)
\citep{Oh15}. However, this  object can be fitted equally  well with a
cored profile; the  difference in the reduced $\chi^{2}$  is only 0.02
for a degree  of freedom (d.o.f) of six  \citep[see][]{Oh15}.  This is
because of the limitation on the spatial resolution and spatial extent
of the observed rotation curve, as  well as different models for cuspy
and cored profiles (see \S~\ref{sec_model}).  For example, the density
slope of a NFW model for a  cuspy profile varies from -1 at the center
to -3 at infinity, while the pseudo-isothermal (ISO) model for a cored
profile has a  slope from 0 at  the center to -2 at  infinity, so that
both  models may  perform  well on  fitting the  data  with a  limited
spatial extent.  In  order to identify a definitive  cuspy dark matter
halo, it  is thus  required to  demonstrate not  only the  validity of
cuspy models but also the invalidity  of cored models. For DDO 101, the
large uncertainty  in its distance  further  challenges the
reliability of  its cuspy  profile \citep{Read16b}.  Studies  of local
dwarf spheroidal galaxies also suggest  cuspy profiles in few objects,
among  which  the highest-probability  one  is  Draco with  a  central
density   slope  of   -1.0   at  5.0-$\sigma$   and  6.8-$\sigma$   by
\citet{Jardel13} and \citet{Hayashi20}, respectively.  However, it has
not been  demonstrated whether a cored  model can fit the  data or not
for this  object, as  done for DDO  101.  And  furthermore, systematic
uncertainties are complicated for these  studies in which the velocity
dispersion among individual stars as a function of the galactic radius
are  used  to  measure  the dark  matter  distribution.   The  orbital
anisotropy, the method to model  stellar orbits, the dark-matter shape
and the limited number of the member stars are all found to affect the
conclusions \citep{Evans09}.  Especially, a  recent study of simulated
galaxies  by  \citet{Chang20} emphasized  the  importance  of a  large
amount  of stars  in order  to unambiguously  measure the  dark matter
distribution.  If the number of stars is less than 10000, an intrinsic
cored profile cannot be differentiated  reliably from a cuspy profile.
For  Draco, there  are only  $\sim$468 member  stars \citep{Walker15}.
Fortunately,   these   systematic  uncertainties   are   significantly
eliminated for HI  interferometric data as seen  in simulated galaxies
\citep{KuziodeNaray09, KuziodeNaray11}.  This is mainly because gas is
collisional and their  dynamics can be relatively  easily described by
titled-ring  models \citep{Begeman89,  DiTeodoro15}.  In  summary, the
above few claimed cuspy profiles of dark halos are not definitive.

There  are  two   solutions  to  the  small   scale  controversies  of
$\Lambda$CDM.  One is to modify the  nature of dark matter so that it
is  no longer cold but instead  self-interacting, fuzzy or
warm.  These modifications  can retain the properties  of the universe
on  large  scales  as  predicted   by  $\Lambda$CDM  but  resolve  its
small-scale  challenges.   Fuzzy  cold  dark  matter,  also  known  as
ultralight  scalar   particles,  has  a  mass   around  10$^{-22}$  eV
\citep[e.g.][]{Hu00, Schive14}.   The uncertainty principle of  its wave
nature  acts on  kpc scales,  smoothing the  density fluctuations  and
preventing the growth  of small halos and formation  of central cusps.
The self-interacting dark matter  transports ``heat'' (higher velocity
dispersion) from the outer region to  the ``cooler'' inner region of a
halo. It  leads to  a constant
density     core      with     isothermal      velocity     dispersion
\citep[e.g.][]{Spergel00, Rocha13, Tulin18}.  If  dark matter is warm,
it decouples from the primordial plasma at relativistic velocity, thus
free streaming out of small density  peaks. As a result, the structure
formation at small scales are suppressed \citep[e.g.][]{Avila-Reese01,
  Lovell14}. Warm dark matter scenario is found to have a ``catch-22''
problem,  i.e.,   if  the  ``cusp-core''  problem   is  resolved,  the
requirement for  the particle mass  cannot form small galaxies  at the
first place, and vice versa \citep[e.g.][]{Maccio12}.

Another  solution is  to keep  cold  dark matter  paradigm but  invoke
efficient gravitational  interaction between dark matter  and baryonic
matter  through   stellar  feedback   \citep{Navarro96,  Governato10}.
Overall,  as  gas falls  into  the  inner  region  of the  halo,  star
formation takes  place and the subsequent  feedback through supernovae
expels an appreciable  amount of gas and stars to  large radii.  These
baryonic matter pulls dark matter particles to migrate outward through
pure gravitational  interaction, lowering the central  density of dark
matter.  The  overall efficiency of  this interaction and  its effects
are still  under investigations \citep{DiCintio14,  Read16a, Tollet16, Bose19,  Benitez-Llambay19,  Read19}.  In addition  to  the  stellar
feedback, other mechanisms have also  been proposed, such as dynamical
friction   between    gas   clouds    and   dark    matter   particles
\citep{Nipoti15}.

\begin{figure}[tbh]
  \begin{center}
    \includegraphics[scale=0.6]{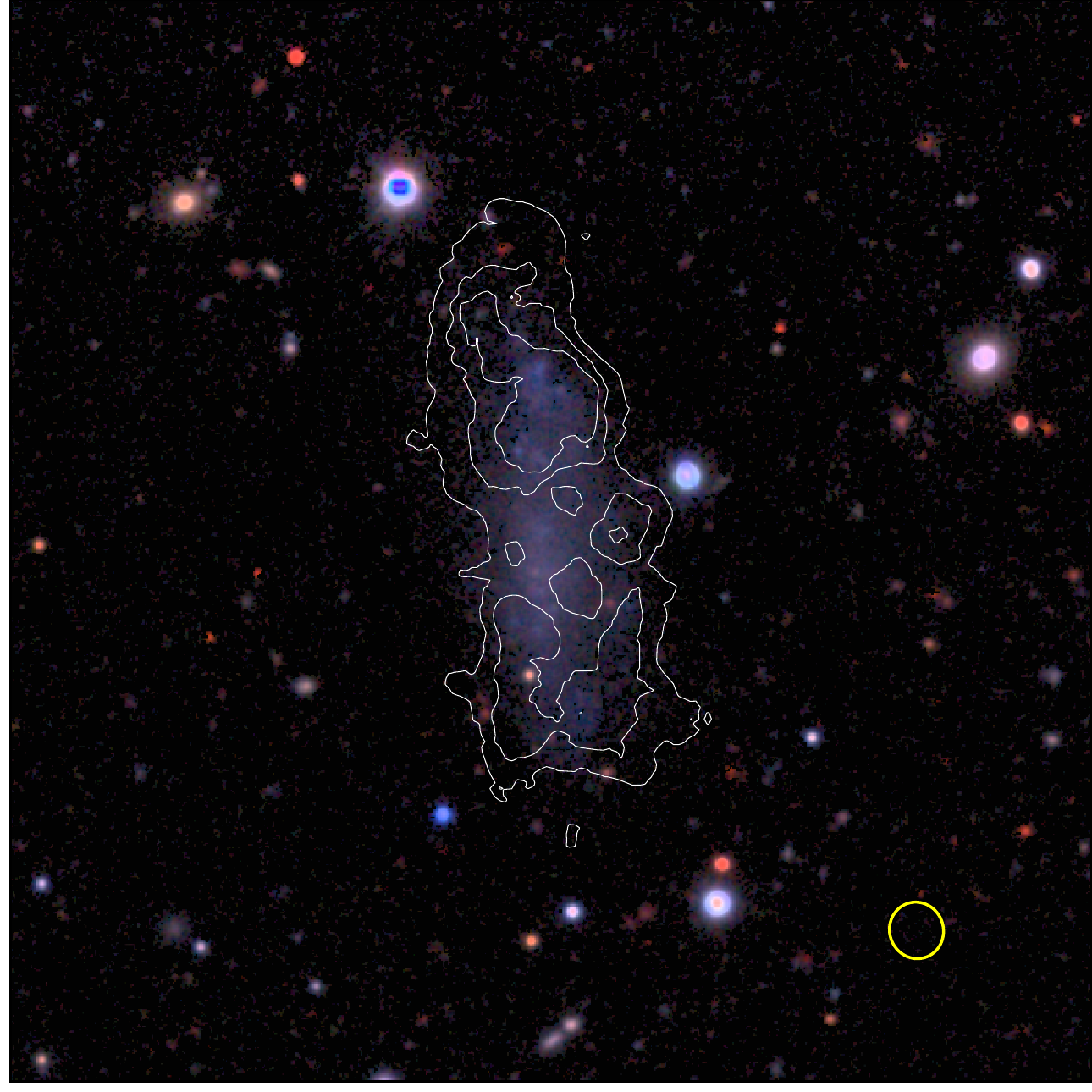}
    \caption{\label{false_color} {\bf  False-color
   image    of  AGC  242019.}  The false-color image of the $g$, $r$
      and $z$ bands. The white contours
of  the  \HI\,   intensity  are  at  levels  of  0.4,  0.6 and 0.8
mJy~\kms.  The  yellow  ellipse indicates  the  \HI\,  beam
size.   }
\end{center}
\end{figure}

A definitive cuspy  profile from observations will  prove the validity
of  the  cold  dark  matter   paradigm  on  subgalactic  scales  while
challenging   other    types   of   dark   matter.     As   shown   in
Figure~\ref{false_color},   the   object   AGC   242019   is   an
ultra-diffuse  galaxy (UDG) identified  by  the Arecibo  Legacy  Fast  ALFA
(ALFALFA) survey of \HI\, galaxies \citep{Leisman17}.  This galaxy has
a stellar mass  of \mstar\,, a HI  mass of \mhi\, and  a star formation
rate  of \sfr\, as listed in Table~\ref{tab_prop}.   Its receding  velocity is  2237$\pm$25 km  s$^{-1}$
after  correcting   for  the   Virgo,  Great  Attractor   and  Shapley
supercluster \citep{Mould00}.  In the  cosmological frame of $h$=0.73,
$\Omega_{0}$=0.27   and  $\Omega_{\Lambda}$=0.73,   the  corresponding
distance is 30.8 Mpc and the uncertainty is estimated to be 5\%, given
that it is an isolated field galaxy (see \S~\ref{sec_sysunc_curve}).

\begin{figure}[tbh]
  \begin{center}
 \includegraphics[scale=0.38]{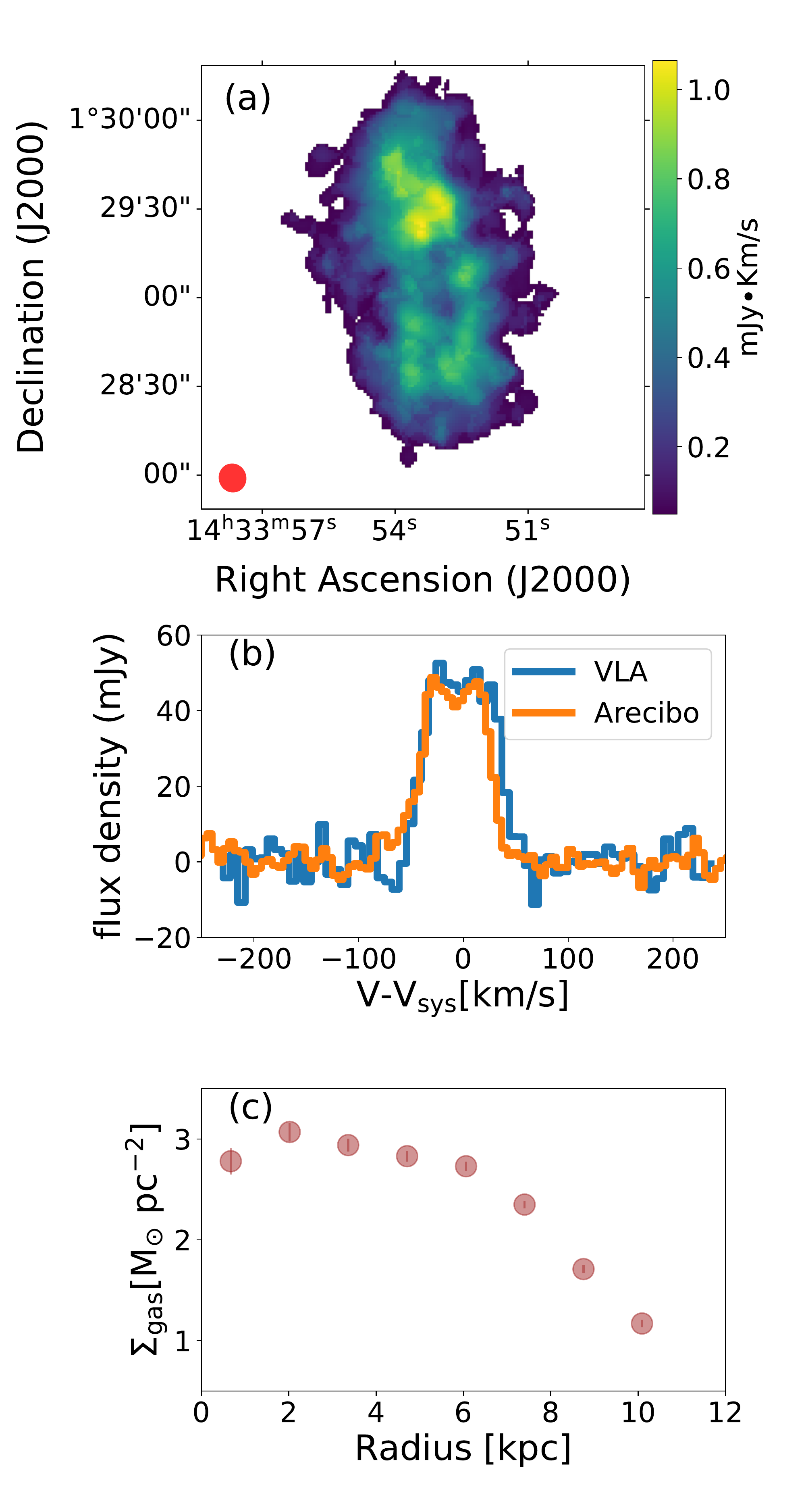}
 \caption{\label{HI_property} {\bf The HI data of AGC 242019.} {\bf a,} The \HI\, intensity.  {\bf b,} The integrated spectra of the VLA compared to the Arecibo's spectrum \citep{Haynes18}. {\bf c,}  The  radial  profile of  the  gas mass  surface
density  corrected  for inclination  and  helium.  }
\end{center}
\end{figure}

\begin{figure*}[tbh]
  \begin{center}
    \includegraphics[scale=1.0]{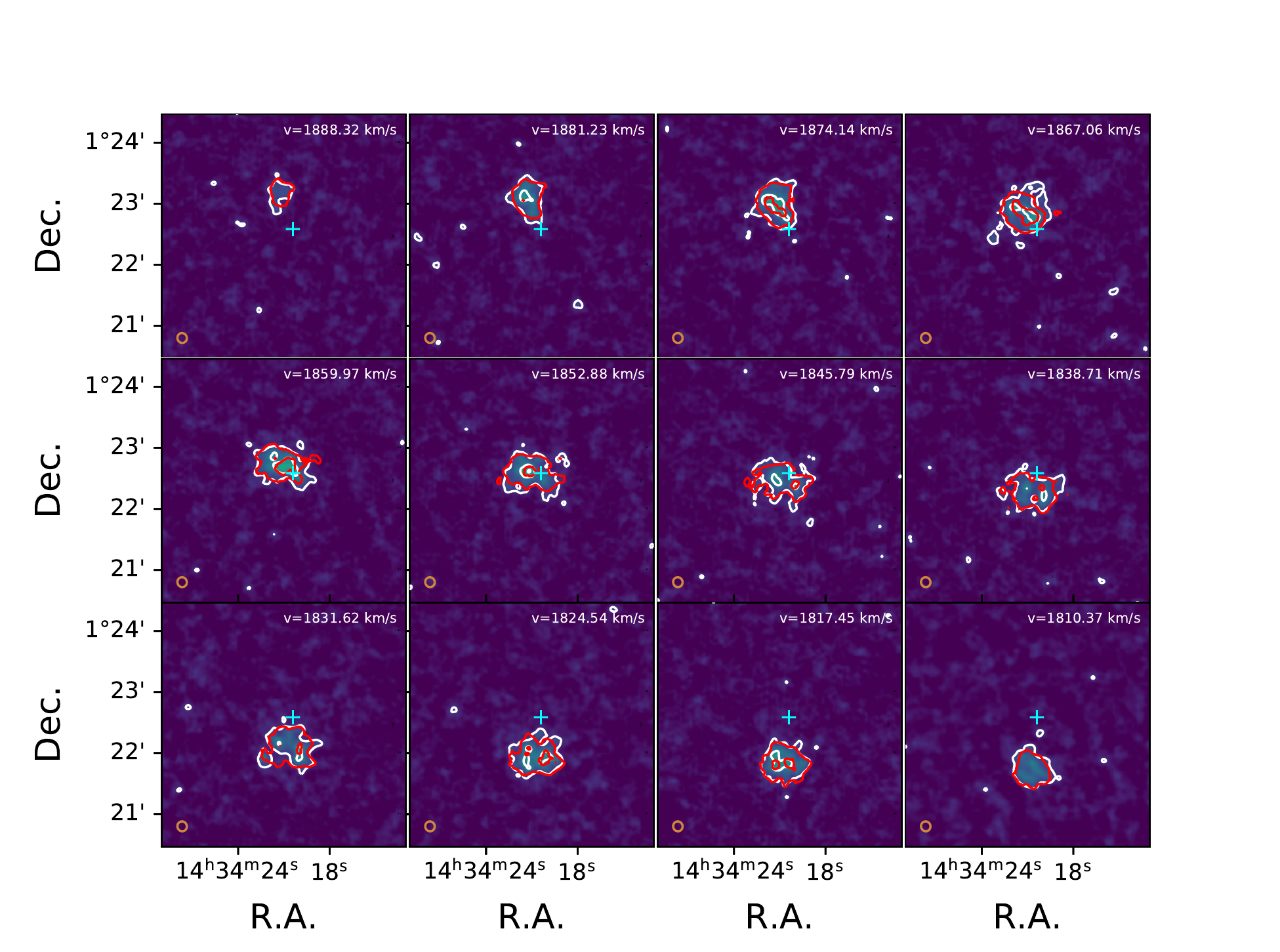}
    \caption{\label{HI_chanmap} {\bf The channel map of the \HI\, data}.
      The contour levels are 1, 3, 4.5 and 6 mJy \kms. The white and red contours
    indicate the observations and best-fitted models, respectively. }
\end{center}
\end{figure*}

\begin{figure*}[tbh]
  \begin{center}
    \includegraphics[scale=0.33]{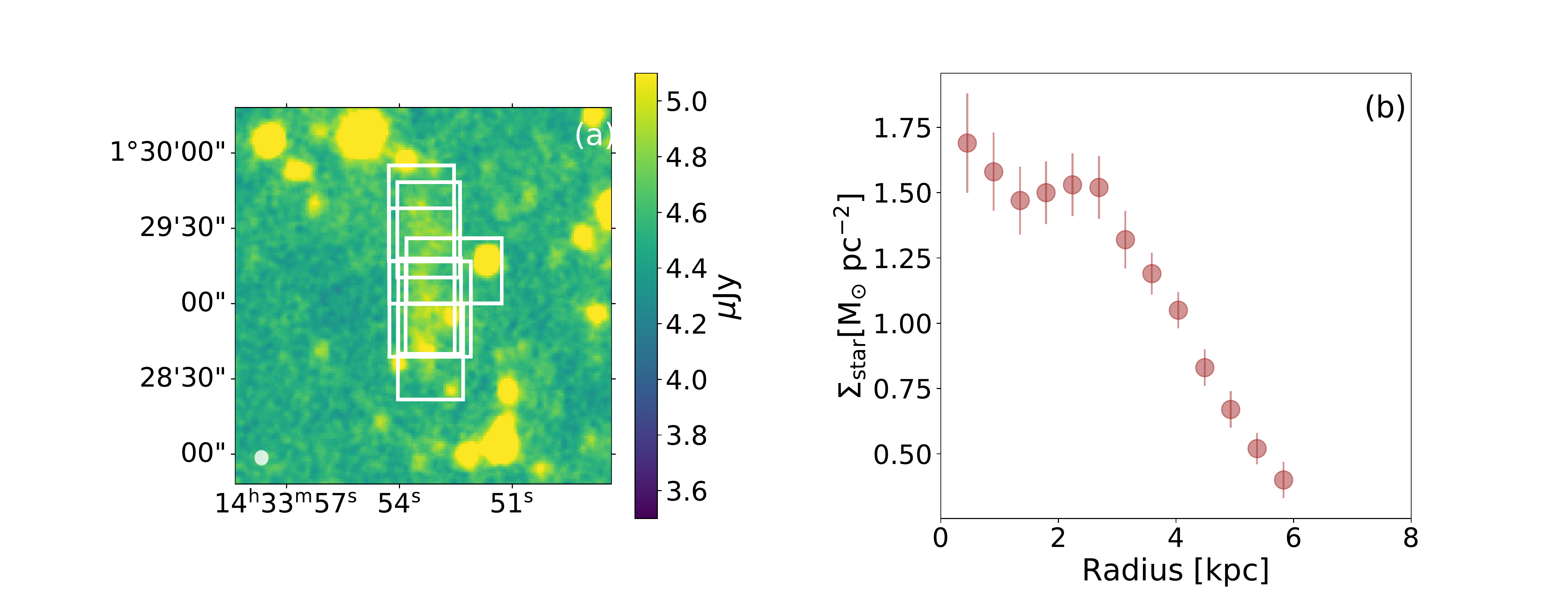}
    \includegraphics[scale=0.33]{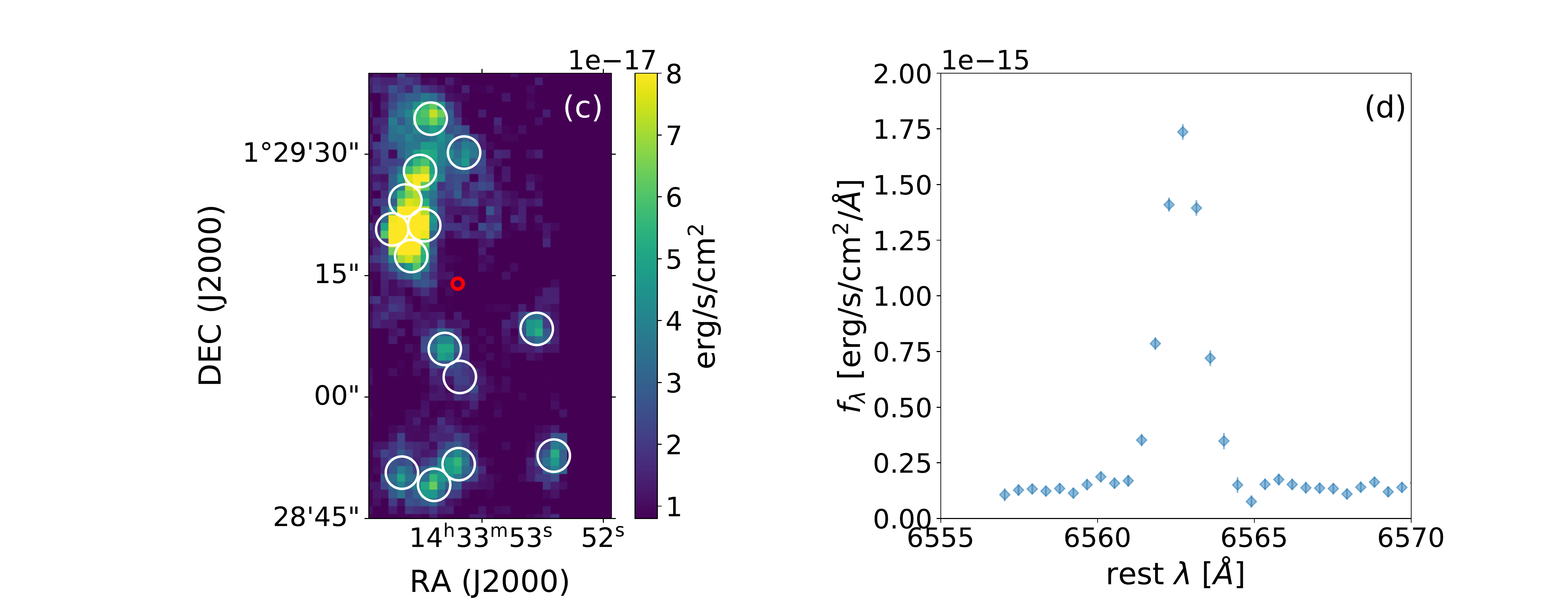}
 \caption{\label{NIROT_property} {\bf The infrared and optical data of
     AGC 242019}. {\bf  a,} The 3.6 $\mu$m flux  density with overlaid
   WiFeS IFU  pointings.  {\bf b,}  The radial profile of  the stellar
   mass  surface  density  corrected  for the  inclination.  {\bf  c,}
   Individual H$\alpha$ clumps for which  the line of sight velocities
   are measured. Each clump has a circular radius of 2.0$\arcsec$. The
   small  red circle  indicates  the dynamical  center.  {\bf d,}  The
   integrated spectrum of H$\alpha$ emissions.  }
\end{center}
\end{figure*}

\begin{figure}[tbh]
  \begin{center}
    \hspace*{-0.7cm}\includegraphics[scale=0.34]{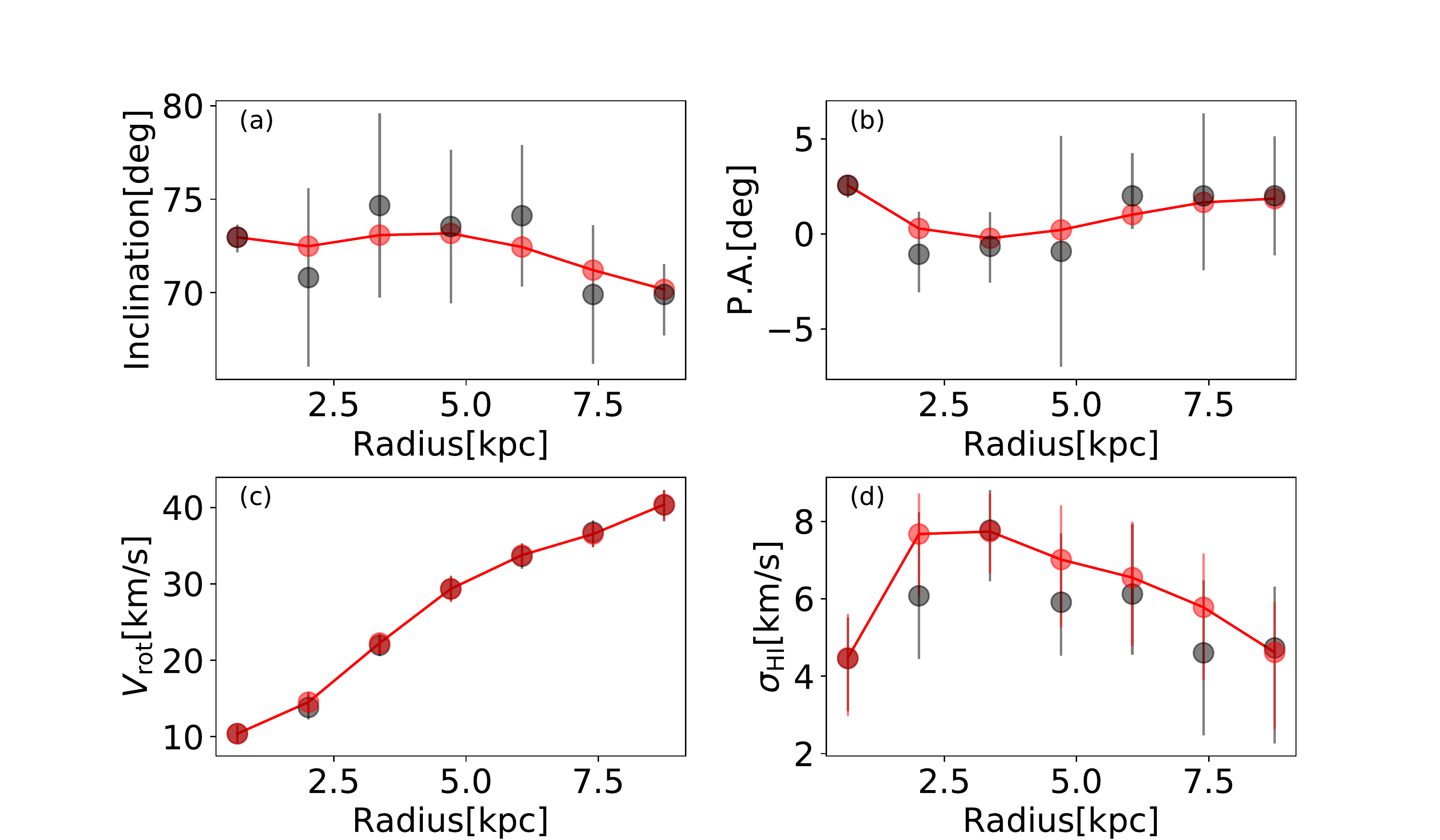}
 \caption{\label{parameter_freeVVPI}{\bf The two-stage operation of \texttt{$^{3D}$Barolo}}. The gray points
 are the first stage, where four parameters of each ring are set as free. The red points are the
 second stage, where the inclination and position angles are regularized by fitting a Bezier function to the
 results from the first stage. }
\end{center}
\end{figure}

\section{Observations and data reduction}

\subsection{Radio interferometric observation of \HI\, gas with the VLA}

Our L-band observations were performed with the Karl G. Jansky Very Large Array
(VLA) through two projects corresponding to the D configuration (19B-072; PI:
Y. Shi) and to the C and B configurations (20A-004; PI: Y. Shi). The
D-configuration observations were performed on \nth{21} and \nth{23} Sep. 2019,
each with a 1.5-hr observing time and a $\sim$ 66 min on-source time. A total
of 27 antennas were employed in both executions, with the first under overcast
weather conditions and the second under clear weather conditions. The projected
baselines of the D-configuration array are in the range of 34--1050 m. The flux
and bandpass calibrator was 3C~295, and the gain calibrator was J\,1419+0628.
We configured the spectrometers with a mixed setup. The spectral line window
that covers the H{\sc i} 21-cm line has a bandwidth of 128 MHz and a channel
width of 32.3 kHz ($\sim$ 6.9 \kms), while the remaining windows cover a
frequency range between 963.0 MHz and 2017.0 MHz to optimize the sensitivity
for the radio continuum.

The C-configuration observations were performed on \nth{04}, \nth{09} and
\nth{13} Feb. 2020, each with a 2-hr observing time and a $\sim$ 90 min
on-source time. Two observations were conducted with 25 antennas, while the
rest were observed with 26 antennas, mostly under overcast or clear weather
conditions. The projected baselines of the C configuration are in the range of
40--3200 m. In the C-configuration observations, the 21-cm emission line was
observed with a spectral line widow that has a channel width of 5.682 kHz
($\sim$ 1.2 \kms) and a bandwidth of 32 MHz. The remaining windows cover a
frequency range between 963.0 MHz and 2017.0 MHz, for optimizing the radio
continuum bandwidth. The flux and bandpass calibrator was 3C~286 and the gain
calibrator was J\,1419+0628.

The B-configuration observation was performed with five executions during July
and August of 2020, each with a 2-hr observing time and a $\sim$ 90 min
on-source time. In most executions, 27 antennas were employed under cloudy
conditions. The projected baselines of the B-configuration are in the range of
230--11000 m. The hardware setup and the calibrators were the same as those
used for the C-configuration observations.

We reduced all data manually with the Common Astronomy Software Applications
(CASA) package \citep{McMullin07}, {\sc v5.6.1}. Both D-configuration
observations were severely affected by radio frequency interferences (RFI).
Therefore we calibrated and flagged the data manually. Approximately 20\% of
the data have to be flagged, across different frequency ranges. The data
obtained on \nth{21} Sep. 2019, including the calibrators, has about three
times higher noise than that estimated from the VLA sensitivity estimator, due
to unknown reasons. However, the fluxes, line profile, and spatial
distributions are highly consistent with the data obtained on \nth{23} Sep.
2019. Whether combine it to the final data does not change the results. All
gain calibrators were checked carefully to ensure a flat bandpass, a
point-source distribution and excellent calibration solutions. The 1.4 GHz flux
densities of 3C~286, 3C~295 and J\,1419+0628 were 15.1$\pm$0.1 Jy, 22.3$\pm$0.1
Jy, and 6.0$\pm$0.1 Jy.

After flagging and calibration, we weighted all dataset with their noise levels
using the \texttt{statwt} task. Then, through the \texttt{uvcontsub} task, we
fitted and subtracted the radio continuum from the visibility data, with
line-free channels on both sides of \HI, using the first-order linear function.
We resampled the C- and B-array data to match with the spectral resolution of
the D-array data, using the \texttt{mstransform} task. The final spectrally
matched line-only data from the D, C, and B configurations were combined
together with the \texttt{concat} task.

In the end, we inverted the visibility data to the image plane and cleaned the
data cube with Briggs' robust parameter of 2.0 using the \texttt{tclean} task.
 The final D+C+B datacube has
a velocity coverage of 500 \kms\, and a channel width of 32 kHz, corresponding
to a velocity resolution of $\sim$ 7 \kms. The r.m.s. noise level reaches 0.26
mJy\,beam$^{-1}$ per channel with a restoring synthesis beam size of 9.85$''$
$\times$9.33$''$ and a position angle of 17.56$^{\circ}$.

The \HI\, intensity map  is shown in Figure~\ref{HI_property}~(a). The
integrated  spectrum  of  the  B+C+D   configuration  has  a  flux  of
3.8$\pm$0.16  \,Jy\,\kms,   which  is   comparable  to  the   flux  of
3.4\,Jy\,\kms\, obtained with the  Arecibo \citep{Leisman17}, as shown
in Figure~\ref{HI_property}~(b).  The  total HI gas mass  is \mhi\,. A
slightly higher flux as seen by VLA in the red wing may be caused by a
small offset of the Arecibo beam from the galaxy center, given a large
size of  the galaxy that is  two arcmins in the  diameter.  The radial
profile   of   the   gas   mass    surface   density   is   shown   in
Figure~\ref{HI_property}~(c).    The   channel   map   is   shown   in
Figure~\ref{HI_chanmap}.

We also tested tilted-ring modeling with the combined data using the B+C
configurations and found essentially no difference, except for a higher noise
level. Though the better velocity resolution leads to better estimates of the
pressure support, which has a very minor contribution to the decomposition of
the rotation curve.

\subsection{Broad-band images}
 
The  co-added images  at 3.6  and 4.5  $\mu$m were  obtained from  the
Wide-field  Infrared Survey  Explorer (WISE)  archive \citep{Wright10}
with spatial  resolutions of  6.0$''$ and 6.8$''$,  respectively.  The
target   is   well   detected   at  3.6   $\mu$m   as   presented   in
Figure~\ref{NIROT_property}~(a), but shows almost  no detection at 4.5
$\mu$m.   To derive  the radial  profile of  the stellar  mass surface
density as presented in  Figure~\ref{NIROT_property}~(b), a few nearby
bright  stars in  the field  were subtracted  using the  stellar point
spread
functions\footnote{http://wise2.ipac.caltech.edu/docs/release/allsky/expsup/sec4\_4c.html}. The
stellar  mass  based on  the  3.6  $\mu$m  image  is estimated  to  be
\mstar\, for a Kroupa stellar initial
mass  function and  a  mass-to-light ratio  $\Upsilon_{3.6{\mu}m}$=0.6
($M_{\odot}/L_{\odot, 3.6{\mu}m}$) (see below).

The optical  $g$ and  $r$ images  were obtained  from the  Dark Energy
Camera Legacy  Surveys \citep{Dey19}.   The far-ultraviolet  image was
retrieved  from  the GALEX  data  archive.   The integrated  flux  was
converted to  a star formation rate  of \sfr\,   for   a   Kroupa  stellar   initial   mass   function
\citep{Leroy08}.

\subsection{Integral field unit observation of H$\alpha$ by ANU 2.3 m }

Integral field  unit observations of  H$\alpha$ were carried  out with
the Wide-Field  Spectrograph (WiFeS)  onboard the  Australian National
University 2.3  m telescope on  the nights of  \nth{21}-\nth{22} Mar.,
\nth{26}  May  and  \nth{20}-\nth{22}   Jul.,  2020.  An  R7000  grism
(5290-7060${\rm \AA}$)  at a resolution  of 7000 was adopted  to cover
the  H$\alpha$  emission   line.   WiFeS  has  a  field   of  view  of
25$\arcsec$$\times$38$\arcsec$.           As         shown          in
Figure~\ref{NIROT_property}~(a),  exposures  were   taken  at  several
positions to  cover the whole optical  extent of the galaxy,  with one
pointing  toward a  nearby  bright star  for astrometric  calibration.
Each exposure  was 30 mins,  and the total on-source  integration time
varied from 2.0 hrs to 3.5 hrs  at each pixel.  For every 1--2 hrs, an
off-target blank  sky and a  standard star were observed.   The seeing
was  between  1.5$''$  and  2.0$''$.  Each  frame  was  first  reduced
following     the    standard     procedure    by     \texttt{pyWiFeS}
\citep{Childress14}, and  then was subtracted  by the median  value of
the sky frame at each wavelength.  Individual frames were aligned with
each other  to produce the final  mosaic image using the  positions of
the H$\alpha$  clumps, as the  continuum emission was too  faint.  The
absolute astrometry was obtained  through alignment with the brightest
star in  the mosaic  field of  view.  Since  H$\alpha$ clumps  are not
perfect point sources, we  estimated the final astrometric uncertainty
to be about 1$''$.  To correct the barycentric velocity offset and any
possible intrinsic instrumental shift, the wavelength solution of each
night was  further cross-calibrated  based on  H$\alpha$ lines  of the
same  clumps observed  during different  nights. The  integrated \ha\,
flux  map is  presented  in  Figure~\ref{NIROT_property}~(c) with  the
integrated spectrum shown in Figure~\ref{NIROT_property}~(d).

\section{Data Analysis}

\begin{figure*}[t]
  \begin{center}
    \includegraphics[scale=0.6]{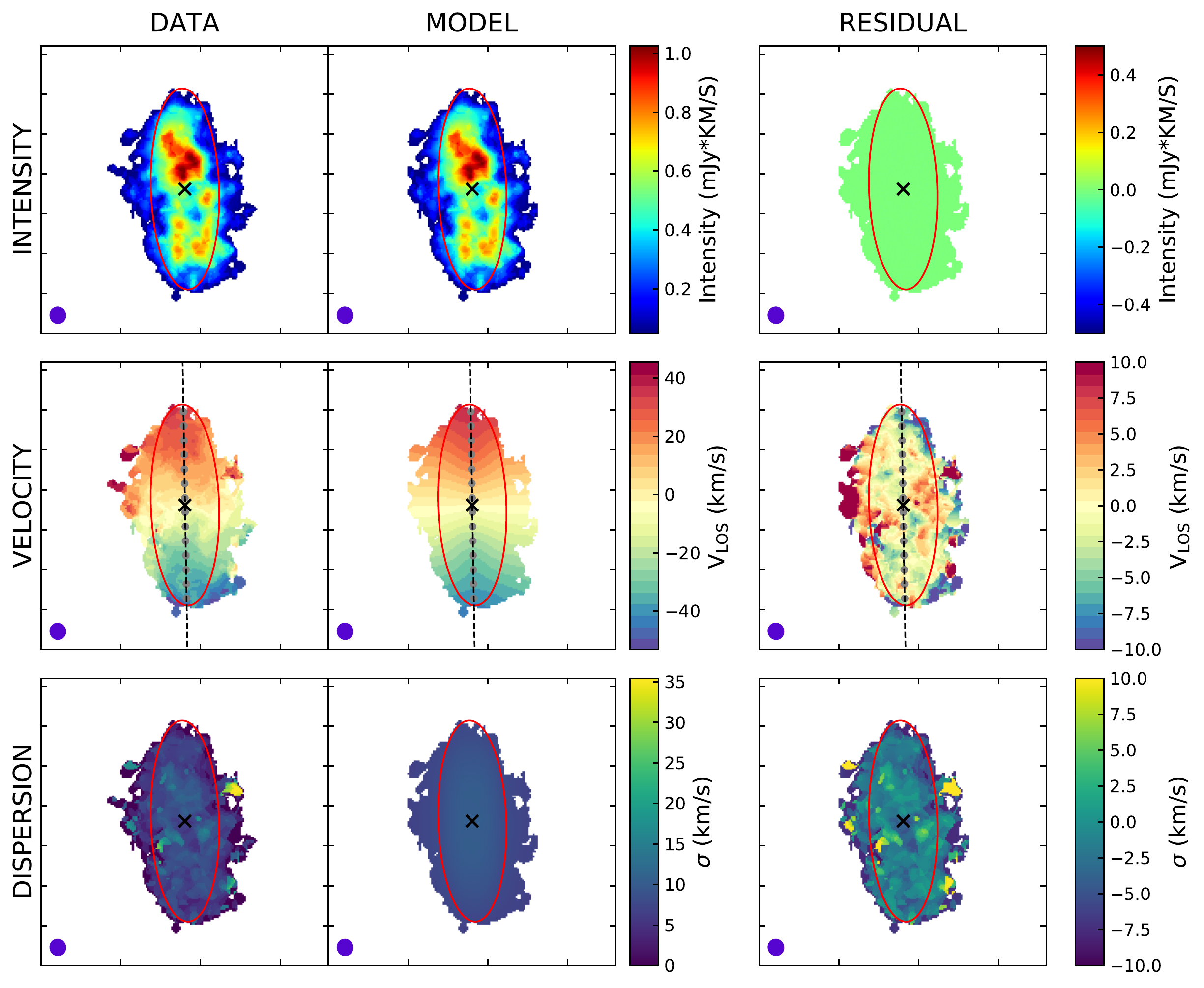}
    \caption{\label{bbarolo_run} {\bf The moment maps and 
        tilted-ring  modeling achieved through \texttt{$^{3D}$Barolo}.} {\bf First row:} the moment-0 map,
      best-fitting   model and residual for the \HI\, intensity. Note that the run of \texttt{$^{3D}$Barolo} adopted the setting
      NORM=LOCAL which normalizes the model's flux to the observed one pixel by pixel. As a result, the flux residual
      is zero. {\bf Second row:} the moment-1 map,
 best-fitting   model and residual for the velocity.  {\bf Third row:} the moment-2 map,
 best-fitting   model and residual for the velocity dispersion. In all panels, the filled ellipse on the left lower corner
    indicates the beam, while the large red ellipse marks the outer boundary of the last ring. } 
\end{center}
\end{figure*}

\begin{figure}[t]
  \begin{center}
    \includegraphics[scale=0.35]{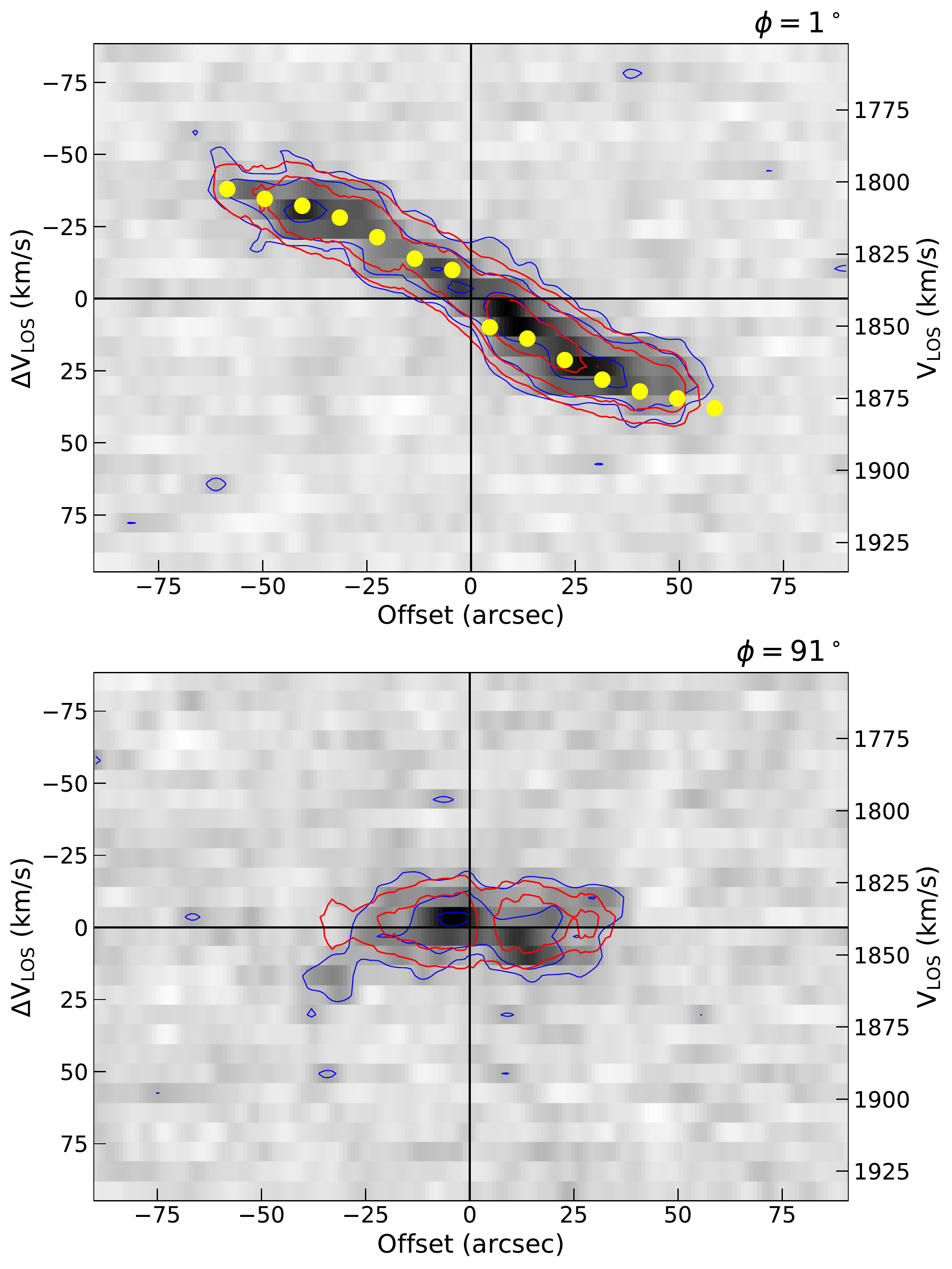}
    \caption{\label{pv_diagram} {\bf The position-velocity diagram of the galaxy along
        the major and minor axes of the H{\sc i} gas disk.} The blue and red contours are the
        observation and best-fitted model, respectively. The yellow symbol is the derived
        line-of-sight velocity of each ring after correcting the beam smearing. The contour levels are
    0.73, 1.44, 2.88 and 5.77 mJy \kms.}  
\end{center}
\end{figure}

\begin{table*}
\footnotesize
\caption{\label{tab_prop} The Properties of AGC 242019}
\begin{tabular}{llllllllllllll}
\hline
Parameters                             &   Mean                             & 1-$\sigma$ error & 3\% HPD$^{\P}$  & 97\% HPD$^{\P}$   \\
\hline 
$M_{\rm star}$  (10$^{8}M_{\odot}$)       &   1.37                              & 0.05              &    -           &   - \\
$M_{\rm HI}$  (10$^{8}M_{\odot}$)         &   8.51                              & 0.36              &    -           &   - \\
SFR         (10$^{-3}$M$_{\odot}$/yr)    &   8.2                               &  0.4              &    -           &   - \\
Dynamical center$^{\ast}$ (J2000)       &  14:33:53.38, 01:29:12.5    & 5.0$''$, 2.2$''$  &    -        & -     \\
$V_{\rm sys}^{\ast}$ (\kms)               &  1840.4                             & 1.9               & -          & -   \\
Distance         (Mpc)                  &   30.8                              & 5\%                & -          & -       \\
log($\Upsilon_{3.6{\mu}m}$ ($M_{\odot}/L_{\odot, 3.6{{\mu}m}}$))    &    -0.22    & 0.1   & -  & -    \\
$R_{200}$ (NFW) (kpc)                   &   65.0               &  7.4        &  54.6       & 74.3       \\
$R_{\rm S}$ (NFW) (kpc)                 &   33.3               &  9.1        &  17.4       & 50.6       \\
concentration (NFW)                    &   2.0                & 0.36        &   1.45      &  2.72      \\
$M_{\rm halo}$ (NFW) (10$^{10}M_{\odot}$)  &   3.5                & 1.2         &  1.5       &   5.8     \\
$V_{\rm norm}$ (ISO) (\kms)             &   16.7                & 1.9         & 13.1        &  20.3      \\
$R_{\rm C}$ (ISO)    (kpc)              &   2.5             &     0.5     &  1.6       &  3.3      \\             
\hline
\end{tabular}
\tablecomments{$^{\ast}$The dynamical center and systematic velocity are obtained by setting them as free parameters when running
  \texttt{$^{3D}$Barolo}. $^{\P}$HPD refers to the Highest Posterior Density. }
\end{table*}

\begin{table*}
\footnotesize
\begin{center}
\caption{\label{tab_ring_result} The results of the tilted-ring modeling}
\begin{tabular}{llllllllllllll}
\hline
Rad  &  $V^{\rm rot}_{\rm obs}$    & $V^{\rm circ}_{\rm obs}$   &  $\sigma^{\rm disp}$ &  $V^{\rm circ}_{\rm gas}$&  $V^{\rm circ}_{\rm star}$ &  $V^{\rm circ}_{\rm dark{\textendash}matter}$  & P.A.      & Inclination \\
\hline
(kpc) &  (\kms)                & (\kms)                 & (\kms)    & (\kms)     & (\kms)     & (\kms)          & ($^{\circ}$)  & ($^{\circ}$) \\
\hline

\multicolumn{7}{c}{H{\sc I}} \\
\hline
0.67 & 10.4$\pm$1.2 & 10.6$\pm$1.2 & 4.5$\pm$1.3 & 0.2$\pm$1.2 & 1.9$\pm$0.5 & 10.4$\pm$1.2 & 2.6 & 73.0 \\
2.02 & 14.5$\pm$1.3 & 15.5$\pm$1.2 & 7.7$\pm$1.3 & 4.1$\pm$0.4 & 4.5$\pm$0.3 & 14.2$\pm$1.4 & 0.3 & 72.5 \\
3.36 & 22.3$\pm$1.4 & 23.5$\pm$1.3 & 7.7$\pm$1.0 & 8.2$\pm$0.3 & 8.3$\pm$0.4 & 20.4$\pm$1.5 & -0.2 & 73.1 \\
4.71 & 29.4$\pm$1.8 & 30.7$\pm$1.7 & 7.0$\pm$1.6 & 11.3$\pm$0.3 & 10.4$\pm$0.5 & 26.6$\pm$1.9 & 0.2 & 73.2 \\
6.06 & 33.8$\pm$1.6 & 35.0$\pm$1.5 & 6.5$\pm$1.6 & 15.0$\pm$0.3 & 10.6$\pm$0.3 & 29.9$\pm$1.8 & 1.0 & 72.4 \\
7.40 & 36.5$\pm$1.6 & 37.5$\pm$1.6 & 5.8$\pm$1.6 & 19.1$\pm$0.4 & 10.1$\pm$0.3 & 30.6$\pm$1.9 & 1.7 & 71.2 \\
8.75 & 40.4$\pm$2.0 & 40.9$\pm$2.0 & 4.6$\pm$1.6 & 22.4$\pm$0.5 & 9.6$\pm$0.4 & 32.8$\pm$2.5 & 1.9 & 70.2 \\
\hline
\multicolumn{7}{c}{\ha} \\
\hline
1.51 &   & 16.8$\pm$1.6 &   & 2.5$\pm$0.7 & 3.5$\pm$0.4 & 16.2$\pm$1.7 &  &  \\
2.78 &   & 22.2$\pm$1.3 &   & 6.7$\pm$0.3 & 6.9$\pm$0.4 & 20.0$\pm$1.4 &  &  \\
3.45 &   & 24.2$\pm$0.8 &   & 8.4$\pm$0.3 & 8.4$\pm$0.4 & 21.0$\pm$1.0 &  &  \\
4.54 &   & 29.6$\pm$2.2 &   & 10.9$\pm$0.3 & 10.3$\pm$0.5 & 25.6$\pm$2.5 &  &  \\
\hline
\end{tabular}
\end{center}
\end{table*}

\subsection{The derivation of the rotation curve of dark matter}

We derived the rotation curve of the dark matter by first obtaining the
total  rotation  curve  from  the  \HI\,  and  H$\alpha$  data  with a
correction   for  the   pressure  support,   and  then   quadratically
subtracting the gas and stellar gravity contributions as detailed below.

\subsubsection{The observed rotation curve from titled-ring fitting to the \HI\, datacube}

We  fitted  the tilted-ring  model  to  the  \HI\, 3-D  datacube  with
\texttt{$^{3D}$Barolo}  \citep{DiTeodoro15}  to  obtain  the  rotation
curve as listed in Table~\ref{tab_ring_result}. The radial width of each ring is set to be 9$''$, which is
roughly the beam size so that each ring is independent. We adopted uniform
weighting (WFUNC=0) and least-squared minimization (FTYPE=1). All other
optional parameters are set as default. The full list of the set-up is
included in Table~\ref{tab_full_par}.

The angular resolution and  the signal-to-noise ratio were good
enough  to set  the rotation  velocity, the  velocity dispersion,  the
position angle and  the inclination angle as free  parameters for each
ring. Before running this setup, we  first ran a model by also setting
the dynamic center and the systematic velocity as free parameters, and
then  fixed  them  to the  mean  values  of  all  rings as  listed  in
Table~\ref{tab_prop}.

The scale  height was fixed to  100 pc, independent of  the radius. If
the height is  varied by a factor of five  (see below), the conclusion
remains  unchanged. We  do notice  that \texttt{$^{3D}$Barolo}  is not
able to remove effects  fully due to the disk height  for a thick disk
\citep{Iorio17}.  When running  \texttt{$^{3D}$Barolo}, we adopted the
``twostage'' fitting method, which allows a second fitting stage after
regularizing   the   first-stage   parameter  sets.    As   shown   in
Figure~\ref{parameter_freeVVPI},   the   radial    profiles   of   the
inclination  and  position angles  from  the  first-stage fitting  are
regularized  by  fitting  a  Bezier   function,  based  on  which  the
second-stage fitting is performed.  The errors of the derived rotation
velocity   and    velocity   dispersion    in   \texttt{$^{3D}$Barolo}
\citep{DiTeodoro15} are  estimated through  a Monte Carlo  method.  We
also ran \texttt{$^{3D}$Barolo} with only the receding and approaching
sides,  respectively.   The  velocities  from  three  runs  are  within
1-$\sigma$  errors, while  the  velocity dispersions  are also  within
errors  except   for  the   innermost  radius,   which  is   shown  in
Figure~\ref{comp_vel_reced_approach}   of   Appendix.  As   shown   in
Figure~\ref{parameter_freeVVPI},  the  derived rotation  velocity  and
velocity dispersion vary well within  the 1-$\sigma$ error bars before
and after regularization.

As shown in Figure~\ref{bbarolo_run}, within the outer boundary of the
last ring,  the residual of  the \HI\,  intensity is dominated  by the
observed flux  noise, and the  residuals of the velocity  and velocity
dispersions show small  amplitudes with medians of 1.9  \kms\, and 1.8
\kms,   respectively.   Figure~\ref{pv_diagram}   further  shows   the
position-velocity diagram along the major and minor axes. Overall, the
observation matches the best-fitted model well.

As gas  is collisional, pressure  support is  a
driver  of gas  motion in addition to  gravity \citep{Bureau02, Oh15, Iorio17, Pineda17}.
A gas disk in equilibrium satisfies:
\begin{linenomath*}
\begin{equation}
V_{\rm circ}^{2} = V_{\rm rot}^{2} + V_{\rm P}^{2},
\end{equation}
\end{linenomath*}
where the circular velocity $V_{\rm circ}$=$\sqrt{R\frac{\partial{\Phi}}{\partial{R}}}$ reflects solely
the effect of the gravitational potential,  $V_{\rm rot}$ is the observed rotation velocity and
$V_{\rm P}$ is the velocity driven by the pressure. $V_{\rm P}$ is related to the gas  density ($\rho$)
and the velocity dispersion ($\sigma_{\rm v}$) following
\begin{linenomath*}
\begin{equation} 
V_{\rm P}^{2}= -R\sigma_{\rm v}^{2}\frac{\partial {\rm ln}({\rho}{\sigma_{\rm v}^{2}})}{\partial R}.
\end{equation}
\end{linenomath*}

Assuming that the scale height is independent of the radius and that the disk is thin, the above
equation becomes
\begin{linenomath*}
\begin{equation}
V_{\rm P}^{2}=-R\sigma_{\rm v}^{2}\frac{\partial {\rm ln}(\sigma_{\rm v}^{2}\Sigma_{\rm obs}{\rm cos} i) }{\partial R},
\end{equation}
\end{linenomath*}
where $\Sigma_{\rm obs}$ is the observed gas mass surface density, and $i$ is the inclination angle.
Following  Equation 3, the pressure support correction is implemented in
\texttt{$^{3D}$Barolo} \citep{Iorio17}.
For AGC 242019, the correction is small, with $|V_{\rm circ} - V_{\rm rot}|$ being
smaller than $\sim$1.0 \kms\, across the
radius. In this case, only the error of $V_{\rm rot}$ is propagated to $V_{\rm circ}$ while ignoring the
error of $V_{\rm A}$. The radial profile of $V_{\rm circ}$ is referred to
as the observed rotation curve.

Note that  our \HI\, data  have a spectral resolution  of 7 \kms\, which is
comparable to or even lower than the derived velocity dispersion shown in
 Figure~\ref{parameter_freeVVPI}. To  check the effect of
the  limited spectral  resolution, we  also performed  tilted-ring
fitting on the  \HI\, datacube from the $B$  and $C$ configurations that
has a spectral resolution of 2.4 \kms. The difference
in the velocity  dispersion between the two data  is within the 1-$\sigma$
error, and the  difference in the derived rotation curve
is within  the 1-$\sigma$
error too.


\subsubsection{The observed rotation curve from the H$\alpha$ map}

The H$\alpha$ emission  is clumpy and sporadic across the  disk.  As a
result,  tilted-ring modeling  cannot be  performed for  the H$\alpha$
data in the same  way as for the \HI\, data.   Instead, we adopted the
same  ring  parameters  from  the \HI\,  data,  including  the  center
position,  inclination  angle,  and  position angle,  to  convert  the
line-of-sight  velocities  into  the  rotation  curve.   As  shown  in
 Figure~\ref{NIROT_property}  (c),   we  used  a  circular
aperture  with a  radius  of  2$''$ to  extract  the  spectrum of  the
H$\alpha$ region and measured  the line-of-sight velocity, whose error
was estimated  through a Monte  Carlo method, by randomly  inserting a
Gaussian error at each data point and iterating one thousand times.

The  velocity was  then converted  into  the rotation  curve with  the
following steps.  1)  We first correct for the  absolute wavelength of
the optical IFU data by comparing  the H$\alpha$ velocity to the \HI\,
velocity at the same position.   2) The line-of-sight velocity is then
converted  into   the  rotation  velocity  by   $V_{\rm  rot}$=$V_{\rm
  l.o.s}$/sin($i$)/cos($\theta$), where  $i$ is the  inclination angle
and $\theta$ is  the azimuthal angle from the major  axis in the plane
of the  un-projected disk.  Here,  the inclination angle  and position
angle  are  also interpolated  based  on  the  result from  the  \HI\,
fitting.  3) The dynamical center from  the \HI\, ring is adopted.  4)
Individual velocities are then rebinned at  a width of 1.0 kpc to have
enough points to  obtain the mean value. Except for  the outermost bin
where  only one  datapoint  is available,  the error  of  the mean  is
obtained through the Monte Carlo method, which fits a mean value after
it randomly inserts  a Gaussian error at each data  point and iterates
one thousand times.  To obtain the pressure support correction for the
H$\alpha$, we assumed that the radial  shape of the H$\alpha$ gas mass
surface density and velocity dispersions are similar to those of \HI\,
\citep{Adams14}.   The velocity  dispersions of  the H$\alpha$  clumps
after correcting  for the thermal  broadening (10 \kms) are  about the
same as that  of \HI\, at similar locations.  As  a result, we adopted
the  $V_{\rm P}$  from the  \HI\, data  to correct  for the  H$\alpha$
rotation velocity.

Since our \ha\, emission is sporadic and the errors of the rotation
curve are only based on those detected regions, we used the H$\alpha$
rotation curve as a sanity check of the \HI\, curve.

\begin{figure}[t]
  \begin{center}
    \includegraphics[scale=0.4]{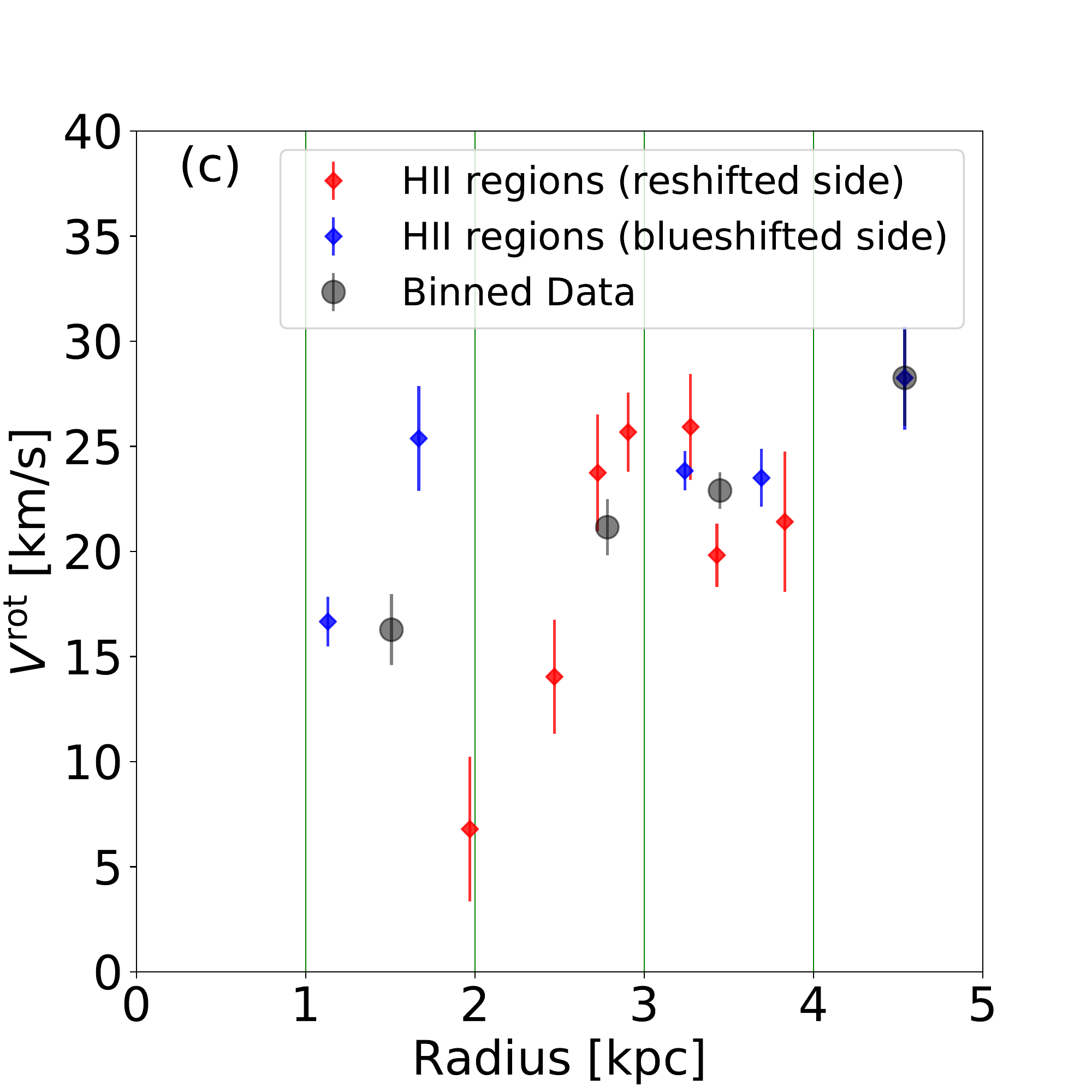}
 \caption{\label{Halpha_data} The rotation
 velocities from the line of sight velocities based on the \HI\,'s ring parameters. These
 velocities are further rebinned with a radial bin of 1.0 kpc. }
\end{center}
\end{figure}

\subsubsection{The contribution to the rotation curve from the gravity of gas}\label{sec_grav_gas}

The radial  profile of the gas  mass surface density derived  from the
tilted-ring modeling  is shown in  Figure~\ref{HI_property}~(c). We used this profile to estimate
the   gas    rotation   curve    using   the    \texttt{ROTMOD}   task
\citep{Begeman89}  in  the  GIPSY  package.  The  gas  mass  has  been
corrected for helium by multiplying it  by a factor of 1.36.   The vertical
distribution  of the  gas disk  is  assumed to  be
``SECH-SQUARED'' with a  scale height of 0.1 kpc.  By  varying the gas
mass surface density with its error, we used the Monte-Carlo method to
estimate the  uncertainty of  the rotation curve.  We then  varied the
scale height from  0.02 kpc to 0.5  kpc to quantify the  effect of the
scale-height  uncertainty on  the rotation  curve.  The  two types  of
errors were summed quadratically to get the final error.

\subsubsection{The contribution to the rotation curve from the gravity of  stars}\label{sec_grav_star}


To estimate the stellar mass distribution, we first derived the radial
profile  of the  3.6  $\mu$m  surface brightness  at  a  bin width  of
3$\arcsec$,  half  of  the  resolution FWHM.  The  ellipse  parameters
derived by the above \HI\, dynamical modeling were used, for which the
inclination angle  and the position  angle were interpolated.  We then
converted  this  brightness  into   the  stellar  mass  profile  using
$\Upsilon_{3.6{\mu}m}$=0.6  $M_{\odot}/L_{\odot,  3.6{\mu}m}$  with  a
Kroupa stellar initial mass function  (see below), as    shown    in
Figure~\ref{NIROT_property} (b).   To extend the profile  to 10 kpc,
an exponential disk was fitted to the part outside a radius of 4 kpc.

To measure the stellar contribution to  the rotation curve, we use the
\texttt{ROTMOD}  task \citep{Begeman89}  in  GIPSY, where  the  vertical
distribution was assumed to be ``SECH-SQUARED'' with a scale height of
0.2 kpc. The errors of the  stellar mass surface density are dominated
by  the subtraction  of  the sky  background and  its  effects on  the
rotation curve  are simulated  with the Monte  Carlo method.   We also
varied  the scale  height  from 0.01  kpc  to 0.5  kpc  to obtain  the
uncertainty  due to  the  scale  height.  The  above  two errors  were
quadratically  added  to represent  the  final  error of  the  stellar
rotation curve.

\subsection{The models of dark matter halo}\label{sec_model}

The  models of  cuspy  dark  matter halo  are  motivated by  numerical
simulations  of dark  matter,  including  a Navarro-Frenk-White  (NFW)
model \citep{Navarro97} and an Einasto model \citep{Einasto65}. The models
of cored dark matter halo  are observatinally motivated, including the
pseudo-isothermal  (ISO)  model   \citep{Begeman91}  and  Burkert  model
\citep{Burkert95}.
 
1) A NFW halo model \citep{Navarro97} has a density profile of
\begin{linenomath*}
\begin{equation}
\rho_{\rm NFW}(R) = \frac{ {\rho_{c}}{\delta_{\rm char} }}{ (R/R_{S})(1+R/R_{S})^{2} },
\end{equation}
\end{linenomath*}
where  $\rho_{c}$=3$H^{2}$/(8{$\pi$}G)   is  the  present  critical
density,  $\delta_{\rm char}$  is dimensionless  density contrast  and
$R_{S}$ is the  scale radius. $R_{S}$ is related  to $R_{200}$ through
the  concentration  $c$=$R_{200}$/$R_{S}$, where  $R_{200}$  is  the
radius within which the halo average  density is 200 times the present
critical   density.    $\delta_{\rm  char}=\frac{200c^{3}g}{3}$   with
$g=\frac{1}{{\rm ln}(1+c)-c/(1+c)}$.  The halo mass with  $R_{200}$ is
$M_{200}=\frac{100H^{2}R_{200}^{3}}{G}$. The inner  density profile of
the  NFW model  shows  a  cusp with  $\rho$  $\propto$ $R^{-1}$.   The
corresponding rotation velocity of the NFW model is

\begin{linenomath*}
\begin{equation}\label{eqn_NFW_v}
 V_{\rm NFW}(R) =  V_{\rm 200} \sqrt{\frac{{\rm ln}(1+cx)-cx/(1+cx)}{x\left[{\rm ln}(1+c)-c/(1+c)\right]}},
\end{equation}
\end{linenomath*}
where $V_{\rm 200}$ is the circular velocity at  $R_{200}$ with
$V_{\rm 200}$=$R_{\rm 200}h$, and $x$=$R$/$R_{200}$.



2) A pseudo-isothermal halo model \citep{Begeman91} is observationally
motivated to describe the presence of a central core:
\begin{linenomath*}
\begin{equation}
\rho_{\rm ISO}(R) = \frac{\rho_{0}}{ 1+(R/R_{C})^{2} },
\end{equation}
\end{linenomath*}
where $\rho_{0}$ is the central density and $R_{C}$ is the core radius of the halo.
The rotation velocity is
\begin{linenomath*}
  \begin{equation}\label{eqn_ISO_v}
    \begin{multlined}
V_{\rm ISO}(R) = \sqrt{4{\pi}G\rho_{0}R_{C}^{2}\left[1-\frac{R_{C}}{R}{\rm arctan}(\frac{R}{R_{C}})\right] } \\
= V_{\rm norm}\sqrt{(\frac{R_{C}}{R_{1}})^{2}\left[1-\frac{R_{C}}{R}{\rm arctan}(\frac{R}{R_{C}})\right] },
\end{multlined}
\end{equation}
\end{linenomath*}
where $V_{\rm norm}$ = $R_{1}\sqrt{4{\pi}G\rho_{0}}$ and $R_{1}$=1 kpc.

3) The Burkert density profile \citep{Burkert95} is described by
\begin{linenomath*}
\begin{equation}
\rho_{\rm Bk} (R) = \frac{\rho_{0}R_{C}^{3}}{(R+R_{C})(R^{2}+R_{C}^{2})},
\end{equation}
\end{linenomath*}
where $\rho_{0}$ and $R_{C}$ are the central core density and core radius,
respectively.

The corresponding circular velocity is given by
\begin{linenomath*}
  \begin{equation}\label{eqn_Bu_v}
V_{\rm Bk} (R) = V_{C}\sqrt{ \frac{2\,{\rm ln}(1+x)+{\rm ln}(1+x^{2})-2\,{\rm arctan}(x)}{x\left[3\,{\rm ln}(2)-2\,{\rm arctan}(1)\right]} },
\end{equation}
\end{linenomath*}
where $x$=$R/R_{c}$ and $V_{C}$ is the circular velocity at the core radius.

4) With the rotation curve of a dark matter halo, the density profile of the halo
can be  derived through \citep{deBlok01}:
\begin{linenomath*}
\begin{equation}
\rho(R) = \frac{1}{4{\pi}G}\left[\frac{2V}{R}\frac{{\partial}V}{{\partial}R} + (\frac{V}{R})^{2}\right]
\end{equation}
\end{linenomath*}

\begin{figure*}[tbh]
  \begin{center}
    \includegraphics[scale=0.5]{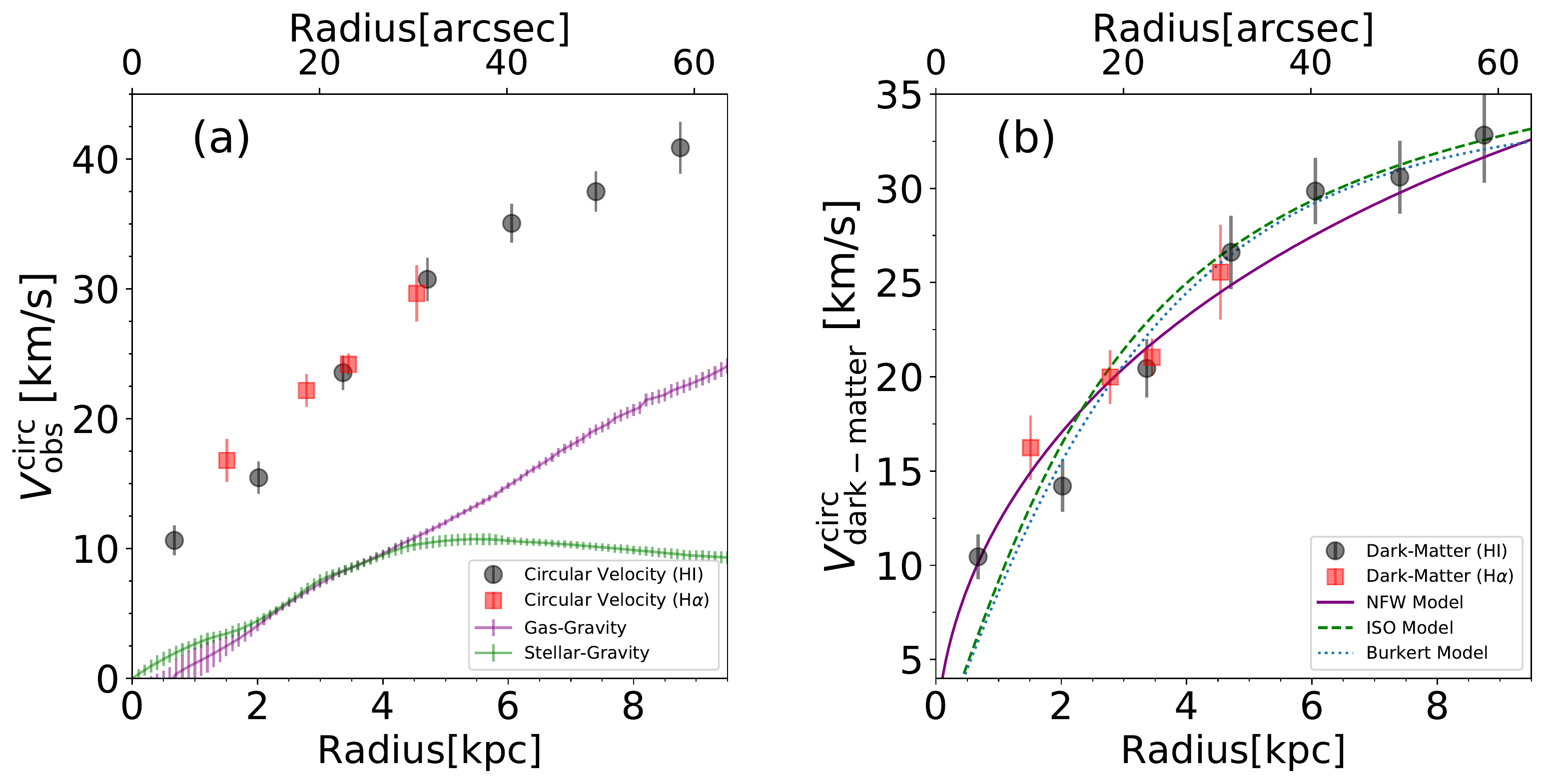}
    \caption{\label{rotation_curve}  {\bf  The rotation   curves  of AGC 242019.} {\bf a,} The observed rotation
      curves from the  \HI\, data (black circles) and  the H$\alpha$ data
      (red  rectangles), along  with the  rotation curves  due to  the
      gas and stellar gravity contributions. {\bf b,} The derived rotation curve
      of dark matter. Different curves indicate the best-fitted NFW,
      ISO and Burkert  models to the H{\sc I} only rotation curve.  }  
\end{center}
\end{figure*}

\begin{figure*}[tbh]
  \begin{center}
    \includegraphics[scale=0.5]{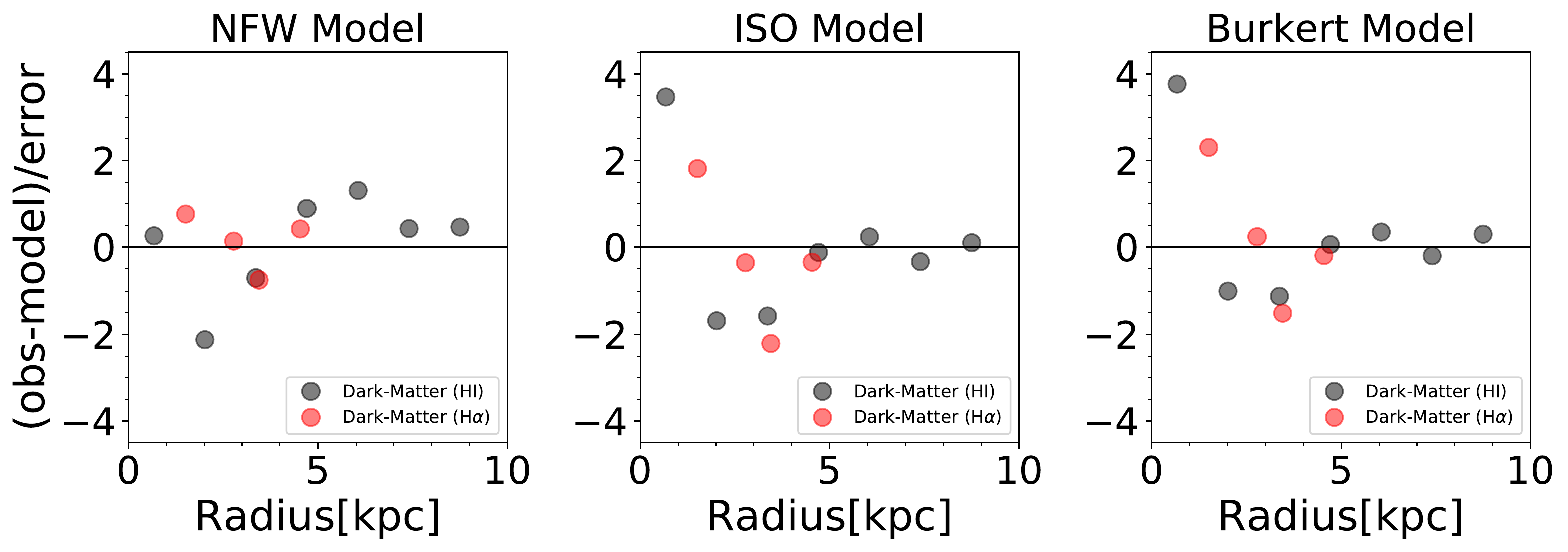}
    \caption{\label{rc_residual}  {\bf  The residuals of the fitting to the dark-matter rotation curve for three models}.  }  
\end{center}
\end{figure*}

\begin{figure}[tbh]
  \begin{center}
    \includegraphics[scale=0.40]{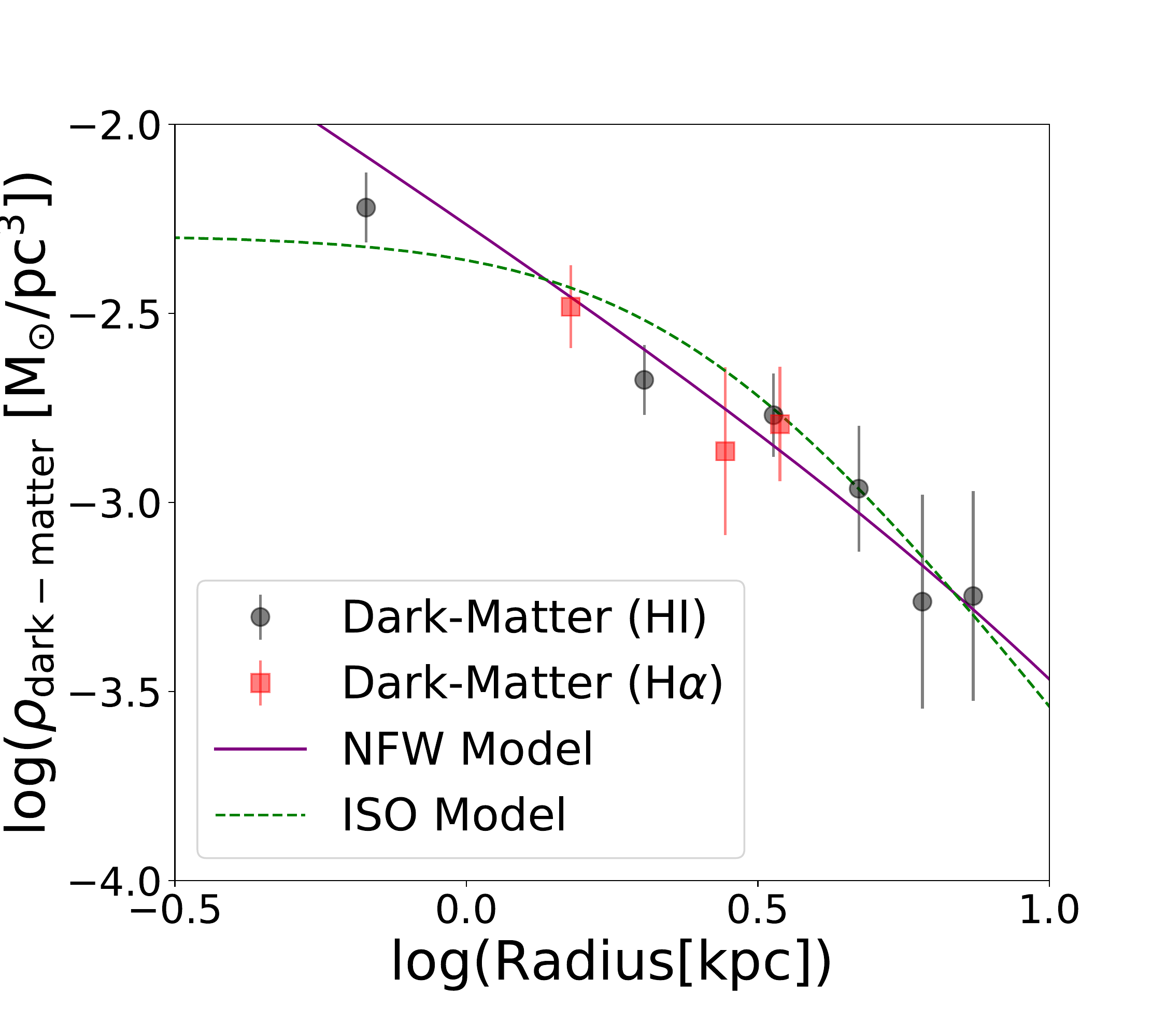}
    \caption{\label{density_profile}  {\bf  The density profile of dark matter of AGC 242019.} Two curves represent
    the best-fitted NFW and ISO models to the H{\sc I} rotation curve. }  
\end{center}
\end{figure}

\begin{figure}[tbh]
  \begin{center}
  \hspace*{-0.5cm}\includegraphics[scale=0.28]{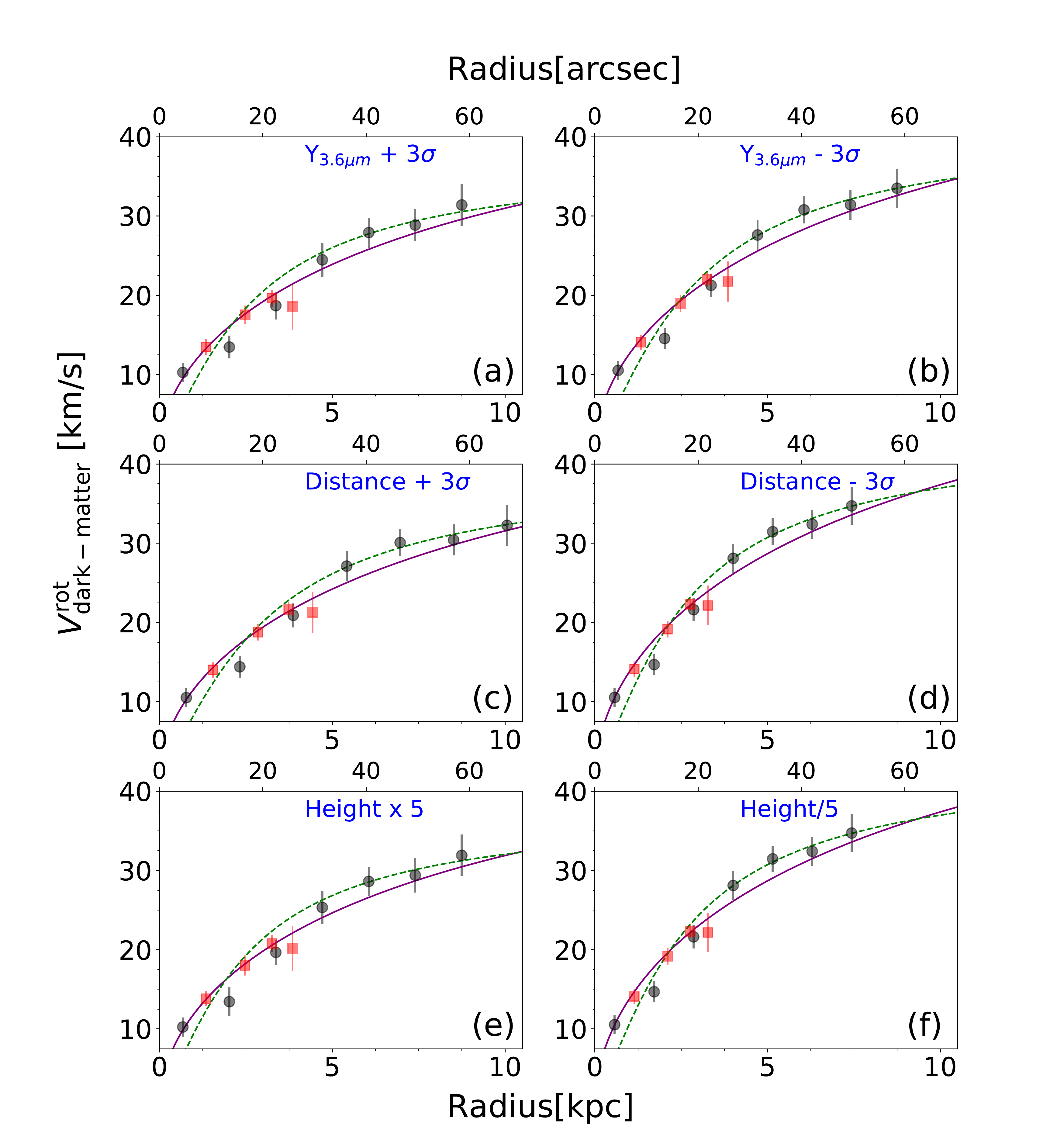}
  \caption{\label{nonfid_rc} {\bf The rotation curves of dark matter by varying different parameters}. {\bf a-b,} The results with
   the mass-to-light ratio in 3.6 $\mu$m varying by $\pm$3-$\sigma$. {\bf c-d,} The results with
   the distance varying by $\pm$3-$\sigma$. {\bf e-f,} The results with
   the scale height of the \HI\, gas disk varying by a factor of five.   }
\end{center}
\end{figure}

\begin{figure*}[tbh]
  \begin{center}
    \includegraphics[scale=0.5]{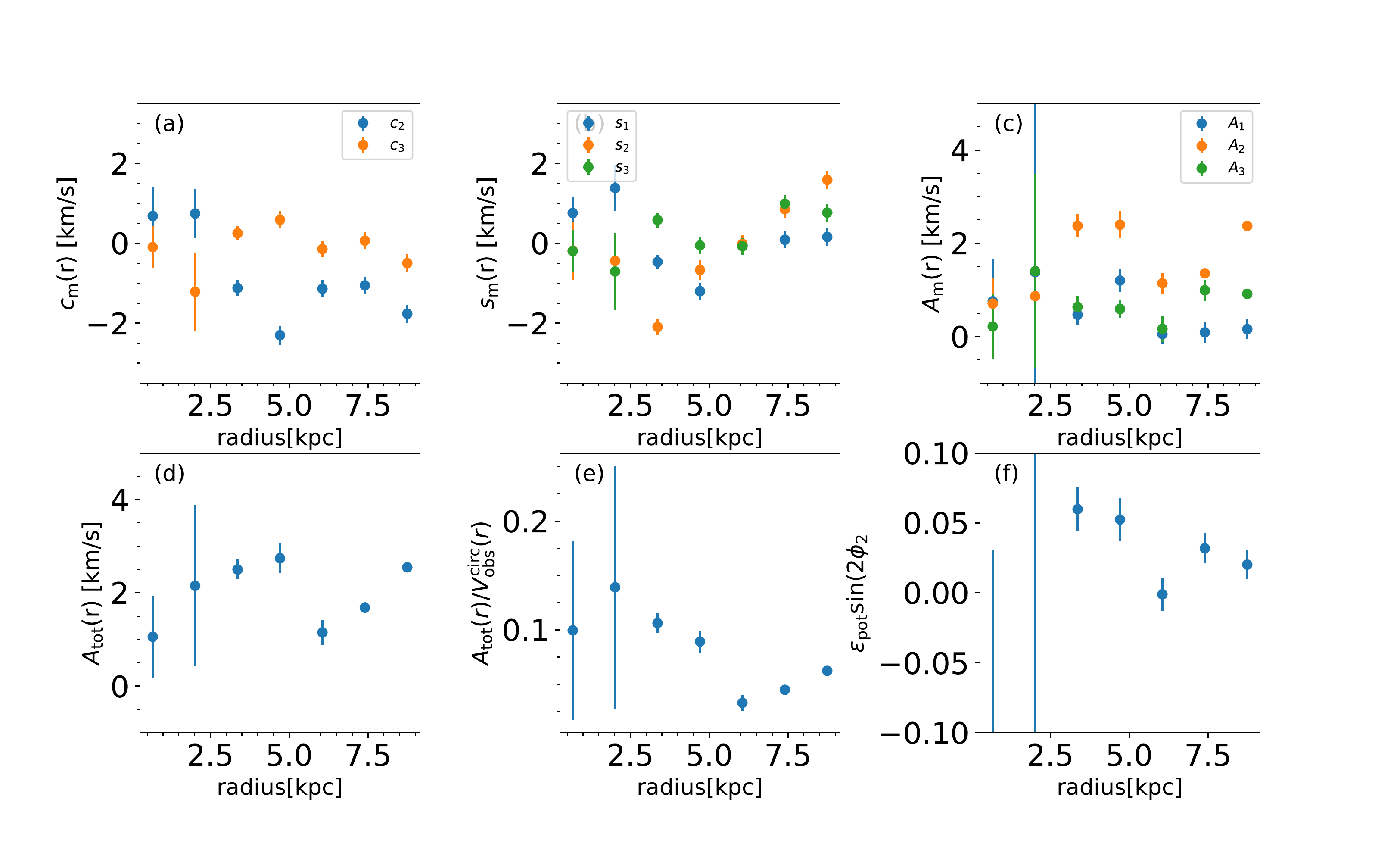}
 \caption{\label{non_circ} {\bf The results of harmonic expansion of non-circular motions.} {\bf a,} The individual $c$
 component up to the harmonic order of 3. {\bf b,} The individual $s$
 component up to the harmonic order of 3. {\bf c,} The amplitude of each harmonic order. {\bf d,} The
 total harmonic amplitude. {\bf e}, The ratio of the total harmonic amplitude to the circular velocity. {\bf f,}
 The elongation of the potential as described by ${\epsilon}_{\rm pot} {\rm sin}2\phi_{2}$. }
\end{center}
\end{figure*}

\subsection{Priors and set-ups of the rotation curve modeling}

\begin{table}
\footnotesize
\caption{\label{tab_prior} The priors for the dark-matter modeling }
\begin{tabular}{llllllllllllll}
\hline
Parameters      & Bounded Normal ($\mu$, $\sigma$)          \\
\hline
NFW model & \\
\hline
$R_{\rm 200}$   &  (75, 150) kpc     \\
$R_{\rm S}$     &  (7.5, 15) kpc    \\
\hline
ISO model       &               \\ 
\hline
$V_{\rm norm}$    &  (25, 50) \kms   \\
$R_{\rm C}$      &   (2, 4) kpc   \\
\hline
Burkert model &      \\
\hline
$V_{\rm C}$    &  (35, 70) \kms   \\
$R_{\rm C}$      &   (2, 4) kpc   \\
\hline
\end{tabular}
\end{table}

We fitted the above three dark-matter  models to the rotation curve of
dark  matter   through  Bayesian   inference  with  the   Python  code
\texttt{PyMC3} \citep{Salvatier16}.  We  adopted the No-U-Turn Sampler
(NUTS) with 5000 samples, 2000 tune, 4 chains and target\_accept=0.99.

As listed in Table~\ref{tab_prior}, the prior of each parameter of a
dark-matter  model is  described with  a bounded  normal distribution,
whose  lower-limit  bound  is  set  to  be  zero.   The  mean  of  the
distribution is set  with some prior knowledge as  detailed below, and
the standard deviation is set to be twice the mean.

The prior $R_{200}$ of  the NFW model  is  set to  have a  mean  of 75  kpc,
corresponding  to a  halo  mass of  5$\times$10$^{10}$ $M_{\odot}$  as
expected  by the  stellar mass  vs. halo  mass relationship  given our
stellar mass of \mstar\, \citep{Santos-Santos16}. At this halo mass, the halo
concentration is about 10 in simulations \citep{Maccio07}, giving
the mean prior  $R_{S}$ of 7.5 kpc.

Since the core radius of a cored dark-matter halo is on a kpc scale
\citep{Oh15}, the mean prior $R_{\rm C}$ of the ISO model is set to be 2 kpc.
The mean prior $V_{\rm norm}$ of the ISO model is thus set to be 25 \kms,
so that the galaxy
is on the baryonic Tully-Fisher relationship \citep{ManceraPina20} given its stellar mass
of \mstar\, and a HI mass of \mhi\,. Similarly, the mean prior $R_{\rm C}$ of the
Burkert model is set to be 2 kpc, and the mean prior $V_{\rm C}$  of the model
is set to be 35 \kms.

The fitting is convergent. By exploring the full posterior distribution,
the  best-fitted parameter  is  given  as the  mean  of the distribution, and
the error is given by the standard deviation, as well as the 3\% and 97\%
Highest Posterior Density (HPD). The percentage of small Pareto shape diagnostic
values ($k$ $<$ 0.7) is also listed. A larger diagnostic value indicates that the model is mis-specified.

\section{Results}

\subsection{The rotation curve of dark matter}

Figure~\ref{rotation_curve} (a)  shows the observed rotation  curve of
AGC 242019.   The \HI\, measurements cover  a radial extent of  9 kpc
while the \ha\,  data cover the central 5 kpc  in radius. Two datasets
are  overall consistent  with each  other in  the overlap  region. The
rotation curve rises all the way up to the last measurable radius.

Figure~\ref{rotation_curve}  (b)  shows  the rotation  curve  of  dark
matter  after quadratically  subtracting the  gas and  stellar gravity
contributions. The  sampling of the  \HI\, curve  is at the beam
size, while the H$\alpha$ curve has a radial bin  of 1.0 kpc. As  shown in the figure, the  \ha\, curve is
overall well consistent  with the \HI\, curve over  the spatial extent
where both  dataset cover.  As  mentioned before, since  the H$\alpha$
emission is  sporadic and the errors  of the \ha\, rotation  curve are
only based  on the detected regions,  its rotation curve is  only used
for the  sanity check of the  \HI\, curve.  We fitted  the \HI\, curve
with   two   spherical   halo    models,   namely,   the   NFW   model
\citep[Equation~\ref{eqn_NFW_v},][]{Navarro97}   and  the   ISO  model
\citep[Equation~\ref{eqn_ISO_v},][]{Begeman91},  through   the  Python
code  \texttt{PyMC3} \citep{Salvatier16}  to represent  the cuspy  and
cored profiles, respectively. The ISO model  is too steep at the inner
radii, whereas  the NFW model matches
the observed  data  in  the overall  fitting, which is also illustrated by
the fitting residuals as shown
in Figure~\ref{rc_residual}. To
quantitatively   discriminate  the   two   models,   we  performed   a
leave-one-out (LOO) predictive check \citep{Vehtari15}.  We found that
the estimated  effective number of parameters  ($p_{\rm loo}$=\pooNFW)
of the NFW  model is smaller than the real  number of free parameters,
i.e.,  two, while  that of  the ISO  model has  $p_{\rm loo}$=\pooISO,
significantly larger than two (see Table~\ref{NFW_ISO_fitting}).  This
quantitatively indicates  that the NFW  model is valid, while  the ISO
model is ruled out. The corresponding difference in the reduced $\chi^{2}$ is 1.9
for d.o.f.=5, equivalent to a Gaussian significance of 3.0-$\sigma$. As listed in Table~\ref{NFW_ISO_fitting}, some Pareto  diagnostic value $k$ of the
ISO model is  larger than 0.7, further indicating that the model is mis-specified. We also
run the fitting by including the \ha\,, and obtained similar difference in $p_{\rm loo}$ values
and $\Delta(\chi^{2}$/d.o.f.)=2.8 for d.o.f.=9 between the two models, equivalent to a Gaussian significance of 5.0-$\sigma$.

The  Burkert model  is  another  observationally-motivated formula  to
describe        a         cored        dark         matter        halo
\cite[Equation~\ref{eqn_Bu_v},][]{Burkert95}.  Its  density profile is
flat toward the center,  which is like the ISO model,  but has a power
law index  of -3.0 toward infinity,  which is like the  NFW model.  As
shown in  Figure~\ref{rotation_curve} (b),  the Burkert model  gives a
bad  fitting too,  with  $p_{\rm loo}$=\pooBu, much larger  than (\pooNFW) that of
the NFW model.

The best-fitted  NFW model has  a halo  scale radius of  $\sim$33 kpc.
Such  a large  cusp is  well spatially  resolved given  that its  size
relative to the \HI\, beam is  24.  This suggests that the presence of
the cusp  is not an  artifact caused  by a limited  spatial resolution
\citep{deBlok01, Oh15}.  The  dark-matter halo has a  mass of \mhalo\,
within $R_{200}$, the radius at which  the average halo density is 200
times the  average cosmic density. Due  to the fact that  the rotation
curve does not  reach the flat part, the constraints  on the $R_{200}$
(or $M_{\rm halo}$)  and $R_{\rm S}$ do  not reach a small  error. But a
small halo concentration of  only \concentration\, is conclusive. This
is much smaller  than the median concentration of 15  at the same halo
mass in the local Universe \citep{Maccio07}. The implication of this small
concentration on the formation of the galaxy is discussed in \S~\ref{sec_udg}.

To constrain the innermost density slope $\alpha_{\rm innermost}$ of
the dark  matter halo, we derived  the density profile of  dark matter
from    the   rotation    curve   \citep{deBlok01}    as   shown    in
Figure~\ref{density_profile}.     Following    \citet{deBlok01}    and
\citet{Oh15}, we  measured $\alpha_{\rm  innermost}$ by  carrying out
the  least-square fitting  to the  density profile  within the  break
radius,  and defined  the error  of $\alpha_{\rm  innermost}$ as  the
difference between  the result including the  first data-point outside
the break radius and the  result including only data-points within the
break radius.  Here  the break radius is the radius  where the density
profile shows  a maximum change  in the  slope.  If adopting  the core
radius of the best-fitted ISO  model as the break radius, $\alpha_{\rm
  innermost}$=-($0.96\pm0.12$). If  adopting the scale radius  of the
best-fitted     NFW    model     as    the     radius,    $\alpha_{\rm innermost}$=-(0.90$\pm$0.08). Here the error is given by the least square fitting since the scale
radius of the NFW model is larger than the last ring. Two   measurements   indicates   a
statistical  significance of  about 10-$\sigma$  that the  dark matter
halo of AGC 242019 is cuspy down to a radius of \innerradius\,.

\begin{table*}
\footnotesize
\begin{center}
\caption{\label{NFW_ISO_fitting} The fitting result of NFW and ISO models to the dark-matter rotation curve with \texttt{PyMC3} }
\begin{tabular}{llllllllllllll}
\hline
Rotation Curves    & \multicolumn{5}{c}{NFW model}   &   &  \multicolumn{5}{c}{ISO model}  \\  
\cline{2-6}     \cline{8-12}
                & $R_{200}$  &  $R_{s}$ & $p_{\rm LOO}$  & $k^{\ast}$ ($<$0.7)  & $\chi^{2}$/d.o.f.$^{\P}$ &  & $V_{\rm norm}$ &  $R_{\rm C}$ &  $p_{\rm LOO}$  & $k^{\ast}$ ($<$0.7)  & $\chi^{2}$/d.o.f.$^{\P}$ \\
\hline
                & (kpc)     &  (kpc)   &              &               &                   &  & (\kms)       &  (kpc)      &               &                &      \\
\hline
HI (fiducial)                          & 65.0$\pm$7.4 & 33.3$\pm$9.1 & 1.6  & 100\%  & 1.60  &  &  16.7$\pm$1.9 & 2.5$\pm$0.5 & 6.3  & 86\%  & 3.51  &  \\
HI+H$\alpha$ (fiducial)                & 64.6$\pm$7.1 & 33.1$\pm$8.7 & 1.2  & 100\%  & 1.90  &  &  17.1$\pm$1.7 & 2.2$\pm$0.4 & 5.8  & 100\%  & 4.74  &  \\
\hline
HI ($\Upsilon_{3.6{\mu}m}$+3$\sigma$)  & 59.5$\pm$7.2 & 31.5$\pm$9.3 & 1.4  & 100\%  & 1.32  &  &  16.3$\pm$2.3 & 2.3$\pm$0.5 & 7.3  & 86\%  & 3.55  &  \\
HI  ($\Upsilon_{3.6{\mu}m}$-3$\sigma$)  & 67.2$\pm$7.7 & 33.6$\pm$9.1 & 1.8  & 100\%  & 1.82  &  & 16.9$\pm$1.8 & 2.5$\pm$0.4 & 6.5  & 86\%  & 3.50  &  \\
\hline
HI (Distance+3$\sigma$)                & 62.1$\pm$6.7 & 33.9$\pm$9.2 & 1.7  & 100\%  & 1.66  &  &  15.2$\pm$1.8 & 2.6$\pm$0.5 & 7.1  & 86\%  & 3.46  &  \\
HI  (Distance-3$\sigma$)                 & 73.8$\pm$8.5 & 33.9$\pm$8.9 & 1.9  & 100\%  & 1.97  &  & 19.5$\pm$2.0 & 2.3$\pm$0.4 & 6.2  & 86\%  & 3.55  &  \\
\hline
HI  (5$\times$Height)                    & 60.7$\pm$7.4 & 30.6$\pm$9.0 & 1.0  & 100\%  & 1.14  &  &  17.5$\pm$2.6 & 2.2$\pm$0.5 & 6.7  & 86\%  & 3.29  &  \\
HI (Height/5)                            & 61.6$\pm$7.4 & 31.4$\pm$9.1 & 1.1  & 100\%  & 1.34  &  &  17.6$\pm$3.0 & 2.2$\pm$0.6 & 9.9  & 86\%  & 3.93  &  \\
\hline
\end{tabular}
\end{center}
\tablecomments{$^{\ast}$The percentage of the Pareto diagnostic values that are ``good'' and ``ok'' ($k$ $<$ 0.7). $^{\P}$The reduced $\chi^{2}$ is given
for the best-fitted result with \texttt{PyMC3}.}
\end{table*}

\begin{table}
\footnotesize
\begin{center}
\caption{\label{den_fit} The fitting result of the Burkert model to the HI's curve of dark matter}
\begin{tabular}{llllllllllllll}
\hline
$V_{C}$ [km/s]    &  $R_{C}$ [kpc]    &     $p_{\rm LOO}$ & $k^{\ast}$ ($<$0.7) & $\chi^{2}$/d.o.f.$^{\P}$  \\
\hline
25.7$\pm$1.9 & 4.4$\pm$0.7 & 4.4  & 86\%  & 3.34  \\
\hline
\end{tabular}
\end{center}
\tablecomments{$^{\ast}$The percentage of the Pareto diagnostic values that are ``good'' and ``ok'' ($k$ $<$ 0.7). $^{\P}$The reduced $\chi^{2}$ is given
for the best-fitted result with \texttt{PyMC3}.}
\end{table}

 \subsection{The systematic uncertainties of the dark-matter rotation curve}\label{sec_sysunc_curve}

The identification of a cuspy profile in AGC 242019 is robust against
systematic uncertainties from several aspects.

{\bf (1) Position and  inclination angles:} Since these two angles
have  been   set  as  free   parameters  during  the   fitting,  their
uncertainties  have  already  been included  in  the derived  rotation
curve.  For comparison, we further estimated the photometric position
and inclination angles based on the $r$-band image.  The galaxy in the
$r$-band is somewhat  asymmetric with clumpy features, but outside the
radius of  8$''$,  the position angle and inclination angles  converge to
(2$\pm$2)$^{\circ}$ and (67$\pm$3)$^{\circ}$,  respectively. These results
are  consistent with the values derived from the dynamic fitting of  the \HI\,
data,  as  shown  in 
Figure~\ref{parameter_freeVVPI} and listed in Table~\ref{tab_ring_result}.

{\bf (2) Mass-to-light ratio in  3.6 $\mu$m:} The rotation velocity due to
stellar gravity is proportional to  the square root of the stellar
mass-to-light ratio.  The overall stellar contribution to the observed
rotation curve  is small, so  our result  is not sensitive  to the
mass-to-light  ratio  $\Upsilon_{3.6{\mu}m}$,  as is detailed  here.   The
mass-to-light ratio in a broad band is obtained by fitting a synthetic
spectrum  to  the  observed  spectrum or a broad-band  spectral  energy
distribution.   The uncertainties  of the  stellar population  synthesis
model,  the   star  formation   history,  the  dust   extinction,  the
metallicity and the initial mass  function all result in the variation
of the  derived mass-to-light  ratio.

With the measured stellar mass and  star formation rate of AGC 242019,
we assumed a low metallicity  with [Fe/H]=-1 and suppressed asymptotic
giant  branch (AGB)  stars, as  seen  in some  low surface  brightness
dwarfs \citep{Schombert19}, to derive  the $\Upsilon_{3.6{\mu}m}$ of our
target   to  be   0.6   \citep{Schombert19}.   The   preceding
assumption about AGB stars causes the mass-to-light ratio to be $\sim$
30\% larger  than that in normal  galaxies.  $\Upsilon_{3.6{\mu}m}$ is
also  color  dependent \citep{Bell03,   Zibetti09,  Jarrett13,  Meidt14,
Shi18, Schombert19, Telford20}.  As the  object is not detected in 4.5
$\mu$m, we used  the radial variation in $g-r$ color  with a median of
0.3  and   standard  deviation  of   0.03.   Converting  to   $B-V$  =
0.98*($g-r$)  +   0.22  \citep{Jester05},  the   variation  in
$\Upsilon_{3.6{\mu}m}$ is expected to have a standard radial deviation
as  small  as  0.02  dex \citep{Schombert19},  and  thus,  its
effects on the rotation curve of dark matter are negligible.

The systematic uncertainties  due to the difference  in star formation
histories, initial mass functions  and stellar population models among
different studies are  much larger even at  a fixed color \citep{Bell03,
Zibetti09, Jarrett13, Meidt14, Schombert19,  Telford20}.  As a result,
we adopted 0.1 dex as the 1-$\sigma$ error to encompass the results in
different   studies.   We   then  varied   $\Upsilon_{3.6{\mu}m}$   by
$\pm$3$\sigma$ to investigate its effect on the rotation curve of dark
matter.  As  listed in   Table~\ref{NFW_ISO_fitting} and shown in Figure~\ref{nonfid_rc}, the
cusp-like NFW model is a more reasonable model than the core-like ISO
model for both $\Upsilon_{3.6{\mu}m}$ values.

{\bf (3) Distance:}  The velocities due to baryonic  gravity vary with
the square root of the distance,  while the observed rotation curve is
independent of the distance. As the  object is an isolated galaxy with
no  close-by companion  that is  brighter  than $m_{r}$  $=$ 17.7  (or
$M_{r}$ $=$ -14.8) within 500 kpc and 1000 \kms\, in the Sloan Digital
Sky Survey, the error in the  distance is dominated by the uncertainty
of the  Hubble constant and  the peculiar velocity.   The heliocentric
velocity  is  2237  \kms\,  after  correcting  for  the  Virgo,  Great
Attractor  and Shapley  supercluster \citep{Mould00}.  We estimated  the
residual  error to  be 100  km/s,  which corresponds  to an  infalling
velocity  toward   an  imaginary  dark  matter   halo  with  10$^{11}$
$M_{\odot}$ at  a separation of  50 kpc.   By adopting a  local Hubble
constant  of  73  \kms\,  with  2.5\%  uncertainty \citep{Riess16},  the
distance thus  has a  1-$\sigma$ uncertainty of  5\%.  We  varied this
distance  by   $\pm$3$\sigma$  to   investigate  its  effect   on  the
dark-matter rotation curve.  As listed in Table~\ref{NFW_ISO_fitting}  and shown in Figure~\ref{nonfid_rc},
the cusp-like NFW  model is again advocated against  the core-like ISO
model for both distances.

{\bf  (4)  The scale  height  of  the gas  disk:}  For  a thick  disk,
emissions from adjacent rings are projected  to be in the same pixels,
which  causes a  beam smearing  and  affects the  measurements of  the
inclination  and position  angles.  By increasing  and decreasing  the
scale height by  a factor of five  to 500 pc and  20 pc, respectively,
the NFW  model is still  much better than the  ISO model, as  shown in
Figure~\ref{nonfid_rc} and Table~\ref{NFW_ISO_fitting}.

{\bf     (5)    The     noncircular    motion:}     As    shown     in
Figure~\ref{bbarolo_run},  the   median  amplitude  of   the  residual
velocity obtained  by \texttt{$^{3D}$Barolo}  fitting is  small, i.e.,
$\sim$2 \kms.   During the  fitting, the line-of-sight  velocities are
assumed  to be  entirely circular  motions, while  noncircular motions
cause the  real circular velocity  to be underestimated.   To quantify
the  amplitude  of  noncircular   motions,  we  carried  out  harmonic
decomposition with the GIPSY task \texttt{RESWRI} \citep{Begeman89}.  As
detailed in previous studies \citep{Schoenmakers97, Trachternach08}, the
line-of-sight velocity of each ring can be decomposed into
\begin{linenomath*}
\begin{equation}
 v_{\rm los}(r) = v_{\rm sys}(r)+\Sigma_{m=1}^{N} (c_{m}(r){\rm cos}m\psi + s_{m}(r){\rm sin}m\psi ),
\end{equation}
\end{linenomath*}
where  $r$ is  the radial  distance of  each ring  from the  dynamical
center,  $v_{\rm  sys}(r)$  is  the system  velocity,  $\psi$  is  the
azimuthal  angle  in the  plane  of  the  disk. $v_{\rm los}(r)  =  v_{\rm
sys}(r)+c_{1}(r){\rm cos}\psi$  corresponds to a pure  circular motion
scenario. In this study, we expanded the velocity up to $m$=3, as has been adopted in other
studies \citep{Schoenmakers97,  Trachternach08}.


To run \texttt{RESWRI}, we used the 2-D velocity field produced by \texttt{$^{3D}$Barolo} as the input,
fixed the dynamical center and system velocity to those determined by \texttt{$^{3D}$Barolo},
and set the rotation velocity, inclination angle and position angle of each ring
as free parameters.  With the derived $c_{m}$ and $s_{m}$, the amplitude of each
noncircular harmonic component with $m>$1 is defined as
\begin{linenomath*}
\begin{equation}
 A_{m}(r) = \sqrt{c_{m}^{2}(r) + s_{m}^{2}(r)}
\end{equation}
\end{linenomath*}
and for $m$=1 where  $c_{1}$ is the circular motion,
\begin{linenomath*}
\begin{equation}
A_{1}(r) = \sqrt{s_{1}^{2}(r) }.
\end{equation}
\end{linenomath*}
The total amplitude of noncircular motion is given by
\begin{linenomath*}
\begin{equation}
 A_{\rm tot}(r) = \sqrt{s_{1}^{2}(r) + c_{2}^{2}(r) + s_{2}^{2}(r) +  c_{3}^{2}(r) + s_{3}^{2}(r) }.
\end{equation}
\end{linenomath*}
The measured harmonic component can be used to quantify the elongation of the potential
$\epsilon_{\rm pot}$ at each radius as follows:
\begin{linenomath*}
\begin{equation}
{\epsilon}_{\rm pot} {\rm sin}2\phi_{2} = (s_{3}-s_{1})\frac{1+2q^{2}+5q^{4}}{c_{1}(1-q^{4})},
\end{equation}
\end{linenomath*}
where $\phi_{2}$ is an unknown angle
between the minor axis of the elongated ring and the observer in the plane of
the ring and  $q$ = cos $i$, with $i$ being the inclination angle of the disk.


 Figure~\ref{non_circ} shows the result of the harmonic decomposition.
 As shown in Figure~\ref{non_circ}  (a), the radial $c_{m}$ fluctuates
 around 0 km/s with amplitudes $\lesssim$  2 km/s.  The radial
 $s_{m}$ shows similar  behaviors with amplitude $\lesssim$ 2 km/s.
 As a  result, the $A_{m}$  of each  harmonic component is  in general
 $\lesssim$ 2  km/s.  Figure~\ref{non_circ}  (d) shows that  the total
 amplitude  $A_{\rm  tot}$  is  only   about  2  km/s.   As  shown  in
 Figure~\ref{non_circ} (c), all the  amplitudes are small fractions of
 the circular  velocities at the  corresponding radii that  are 
 $<$ 15\%.

Compared   to  other   galaxies  with   similar  absolute   magnitudes
\citep{Trachternach08}, AGC 242019 is indeed a galaxy without stronger
noncircular motions. Compared to  simulated galaxies where noncircular
motions result  in noticeable underestimations of  circular velocities
\citep{Oman19}, the  noncircular amplitude of $\sim$ 2  km/s in AGC
242019  is  much  less  than  the  simulated  values  of  20-30  km/s.
Therefore, we expect that the  noncircular motion of AGC 242019 should
not affect  the derived rotation  curve of dark matter.   Finally, the
derived  ${\epsilon}_{\rm  pot}  {\rm   sin}2\phi_{2}$,  as  shown  in
Figure~\ref{non_circ}   (f),   suggests  a   spherical   gravitational
potential.

{\bf  (6) The  triaxiality of  a dark  matter halo:}  In this  study a
spherical dark matter halo is assumed,  while a halo has been found to
be   moderately   triaxial   in   numerical   simulations \citep{Jing02,
Bailin05}.  However, a typical triaxial mass distribution results
in only a  small deviation  in the density  from the  spherical assumption.
Within the scale  radius of the halo, the difference  is only 10-20\%,
which is much smaller than the required  variation of a factor of 3  to decrease  the  inner slope  by 0.5  \citep{Knebe06}.
Numerical modeling  of the rotation  curve further suggests  that the
halo triaxiality cannot significantly change the shape of  the curve 
to make an intrinsic  cusp  to be a core (or vice versa) in the observed
data \citep{KuziodeNaray09, KuziodeNaray11}.

{\bf (7) Beam  smearing:} $^{3D}$Barolo takes the beam
smearing into account  in  the  tilted-ring  fitting \citep{DiTeodoro15}. $^{3D}$Barolo   first
builds  a gas  disk in  three  spatial dimensions  and three  velocity
dimensions, and  then convolves this  artificial disk to  the observed
spatial  resolution for comparison  with the  observed  3-D datacube  to
derive   the  best-fitting   parameters.   It   has  been   shown  that
$^{3D}$Barolo  is able  to recover  the  rotation curve  even at a  low
spatial  resolution, i.e., two resolution  elements  over   the  semimajor
axis \citep{DiTeodoro15}. The semimajor axis of AGC 242019 is resolved
into $\sim$ 7 beams in the \HI\, map.  In addition, the H$\alpha$ map has
a rebinned spatial resolution (4$''$  in diameter) that is two times
higher than the \HI\, map.  Although our H$\alpha$ clumps are mainly distributed
along the major axis with a  narrower spatial extent, the overall shape of
the  H$\alpha$-derived   curve  is  consistent  with   the  \HI\, curve,
demonstrating that the beam smearing effect on the \HI\,'s rotation curve
has been  largely removed by $^{3D}$Barolo \citep{DiTeodoro15}.

{\bf (8)  The contributions  from  molecular  gas and ionized gas:} The
very low  surface density  of AGC  242019 results  in a  low mid-plane
pressure with a $P_{\rm ext}/k$ $\lesssim$ 10$^{3}$  cm$^{-3}$ K, which
gives a molecular-to-atomic  gas ratio of $\lesssim$ 5\% (see their Figure 3
and Equation 12 in
studies of nearby galaxies \citep{Blitz06, Leroy08}).  We also checked
that the atomic gas alone is sufficient  to place the galaxy in the extended
Schmidt law \citep{Shi11, Shi18}, consistent with a negligible fraction of
molecular gas.

  The ionized
gas mass  can be  estimated from the  H$\alpha$ luminosity  by $M_{\rm
HII}=m_{\rm        p}\frac{L_{        Ha}}{3.56{\times}10^{-25}N_{e}}$
\citep{Osterbrock06}, where  $m_{\rm p}$  is the  proton mass,
$L_{  Ha}$  is the  H$\alpha$  luminosity  in  erg/s, and $N_{e}$  is  the
electron volume density in cm$^{-3}$. By assuming a low $N_{e}$ of 100
cm$^{-3}$, we found that, even at  the peak spaxel  as shown in
  Figure~\ref{NIROT_property}~(c), an ionized gas  mass surface
density of  0.025 $M_{\odot}$/pc$^{2}$ is  too small to  affect the
derived rotation curve of dark matter.

\section{Discussions}\label{discussions}

Through measurements  of the  dynamics of atomic  and ionized  gas, we
demonstrate that the dark matter halo of AGC 242019 can be well fitted
by the cuspy profile as described the NFW model, while excluding cored
models including  ISO and Burkert  ones.   We here discuss its constraints
on the alternatives of standard cold dark matter, implications for the
role of feedback and implications for formation of UDGs.

\subsection{Implications for the alternatives of standard cold dark matter}

\subsubsection{Fuzzy cold dark matter}

Through numerical simulations, a halo of fuzzy cold dark matter  is found to be
composed of a soliton core superposed on an extended halo \citep{Hu00, Schive14}.
The latter can be
represented by the NFW model, while the former can
be approximately described  by:
\begin{equation}
\rho_{c}(r) \approx \frac{1.9(m_{\psi}/10^{-23} {\rm eV} )^{-2}(R_{c}/{\rm kpc})^{-4}}{\left[1+9.1\times10^{-2}(r/R_{c})^{2}\right]^{8}} {\rm M_{\odot} pc^{-3}},
\end{equation}
where $m_{\psi}$ is the particle mass  and $R_{c}$ is the core radius.
The soliton core is linked to the total halo through the core-halo
relationship \citep{Schive14}, which gives
\begin{equation}
R_{c} = 1.6(m_{\psi}/10^{-22} {\rm eV})^{-1}(M_{\rm h}/10^{9} M_{\odot})^{-1/3} {\rm kpc}.
\end{equation}

As  shown in  Figure~\ref{rotation_curve} (b),  a NFW  model fits  the
density  profile  of AGC  242019  very  well, which  shows  negligible
residual density for the soliton core to account for. As a result, the
possible contribution to the dark-matter density from the soliton core
should not exceed  the observed error at all radii,  which can be used
to  constrain  the $m_{\psi}$.   For  each  radius, we  estimated  the
density  of the  soliton  core  as a  function  of  the particle  mass
$m_{\psi}$ with the best-fitted $M_{\rm h}$=\mhalo\, through the above
two equations.  It is found that the measurements at the innermost two
radii  give the  strongest  constraints on  $m_{\psi}$.  The  3-$\sigma$ observed errors
at two  radii constrain  the $m_{\psi}$ range  to be  \fuzzylower\, or
\fuzzyupper\,.  Compared  to the constraint from  Ly$\alpha$ forest in
which  $m_{\psi}$  $<$  1.0$\times$10$^{-22}$  eV  or  $m_{\psi}$  $>$
23$\times$10$^{-22}$ eV, the dynamics of  AGC 242019 gives a factor of
about  20 times  smaller constraint  on the  upper-bound of  the lower
range.   As shown  in
Figure~\ref{test_CDM_alternative}~(a), if  adopting  the  3\% HPD  halo  mass  of  \mhalothreehpd\,,
$m_{\psi}$  \fuzzylowerthreehpd\, or  $m_{\psi}$ \fuzzyupperthreehpd\,
whose upper-bound of the lower range  is still 10 times lower than the
constraint  by  the  Ly$\alpha$  forest.   If  somehow  our  error  is
underestimated by  a factor of  2, the  upperbound of the  lower range
only increases  by a factor of  $\sim$ 1.6, and the  lowerbound of the
upper range decreases by a factor of $\sim$ 1.2.
Therefore, the constraint by AGC 242019, along with the one from the Ly$\alpha$ forest, is
 inconsistent with the  typical $m_{\psi}$ of
$\sim$ 10$^{-22}$ eV that is required to explain the dynamics of other
galaxies with  cored dark-matter  halos \citep{Hu00,  Schive14}. It is thus  found that there  is no
$m_{\psi}$ value that can reconcile all the observational facts.

\subsubsection{Self-interacting Dark Matter}

Self-interacting dark matter transmit the kinetic energy from the outer part inward
to form a constant density core. For the interaction to be efficient, the
scattering rate per particle should be important, that is
at least once  over the galaxy age  \citep{Spergel00, Rocha13, Tulin18}:
\begin{equation}
 {\Gamma}(r)t_{\rm age} {\approx}  \rho(r)(\sigma/m)v_{\rm rms}(r)t_{\rm age} {\sim} 1,
\end{equation}
where is ${\Gamma}(r)$ is the scattering rate per particle, $\rho(r)$ is dark
matter density at a radius of $r$, $\sigma/m$ is the cross section per particle mass
and $v_{\rm rms}$ is the relative velocity of dark matter particles. The above
equation can be re-written as:
\begin{equation}
\frac{\sigma/m}{\rm 1\;cm^{2}/g} {\approx}   (\frac{\rm 9.2\;Gyr}{t_{\rm age}})(\frac{\rm 0.01\;M_{\odot}/pc^{3}}{\rho(r)})(\frac{\rm 50\;km/s}{v_{\rm rms}}).
  \end{equation}

For AGC 242019, the NFW model fits the rotation curve well down to the
innermost  radius. This sets  the  upperlimit  to  the radius  of  a
possible density  core and  the lowerlimit to  the above  $\rho(r)$ if
dark matter is self-interacting in AGC 242019. If the halo forms around $z$=2, we got $t_{\rm age}$ = 10 Gyr.
The 
$v_{\rm rms}$ is set to be the virial velocity at the virial radius, which is \Vvir\,.  we have 
$\sigma$/m \sidm\, for AGC 242019.

\begin{figure*}[tbh]
  \begin{center}
    \includegraphics[scale=0.39]{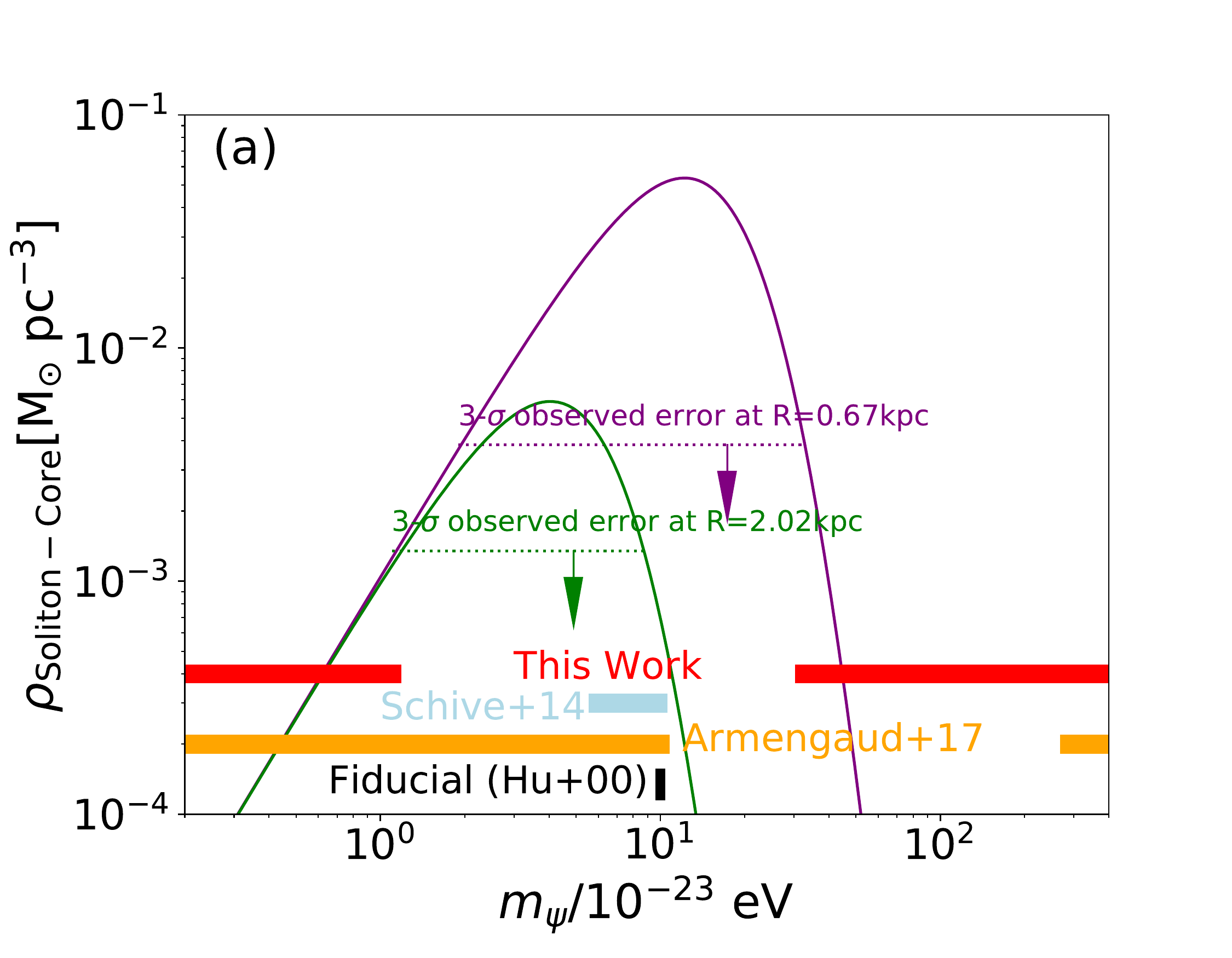}
    \includegraphics[scale=0.39]{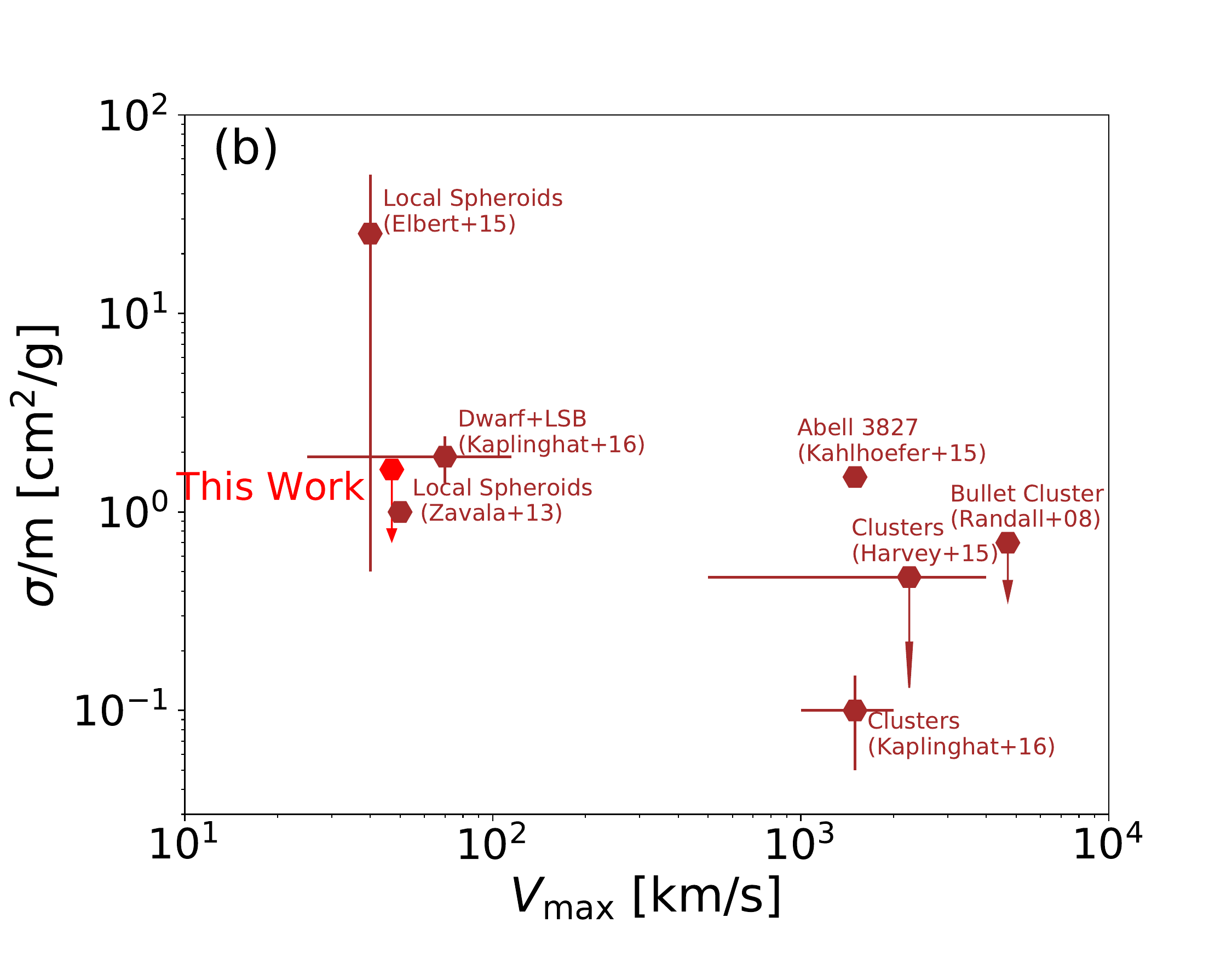}
    \hspace*{-0.38cm}\includegraphics[scale=0.39]{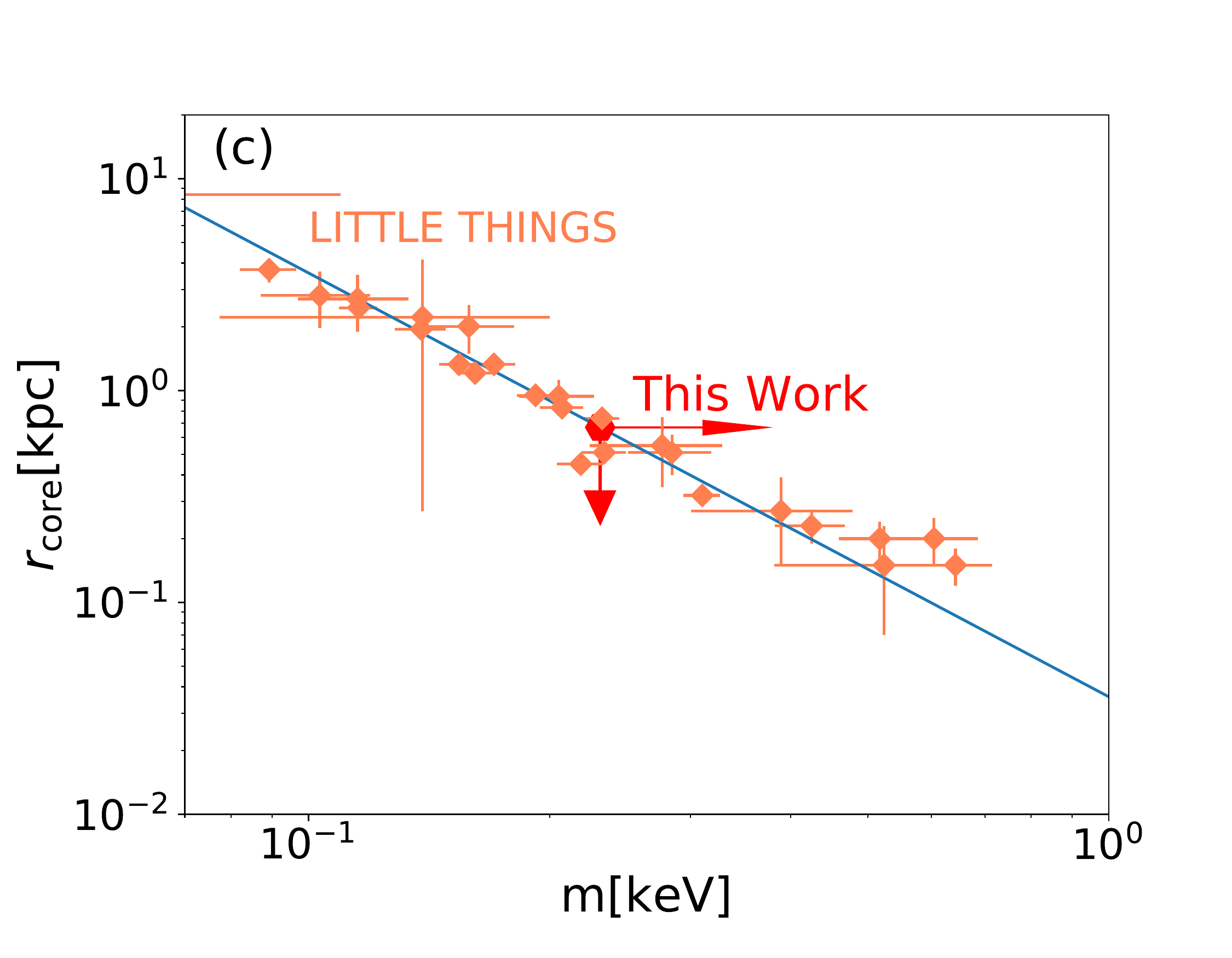}
    \hspace*{-0.38cm}\includegraphics[scale=0.39]{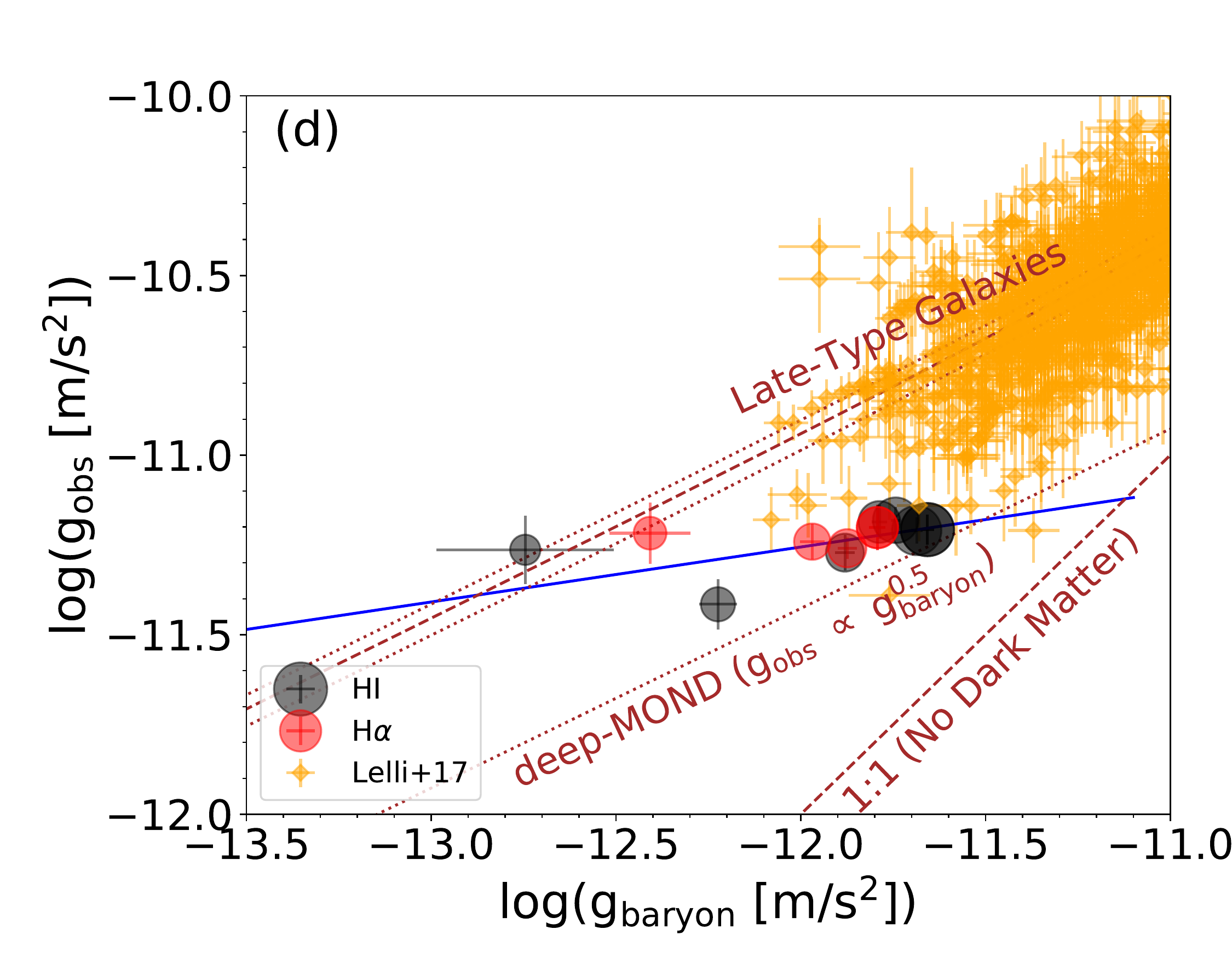}
    \caption{\label{test_CDM_alternative}    {\bf    Test    of    the
        alternatives of  standard cold  dark matter with  AGC 242019}.
      {\bf a,} The test of the fuzzy dark matter. Two curves represent
      the soliton core  mass density as a function of  the dark matter
      particle  mass  at the  innermost  and  second innermost  rings,
      respectively.  The  dotted  lines are  the  observed  3-$\sigma$
      errors  at  the corresponding  two  radii.  Color bars  are  the
      constraint  of  the  particle  mass by  different  studies  (see
      text). {\bf  b,} The test  of the self-interacting  dark matter.
      Brown symbols are the constraints in the literature. The density
      of the innermost ring of AGC 242019 results in the upperlimit of
      the cross section of self-interacting  dark matter. {\bf c,} The
      test of  warm dark matter. The  radius of the innermost  ring of
      AGC 242019  marks the  lowerbound of the  particle mass  of warm
      dark matter. {\bf d,} The  test of MOND through the relationship
      between the observed radial acceleration and the baryonic radial
      acceleration.  The black/red symbols are the  results of AGC 242019, where
      a  larger symbol  size corresponds  to  the ring  with a  larger
      radius.   The blue  solid line  is the  best linear  fit to  the
      observations.  Lines labeled with ``Late-Type Galaxies'' and orange symbols are the
      best-fitted line plus its scatter and individual late-type galaxies in \citet{Lelli17}.
      The  dotted line  labeled
      with   ``deep-MOND''  is   the  MOND   prediction  in   the  low
      acceleration  regime  with a  slope  of  0.5.  The  dashed  line
      labeled with ``No Dark Matter'' is the no dark matter line.  }
\end{center}
\end{figure*}

As shown  in Figure~\ref{test_CDM_alternative}~(b), existing studies  prefer somewhat
larger  $\sigma/m$  on  galaxy scales  \citep{Elbert15,  Kaplinghat16,
  Zavala13} and smaller values  on cluster scales \citep{Kahlhoefer15,
  Randall08, Harvey15,  Kaplinghat16}. Such  a velocity  dependence of
the cross  section seems  to reconcile  results over  different scales
\citep{Kaplinghat16}.   However,  the cuspy  dark  matter  halo of  AGC
242019  may  challenge  this   simple  picture,  whose  upperlimit  to
$\sigma/m$ is  somewhat in tension with the  lowerbound of the
$\sigma/m$ range as required by cored halos of other dwarf
galaxies.

\subsubsection{Warm Dark Matter}

In warm dark matter, the density core has a size that can be approximately
by \cite{Hogan00, Maccio12}
\begin{equation}
 r^{2}_{\rm core} = \frac{\sqrt{3}}{4\pi G Q_{\rm max}}\frac{1}{<v_{\rm rms}^{2}>^{0.5}},
\end{equation}
where $v_{\rm rms}$ is the velocity dispersion (i.e. the mass) of the halo. $Q_{\rm max}$
is the maximum phase density as given by
\begin{equation}
 Q_{\rm max} = 1.64\times10^{-3} (\frac{\rho_{\rm L}}{\rho_{\rm cr}})(\frac{m}{\rm keV})^{4}\frac{\rm M_{\odot}\;pc^{-3}}{\rm (km\;s^{-1})^{3}},
\end{equation}
where $m$ is the mass of warm dark matter, $\frac{\rho_{\rm L}}{\rho_{\rm cr}}$ is the local density
relative to the critical density.
\begin{equation}
\frac{m}{\rm keV} = 0.37 \frac{1}{(\rho_{\rm L}/\rho_{\rm cr})^{0.25}}*(\frac{\rm kpc}{r_{\rm core}})^{0.5}(\frac{\rm km/s}{<v_{\rm rms}^{2}>^{0.5}})^{0.25}.
\end{equation}
  
As shown  in Figure~\ref{test_CDM_alternative}~(c), the innermost radius of the cuspy halo
of AGC 242019 leads to $m$ 
\wmdm\,. The figure also includes the constraints from the LITTLE THINGS galaxies \citep{Oh15} that are derived from the
core radius and the velocity at the outermost measurable radius listed in their Table 2. It seems that there is no consistent mass range that can explain all observational facts.   Warm dark  matter is  known to  have a  ``catch-22''
problem: the requirement for the particle mass to solve the cuspy-core
problem will  result in no  formation of  small galaxies at  the first
place. Our  discovery of a cuspy  dark matter halo in  AGC 242019 will
further challenge  the warm dark  matter scenario, such that,  even on
kpc scales, the  required particle mass cannot reconcile  with the constraint that
accounts for cored halos in some other dwarf galaxies.

\subsubsection{The modified Newtonian dynamics}

Unlike massive disk galaxies whose baryonic disk surface density rises
exponentially toward their  galactic centers, AGC 242019  shows a much
flatter  profile with  a  density deficit  in the  central
region.  Such  a distinct spatial  offset between the  baryonic matter
and the dynamical mass leads to an increasing baryonic matter relative
to  dark  matter   at  larger  radii,  in  contrast   to  galaxies  in
general.  This  can  be  quantified by  the  logarithmic  relationship
between  the  observed radial  acceleration  and  the baryonic  radial
acceleration as shown in  Figure~\ref{test_CDM_alternative}~(d). From the  inner (smaller  symbols) to  larger
radii (larger  symbols), the data  are more  closer to no  dark matter
line,  demonstrating the  increasing baryonic  matter relative  to the
dark matter at larger radii.

The modified Newtonian dynamics (MOND) paradigm \citep{Milgrom83} has
been proposed as an alternative to dark matter theory for interpreting
dynamical features.  However, MOND cannot  explain the dynamics of AGC
242019 as specified  by the radial acceleration  relationship shown in
Figure~\ref{test_CDM_alternative}~(d).  The  relationship  has a  slope  of
\radaccslope\, as given by the linear fitting of the data with errors
on both axes  \citep{Kelly07}.  However, the MOND only  allows a slope
ranging  from  1.0  in  the  classical   regime  to  0.5  at  the  low
acceleration limit by adjusting its fundamental constant $a_{0}$.  The
slope of  AGC 242019 is thus  \radaccslopesigma-$\sigma$ below the threshold  in the
MOND, although  it lies on the
the  extrapolation  of  the relationship defined by late-type  galaxies  \citep{Lelli17}.

\begin{figure}[tbh]
  \begin{center}
    \includegraphics[scale=0.4]{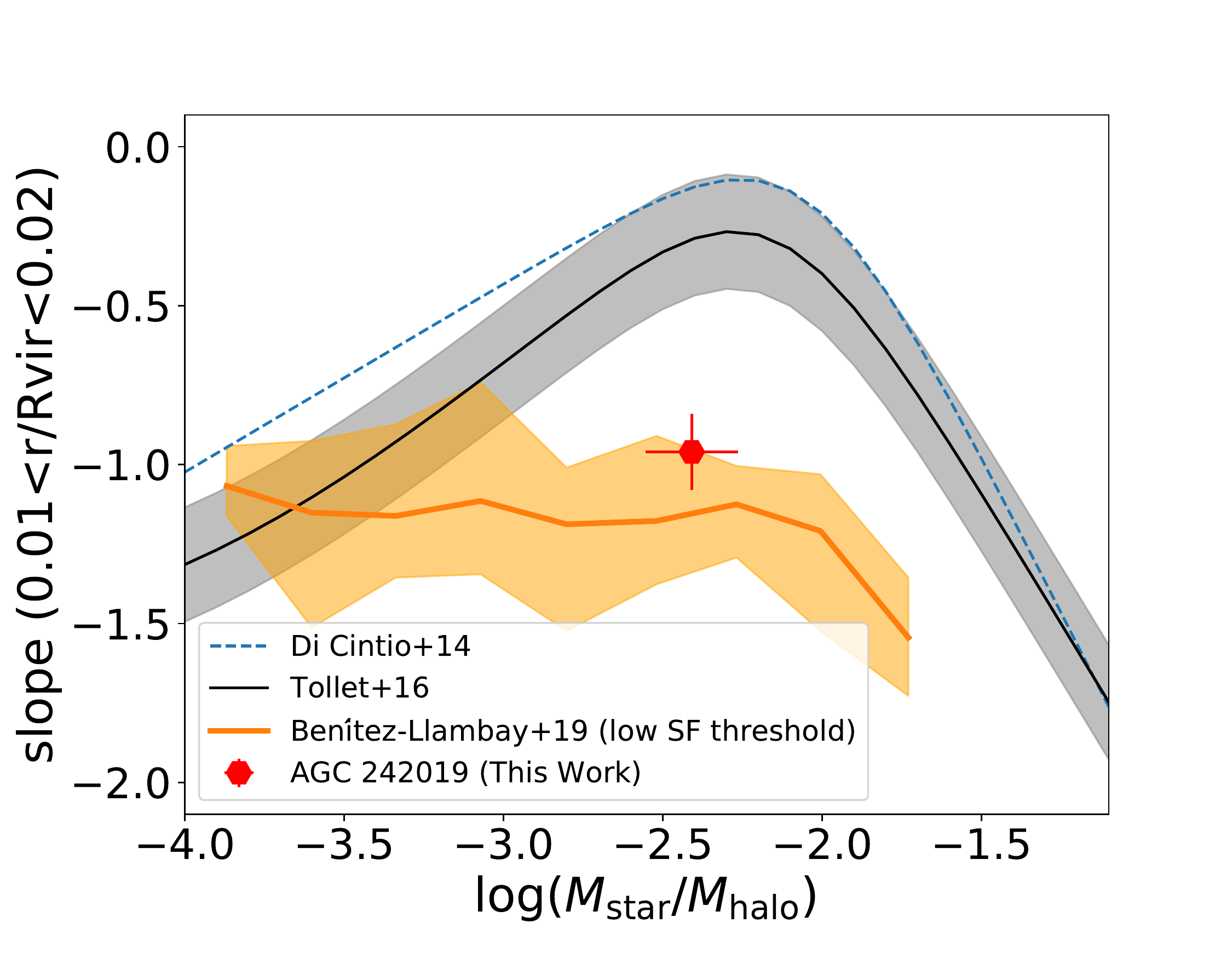}
 \caption{\label{test_feedback}  {\bf Implications  of AGC  242019 for
     the  stellar feedback  model.}  Three  lines represent  different
   models about the dark-matter central density slope as a function of
   the  $M_{\rm  star}/M_{\rm halo}$  ratio.  The  red diamond  symbol
   represents AGC 242019 in this study. }
  \end{center}
\end{figure}

\begin{figure}[tbh]
  \begin{center}
    \includegraphics[scale=0.4]{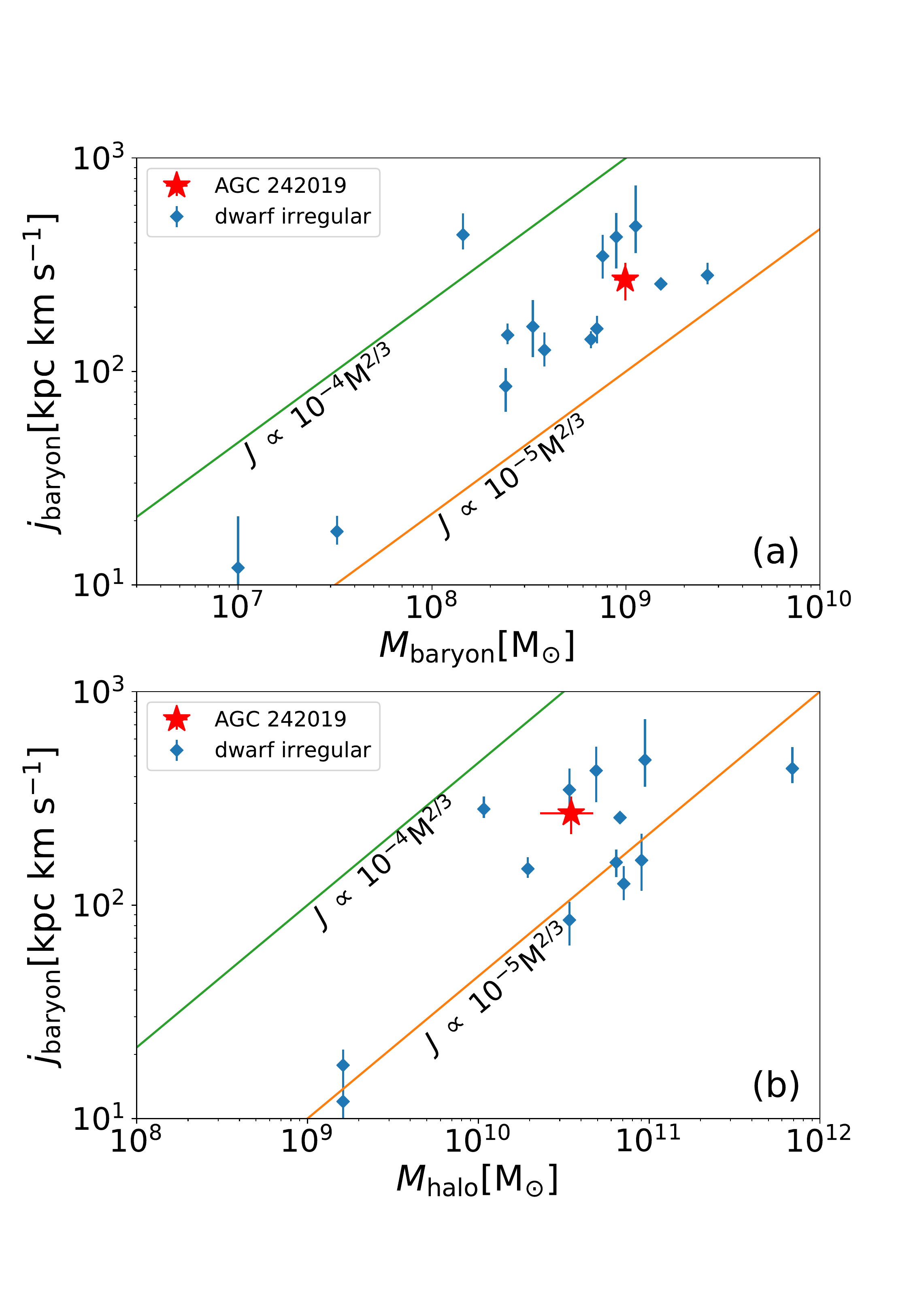}\\
     \vspace*{-1.0cm}\includegraphics[scale=0.4]{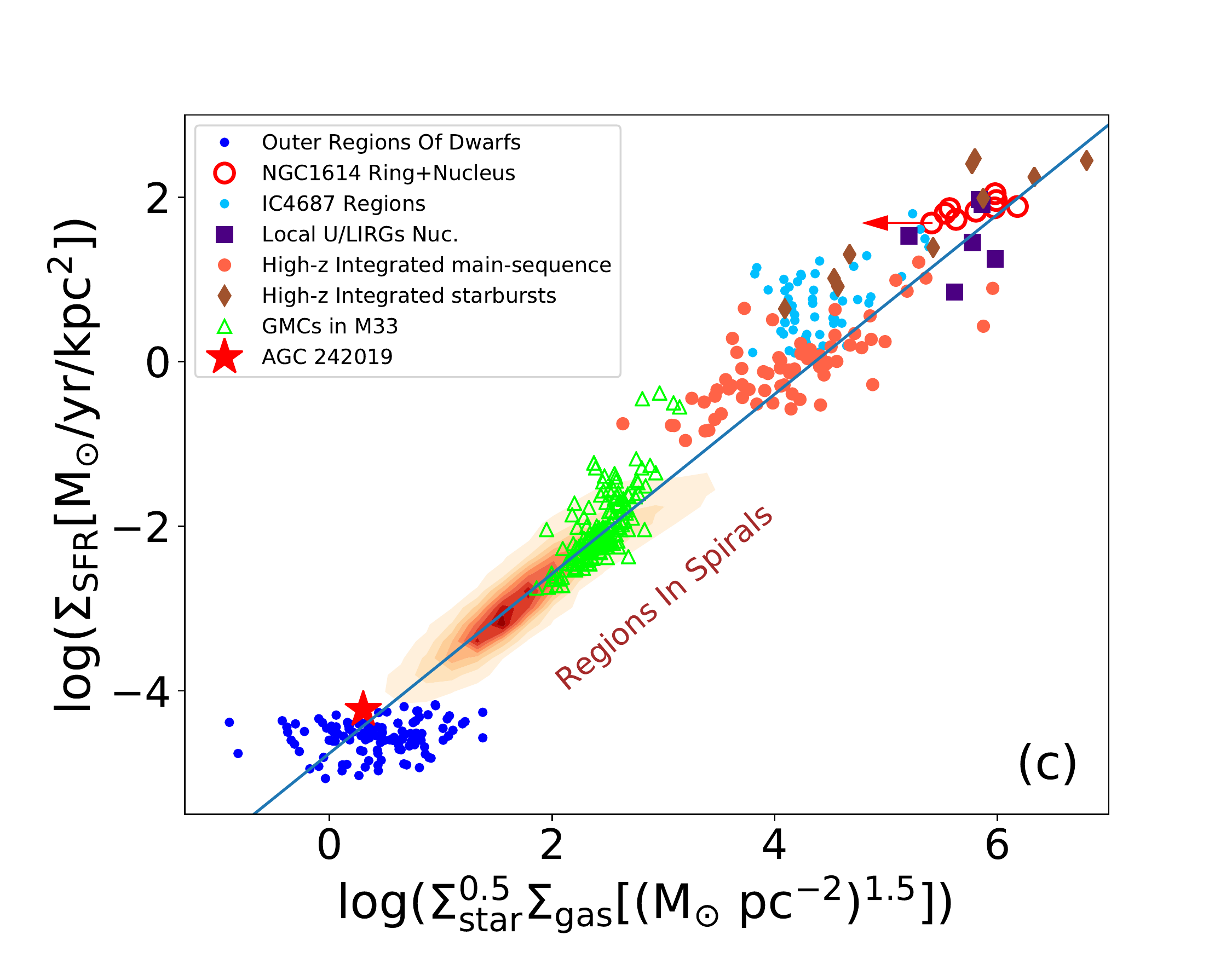}
      \caption{\label{test_UDG} {\bf The implication of AGC 242019 for the UDG formation.}
          {\bf a,} The specific angular momentum of the baryons as a function of
          the baryonic mass, where red
          star symbol is AGC 242019 as compared to dwarf irregulars \citep{Butler17}.
{\bf b,} the same as {\bf a,} but versus the halo mass. 
          {\bf c,} The location of AGC 242019 in the extended
          Schmidt law \citep{Shi18}. }
\end{center}
\end{figure}

\subsection{Implications for the role of feedback on producing cored
  dark matter halo}\label{sec_feedback}

It  is found  that the  effect of  baryonic feedback  on the  density
profile  of dark  matter  depends on  the  stellar-to-halo mass  ratio
($M_{*}/M_{\rm  halo}$)  \citep[e.g.][]{DiCintio14,  Tollet16}, as shown in Figure~\ref{test_feedback}.   In  the  low
$M_{*}/M_{\rm halo}$ regime  ($<$ 0.01\%), the feedback  is not strong
enough to expel baryons and no  dark-matter core  forms. In
the  high  $M_{*}/M_{\rm  halo}$  regime ($>$  3\%),  the  significant
contribution  to  the  potential  from  baryonic  matter  cancels out  the
feedback effect  and even produces  a cuspier profile.   In between, the
stellar  feedback can  efficiently  alter the  density  profile and  a
flattest density profile forms  around $M_{*}/M_{\rm halo}$=0.6\%. The above
result is independent of model parameters such as star
formation threshold, initial mass function, supernovae energy etc. However,
AGC 242019 has  a stellar  mass of  \mstar\, and a  halo mass  of \mhalo\,,
leading to  $M_{*}/M_{\rm halo}$ of \mstarmhalo\,, which is close to the
ratio  where a  flattest slope  should form  in simulations. AGC
242019 is thus inconsistent with the above  $M_{*}/M_{\rm halo}$ dependence
of the inner dark-matter profile.

Some  other  studies  emphasize   the  importance  of  star  formation
threshold    on     the    effect    of    the     stellar    feedback
\cite[e.g.][]{Governato10,  Benitez-Llambay19}. If  the threshold  for
gas to form stars is high, a large amount of gas can accumulate in the
center  of a  halo and  dominate the  potential before  star formation
takes  place.   The  subsequent feedback-driven  massive  outflows  or
repeated multiple outflows alter the  orbits of dark matter to produce
a  dark-matter  core.  On  the  other  hand,  for low  star  formation
threshold,  gas   is  expelled  by  feedback   before  it  contributes
significantly  to the  potential.   A dark-matter  cuspy profile  thus
preserves. This  scenario at  least partly  explains the  formation of
dark-matter  cores  in  some simulations  \citep{DiCintio14,  Fitts17,
  Maccio17}  while not  in others  \citep{Bose19}.  The  simulation by
\cite{Benitez-Llambay19} predicts an  intact cuspy dark-matter density
profile  independent of  the  $M_{*}/M_{\rm halo}$  if star  formation
threshold is low.  As  shown in Figure~\ref{test_feedback}, AGC 242019
is consistent with their prediction.  As a UDG, AGC 242019 does have a
low gas and  stellar mass surface density with  ongoing star formation
as revealed by the GALEX far-UV  image. This indicates that on sub-kpc
scales star formation in AGC 242019 can proceed at a very low gas mass
surface  density that  is on  the order  of 1  M$_{\odot}$/pc$^{2}$ or
0.4$\times$(100pc/$h$) cm$^{-3}$, where $h$ is the scale height of the
gas  disk.  However,  its   star  formation  efficiency  (SFR/gas=0.03
Gyr$^{-1}$)  is much  lower than  that in  spiral galaxies  ($\sim$0.3
Gyr$^{-1}$) \citep{Leroy08, Shi11}, inconsistent  with that adopted in
the simulations.   Therefore, low star formation  threshold should not
be simply the physical cause for a cuspy profile in AGC 242019.

Besides the star  formation threshold, the duration  of star formation
may   be   also   important   as  recognized   in   some   simulations
\citep{Read16a}.  As  long as  star  formation  proceeds long  enough, e.g.,
$\sim$10 Gyr  for a  halo mass  of 10$^{9}$  M$_{\odot}$ and a longer
timescale for a larger halo, a  halo core always  form. AGC  242019 has
a halo  mass of  \mhalo\,, and  its low
concentration implies late  formation time (see next  section), two of
which may explain its cuspy profile given the above scenario.

\subsection{Implications for formation of UDGs}\label{sec_udg}

 UDGs are  low-stellar-mass dwarfs  but with  sizes typical  of spiral
 galaxies  \citep{Abraham14,  Leisman17}.   They  were  found  in  all
 environments  including galaxy  clusters,  galaxy  groups and  field.
 Many mechanisms have  been proposed to understand  their origin: they
 may be  normal dwarf galaxies but  experience star-formation feedback
 that re-distributes gas and stars to larger radii \citep{Governato10,
   DiCintio14, Jiang19}; they may live in a high-spin dark matter halo
 with extended  gas distributions  and low efficiencies  in converting
 gas into stars \citep{Amorisco16}; some environmental effects such as
 ram-pressure  stripping or  tidal puffing  may be  also important  in
 formation of UDGs \citep{Yozin15, Jiang19}.

The dark matter  halo of AGC 242019  has a mass of  \mhalo\,, which is
typical of  a halo hosting a  dwarf. This suggests that  AGC 242019 is
not  a failed  massive galaxy,  unlike  other UDGs  found in  clusters
\citep{vanDokkum16,  Beasley16}.   Although   UDGs  may  have  diverse
origins,  our  measurements are   more  reliable. In  studies  of
\citet{vanDokkum16} and \citet{Beasley16}, the  halo mass was inferred
from 1-2 velocity data-points by  assuming a halo shape especially the
concentration.

The cuspy halo  of AGC 242019 also suggests that  the feedback has not
been  strong over  its history  to expel  a large  amount of  baryonic
matter to large distances. Otherwise, a cored halo should have already
formed like  in other dwarf  galaxies as suggested by  feedback models
\citep{Navarro96, Governato10}.  AGC 242019  thus has experienced weak
feedback  over its  history.  This  seems  to be  consistent with  the
deviation of  UDGs from the  Tully-Fisher relationship: gas  and stars
are not expelled out of the disk so that a UDG contain more baryons at
a given  maximum circular  velocity that  roughly represents  the halo
mass \citep{ManceraPina20}. AGC 242019 has a maximum circular velocity
of  \Vvir\, and  a baryonic  mass of  \mbaryon\,, placing  it slightly
above  the Tully-Fisher relationship  too. With  the accurate measurement  of the
halo mass, AGC 242019 is found to  be off both the $M_{*}$ vs. $M_{\rm
  halo}$  and  $M_{\rm  baryon}$   vs.  $M_{\rm  halo}$  relationships
\citep{Santos-Santos16}. The low SFR of  AGC 242019 also suggests weak
ongoing stellar  feedback as implied  by the relationship  between the
SFR and the ionized  gas velocity dispersion \citep[e.g.][]{Yu19}. Our
regular  velocity field  with small  non-circular motion  as shown  in
\S~\ref{sec_sysunc_curve}  is consistent  with  weak ongoing  feedback
too.

The halo of AGC 242019  has a small concentration of \concentration\,,
which is much smaller than the  median concentration of 15 at the same
halo  mass  in  the  local   Universe  as  expected  from  simulations
\citep{Maccio07}.   This difference  is most  likely because  that AGC
242019 is  an isolated halo that  origins from a tiny  initial density
peak in  the early time and  collapses recently. It is  found that the
halo concentration decreases with  later formation time \citep{Zhao03,
  Lu06}. A ``young'' halo thus  suggests late formation of AGC 242019,
which seems to be consistent with the findings of UDGs in cosmological
simulations \citep{Rong17}.

The specific angular  momentum can be derived from  the rotation curve
combined with  the stellar  and gas mass  profiles. The  mass profiles
were extended  to 15 kpc  following the same procedure  that estimates
the    baryonic   contribution    to   the    rotation   curve    (see
\S~\ref{sec_grav_gas} and \S~\ref{sec_grav_star}).  The rotation curve
is extended  to 15 kpc too  by combining the best-fitted  NFW rotation
curve    plus    the    baryonic   contribution.     As    shown    in
Figure~\ref{test_UDG}~(a), it is found that  AGC 242019 has an angular
momentum that is consistent with  dwarf irregular galaxies at the same
baryonic  mass \citep{Butler17}.   However, as  discussed before,  AGC
242019 has a higher baryon/halo mass  ratio as compared to galaxies in
general.  We derived  the halo  masses of  dwarf irregulars  following
\citet{Butler17}  and  compared them  with  AGC  242019. As  shown  in
Figure~\ref{test_UDG}~(b),  AGC 242019  still has  a similar  specific
angular momentum  at a given halo  mass as compared to  the average of
dwarf irregulars.  The  result may be against the model that  a UDG forms
in a high-spin dark matter halo \citep{Amorisco16}.

As shown in \S~\ref{sec_feedback}, AGC 242019 has a low star formation
efficiency,  consistent  with  the model  of  \citet{Amorisco16}.   If
placing it on the extended star formation law \citep{Shi11, Shi18}, we
found   that    AGC   242019   follows    the   law   as    shown   in
Figure~\ref{test_UDG} (c).  This suggests  that its low star formation
efficiency is regulated by its low stellar mass surface density.

In summary, AGC 242019 forms in  a dwarf-size halo with late formation
time. It has  weak feedback, low star formation  efficiency and a normal
specific angular momentum of baryons at a given halo mass.

\section{Conclusions}\label{sec:conclusion}

We have  carried out  the spatially-resolved  mapping of  gas dynamics
toward a  nearby UDG, AGC  242019. It is found  that AGC 242019  has a
cuspy dark  matter halo at  a high confidence, which  demonstrates the
validity of the  cold dark matter paradigm on  subgalactic scales. Our
main conclusions are
  
(1)  AGC  242019  has  an  overall  regular  velocity  field.   After
subtracting  the baryonic  contribution,  the rotation  curve of  dark
matter  is well  fitted  by  the cuspy  profile  as  described by  the
Navarro-Frenk-White (NFW)  model, while  the cored  profiles including
both the  pseudo-isothermal and  Burkert models  are excluded.  The   result  is   robust  against   various  systematic
uncertainties.
  
(2) The central  density slope of
dark matter halo is found to be \innerslope\, at the innermost radius of \innerradius.

  (3) AGC  242019 poses  challenges to  alternatives of  standard cold
  dark matter by  constraining the particle mass of  fuzzy dark matter
  to  be  \fuzzylowerthreehpd,   or  \fuzzyupperthreehpd\,,  the   cross  section  of
  self-interacting dark matter  to be \sidm, and the  particle mass of
  warm dark matter to be \wmdm, all of which are in tension with other
  constraints.

  (4) AGC 242019 lies on  the extrapolation of the radial acceleration
  relationship as defined by spirals  and dwarf galaxies. However, the
  slope of the  relationship defined by AGC  242019 is \radaccslope\,,
  3.2-$\sigma$  below the  threshold (0.5)  of the  modified Newtonian
  dynamics.

  (5) In the  cold dark matter paradigm, the cuspy  halo of AGC 242019
  thus supports the  feedback scenario that transforms  cuspy halos to
  cored  halos as  frequently seen  in other  galaxies.  However,  the
  detailed physical process is unclear.   The cuspy halo of AGC 242019
  is inconsistent with the stellar-to-halo-mass-ratio dependent model,
  while consistent with  the star-formation-threshold dependent model.
  But  even for  the  later, the  observed  star formation  efficiency
  (SFR/gas) is much lower than what  is adopted in simulations. It may
  be consistent with the scenario  that the duration of star formation
  is the key driver.
   
  (6) As a UDG, AGC 242019 has  a halo mass of \mhalo\,, implying its
  formation in a dwarf-size halo. The cuspy halo further suggests weak
  feedback over  the history. The  small concentration of its  halo is
  consistent with  late formation time. Its  specific angular momentum
  of baryons is  consistent with the average of dwarf  irregulars at a
  given halo/baryonic  mass.  Its star  formation efficiency (SFR/gas)  is low,
  probably due to the low stellar mass surface density.

  \clearpage
  
\appendix
\restartappendixnumbering
 
\section{The full setup to run \texttt{$^{3D}$Barolo}}

In Table~\ref{tab_full_par}, the full list of the parameters to run \texttt{$^{3D}$Barolo}
is listed.
 
\begin{table}
\footnotesize
\caption{\label{tab_full_par} The full parameter list for $^{3D}$Barolo }
\begin{tabular}{llllllllllllll}
\hline
Parameters                   & Values               \\
\hline
Checking for bad channels in the cube........[checkChannels]  &  false               \\
Using Robust statistics?...................[flagRobustStats]  &  true                \\
Writing the mask to a fitsfile....................[MAKEMASK]  &  false               \\
Searching for sources in cube?......................[SEARCH]  &  false               \\
Smoothing the datacube?.............................[SMOOTH]  &  false               \\
Hanning smoothing the datacube?....................[HANNING]  &  false               \\
Writing a 3D model?.................................[GALMOD]  &  false               \\
Fitting a 3D model to the datacube?..................[3DFIT]  &  true                \\
   Number of radii..................................[NRADII]  &  7                   \\
   Separation between radii (arcsec)................[RADSEP]  &  9                   \\
   X center of the galaxy (pixel).....................[XPOS]  &  415.19              \\
   Y center of the galaxy (pixel).....................[YPOS]  &  405.49              \\
   Systemic velocity of the galaxy (km/s).............[VSYS]  &  1840.46             \\
   Initial global rotation velocity (km/s)............[VROT]  &  30                  \\
   Initial global radial velocity (km/s)..............[VRAD]  &  -1                  \\
   Initial global velocity dispersion (km/s).........[VDISP]  &  5                   \\
   Initial global inclination (degrees)................[INC]  &  69.90               \\
   Initial global position angle (degrees)..............[PA]  &  2.0025              \\
   Scale height of the disk (arcsec)....................[Z0]  &  0.7                 \\
   Global column density of gas (atoms/cm2)...........[DENS]  &  -1                  \\
   Parameters to be minimized.........................[FREE]  &  VROT,VDISP,PA,INC   \\
   Type of mask.......................................[MASK]  &  SEARCH              \\
   Side of the galaxy to be used......................[SIDE]  &  B                   \\
   Type of normalization..............................[NORM]  &  LOCAL               \\
   Layer type along z direction......................[LTYPE]  &  gaussian            \\
   Residuals to minimize.............................[FTYPE]  &  chi-squared         \\
   Weighting function................................[WFUNC]  &  uniform             \\
   Weight for blank pixels.........................[BWEIGHT]  &  1                   \\
   Minimization tolerance..............................[TOL]  &  0.001               \\
   What side of the galaxy to be used.................[SIDE]  &  B                   \\
   Two stages minimization?.......................[TWOSTAGE]  &  true                \\
     Degree of polynomial fitting angles?............[POLYN]  &  bezier              \\
   Estimating errors?...........................[FLAGERRORS]  &  true                \\
   Redshift of the galaxy?........................[REDSHIFT]  &  0                   \\
   Computing asymmetric drift correction?...........[ADRIFT]  &  true                \\
   Overlaying mask to output plots?...............[PLOTMASK]  &  false               \\
   RMS noise to add to the model..................[NOISERMS]  &  false               \\
   Using cumulative rings during the fit?.......[CUMULATIVE]  &  false               \\
Full parameter space for a pair of parameters.....[SPACEPAR]  &  false               \\
Generating a 3D datacube with a wind model?........[GALWIND]  &  false               \\
Fitting velocity field with a ring model?............[2DFIT]  &  false               \\
Deriving radial intensity profile?.................[ELLPROF]  &  false               \\   

\hline
\end{tabular}
\end{table}

\section{Different runs of \texttt{$^{3D}$Barolo}}

\begin{figure}[tbh]
  \begin{center}
    \includegraphics[scale=0.4]{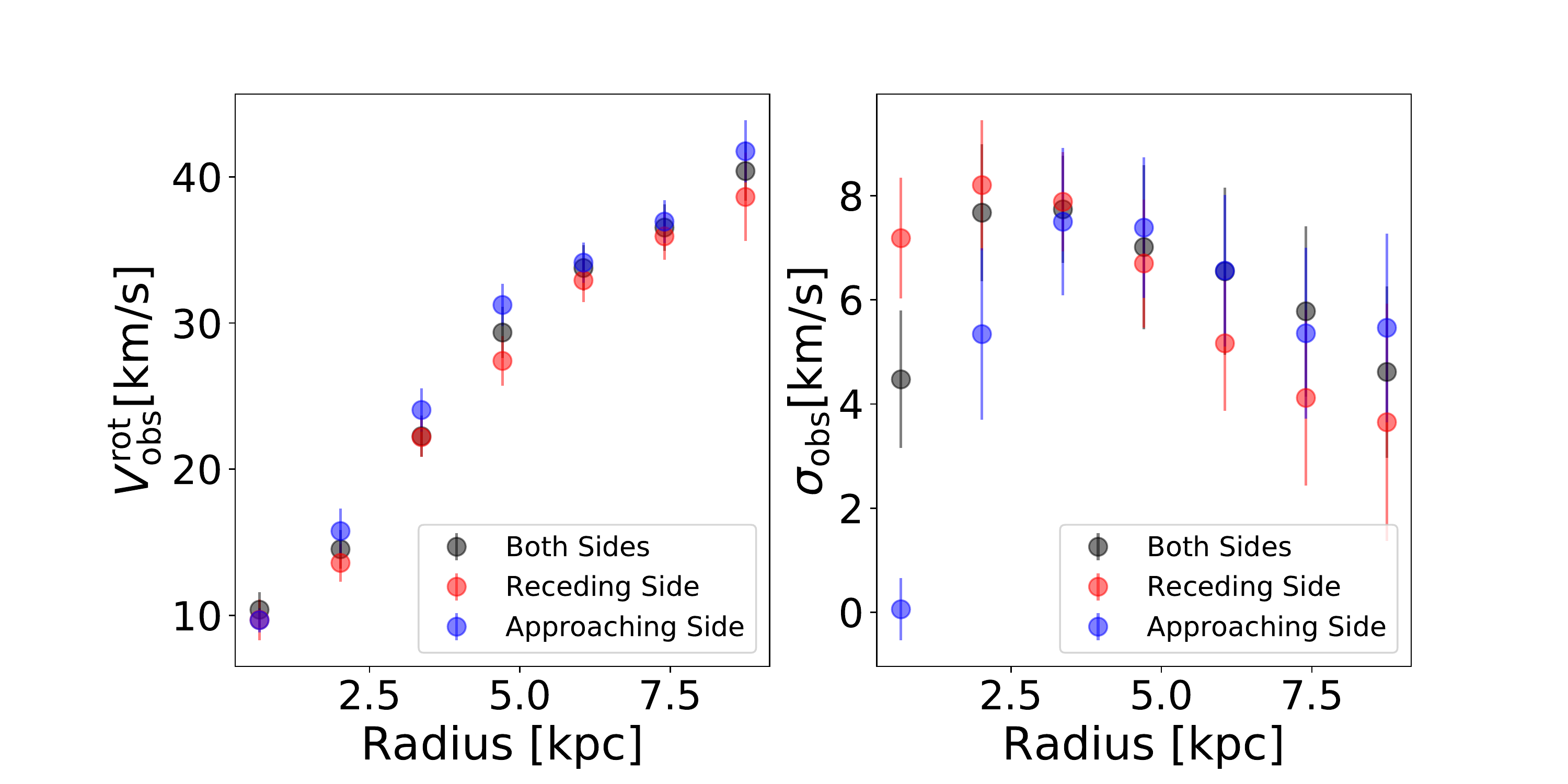}\\
     \caption{\label{comp_vel_reced_approach} The comparison in the
       derived rotation velocities and velocity dispersions
       by using both sides, only the receding side
       and only the approaching side of HI data, respectively. }
\end{center}
\end{figure}

In Figure~\ref{comp_vel_reced_approach}, we compared the derived
rotation velocities (before the pressure support correction) and velocity dispersions
by running \texttt{$^{3D}$Barolo} with both sides of the gas disk,
only the receding side and only the approaching side, respectively.

\acknowledgments

We thanks a lot the referee for his/her very helpful and constructive
comments that improve the paper significantly.
Y.S. acknowledges  the support from  the National Key R\&D  Program of
China (No.  2018YFA0404502, No.  2017YFA0402704), the National Natural
Science  Foundation  of  China  (NSFC grants  11825302,  11733002  and
11773013), and the Tencent Foundation  through the XPLORER PRIZE.  The
National Radio  Astronomy Observatory  is a  facility of  the National
Science Foundation operated under  cooperative agreement by Associated
Universities, Inc.  This  publication makes use of  data products from
the Wide-field Infrared  Survey Explorer, which is a  joint project of
the  University of  California, Los  Angeles, and  the Jet  Propulsion
Laboratory/California Institute of Technology,  funded by the National
Aeronautics and Space Administration.

\bibliography{ms}{}

\begin{thebibliography}{}
\expandafter\ifx\csname natexlab\endcsname\relax\def\natexlab#1{#1}\fi
\providecommand{\url}[1]{\href{#1}{#1}}
\providecommand{\dodoi}[1]{doi:~\href{http://doi.org/#1}{\nolinkurl{#1}}}
\providecommand{\doeprint}[1]{\href{http://ascl.net/#1}{\nolinkurl{http://ascl.net/#1}}}
\providecommand{\doarXiv}[1]{\href{https://arxiv.org/abs/#1}{\nolinkurl{https://arxiv.org/abs/#1}}}

\bibitem[{{Abraham} \& {van Dokkum}(2014)}]{Abraham14}
{Abraham}, R.~G., \& {van Dokkum}, P.~G. 2014, \pasp, 126, 55,
  \dodoi{10.1086/674875}

\bibitem[{{Adams} {et~al.}(2014){Adams}, {Simon}, {Fabricius}, {van den Bosch},
  {Barentine}, {Bender}, {Gebhardt}, {Hill}, {Murphy}, {Swaters}, {Thomas}, \&
  {van de Ven}}]{Adams14}
{Adams}, J.~J., {Simon}, J.~D., {Fabricius}, M.~H., {et~al.} 2014, \apj, 789,
  63, \dodoi{10.1088/0004-637X/789/1/63}

\bibitem[{{Amorisco} \& {Loeb}(2016)}]{Amorisco16}
{Amorisco}, N.~C., \& {Loeb}, A. 2016, \mnras, 459, L51,
  \dodoi{10.1093/mnrasl/slw055}

\bibitem[{{Avila-Reese} {et~al.}(2001){Avila-Reese}, {Col{\'\i}n},
  {Valenzuela}, {D'Onghia}, \& {Firmani}}]{Avila-Reese01}
{Avila-Reese}, V., {Col{\'\i}n}, P., {Valenzuela}, O., {D'Onghia}, E., \&
  {Firmani}, C. 2001, \apj, 559, 516, \dodoi{10.1086/322411}

\bibitem[{{Bailin} \& {Steinmetz}(2005)}]{Bailin05}
{Bailin}, J., \& {Steinmetz}, M. 2005, \apj, 627, 647, \dodoi{10.1086/430397}

\bibitem[{{Beasley} {et~al.}(2016){Beasley}, {Romanowsky}, {Pota}, {Navarro},
  {Martinez Delgado}, {Neyer}, \& {Deich}}]{Beasley16}
{Beasley}, M.~A., {Romanowsky}, A.~J., {Pota}, V., {et~al.} 2016, \apjl, 819,
  L20, \dodoi{10.3847/2041-8205/819/2/L20}

\bibitem[{{Begeman}(1989)}]{Begeman89}
{Begeman}, K.~G. 1989, \aap, 223, 47

\bibitem[{{Begeman} {et~al.}(1991){Begeman}, {Broeils}, \&
  {Sanders}}]{Begeman91}
{Begeman}, K.~G., {Broeils}, A.~H., \& {Sanders}, R.~H. 1991, \mnras, 249, 523,
  \dodoi{10.1093/mnras/249.3.523}

\bibitem[{{Bell} {et~al.}(2003){Bell}, {McIntosh}, {Katz}, \&
  {Weinberg}}]{Bell03}
{Bell}, E.~F., {McIntosh}, D.~H., {Katz}, N., \& {Weinberg}, M.~D. 2003, \apjs,
  149, 289, \dodoi{10.1086/378847}

\bibitem[{{Ben{\'\i}tez-Llambay} {et~al.}(2019){Ben{\'\i}tez-Llambay}, {Frenk},
  {Ludlow}, \& {Navarro}}]{Benitez-Llambay19}
{Ben{\'\i}tez-Llambay}, A., {Frenk}, C.~S., {Ludlow}, A.~D., \& {Navarro},
  J.~F. 2019, \mnras, 488, 2387, \dodoi{10.1093/mnras/stz1890}

\bibitem[{{Blitz} \& {Rosolowsky}(2006)}]{Blitz06}
{Blitz}, L., \& {Rosolowsky}, E. 2006, \apj, 650, 933, \dodoi{10.1086/505417}

\bibitem[{{Bose} {et~al.}(2019){Bose}, {Frenk}, {Jenkins}, {Fattahi},
  {G{\'o}mez}, {Grand }, {Marinacci}, {Navarro}, {Oman}, {Pakmor}, {Schaye},
  {Simpson}, \& {Springel}}]{Bose19}
{Bose}, S., {Frenk}, C.~S., {Jenkins}, A., {et~al.} 2019, \mnras, 486, 4790,
  \dodoi{10.1093/mnras/stz1168}

\bibitem[{{Bureau} \& {Carignan}(2002)}]{Bureau02}
{Bureau}, M., \& {Carignan}, C. 2002, \aj, 123, 1316, \dodoi{10.1086/338899}

\bibitem[{{Burkert}(1995)}]{Burkert95}
{Burkert}, A. 1995, \apjl, 447, L25, \dodoi{10.1086/309560}

\bibitem[{{Butler} {et~al.}(2017){Butler}, {Obreschkow}, \& {Oh}}]{Butler17}
{Butler}, K.~M., {Obreschkow}, D., \& {Oh}, S.-H. 2017, \apjl, 834, L4,
  \dodoi{10.3847/2041-8213/834/1/L4}

\bibitem[{{Carignan} \& {Beaulieu}(1989)}]{Carignan89}
{Carignan}, C., \& {Beaulieu}, S. 1989, \apj, 347, 760, \dodoi{10.1086/168167}

\bibitem[{{Chang} \& {Necib}(2020)}]{Chang20}
{Chang}, L.~J., \& {Necib}, L. 2020, arXiv e-prints, arXiv:2009.00613.
\newblock \doarXiv{2009.00613}

\bibitem[{{Childress} {et~al.}(2014){Childress}, {Vogt}, {Nielsen}, \&
  {Sharp}}]{Childress14}
{Childress}, M., {Vogt}, F., {Nielsen}, J., \& {Sharp}, R. 2014, {PyWiFeS: Wide
  Field Spectrograph data reduction pipeline}.
\newblock \doeprint{1402.034}

\bibitem[{{de Blok} {et~al.}(2001){de Blok}, {McGaugh}, {Bosma}, \&
  {Rubin}}]{deBlok01}
{de Blok}, W.~J.~G., {McGaugh}, S.~S., {Bosma}, A., \& {Rubin}, V.~C. 2001,
  \apjl, 552, L23, \dodoi{10.1086/320262}

\bibitem[{{Dey} {et~al.}(2019){Dey}, {Schlegel}, {Lang}, {Blum}, {Burleigh},
  {Fan}, {Findlay}, {Finkbeiner}, {Herrera}, {Juneau}, {Landriau}, {Levi},
  {McGreer}, {Meisner}, {Myers}, {Moustakas}, {Nugent}, {Patej}, {Schlafly},
  {Walker}, {Valdes}, {Weaver}, {Y{\`e}che}, {Zou}, {Zhou}, {Abareshi},
  {Abbott}, {Abolfathi}, {Aguilera}, {Alam}, {Allen}, {Alvarez}, {Annis},
  {Ansarinejad}, {Aubert}, {Beechert}, {Bell}, {BenZvi}, {Beutler}, {Bielby},
  {Bolton}, {Brice{\~n}o}, {Buckley-Geer}, {Butler}, {Calamida}, {Carlberg},
  {Carter}, {Casas}, {Castander}, {Choi}, {Comparat}, {Cukanovaite}, {Delubac},
  {DeVries}, {Dey}, {Dhungana}, {Dickinson}, {Ding}, {Donaldson}, {Duan},
  {Duckworth}, {Eftekharzadeh}, {Eisenstein}, {Etourneau}, {Fagrelius},
  {Farihi}, {Fitzpatrick}, {Font-Ribera}, {Fulmer}, {G{\"a}nsicke},
  {Gaztanaga}, {George}, {Gerdes}, {Gontcho}, {Gorgoni}, {Green}, {Guy},
  {Harmer}, {Hernand ez}, {Honscheid}, {Huang}, {James}, {Jannuzi}, {Jiang},
  {Joyce}, {Karcher}, {Karkar}, {Kehoe}, {Kneib}, {Kueter-Young}, {Lan},
  {Lauer}, {Le Guillou}, {Le Van Suu}, {Lee}, {Lesser}, {Perreault Levasseur},
  {Li}, {Mann}, {Marshall}, {Mart{\'\i}nez-V{\'a}zquez}, {Martini}, {du Mas des
  Bourboux}, {McManus}, {Meier}, {M{\'e}nard}, {Metcalfe},
  {Mu{\~n}oz-Guti{\'e}rrez}, {Najita}, {Napier}, {Narayan}, {Newman}, {Nie},
  {Nord}, {Norman}, {Olsen}, {Paat}, {Palanque-Delabrouille}, {Peng},
  {Poppett}, {Poremba}, {Prakash}, {Rabinowitz}, {Raichoor}, {Rezaie},
  {Robertson}, {Roe}, {Ross}, {Ross}, {Rudnick}, {Safonova}, {Saha},
  {S{\'a}nchez}, {Savary}, {Schweiker}, {Scott}, {Seo}, {Shan}, {Silva},
  {Slepian}, {Soto}, {Sprayberry}, {Staten}, {Stillman}, {Stupak}, {Summers},
  {Sien Tie}, {Tirado}, {Vargas-Maga{\~n}a}, {Vivas}, {Wechsler}, {Williams},
  {Yang}, {Yang}, {Yapici}, {Zaritsky}, {Zenteno}, {Zhang}, {Zhang}, {Zhou}, \&
  {Zhou}}]{Dey19}
{Dey}, A., {Schlegel}, D.~J., {Lang}, D., {et~al.} 2019, \aj, 157, 168,
  \dodoi{10.3847/1538-3881/ab089d}

\bibitem[{{Di Cintio} {et~al.}(2014){Di Cintio}, {Brook}, {Macci{\`o}},
  {Stinson}, {Knebe}, {Dutton}, \& {Wadsley}}]{DiCintio14}
{Di Cintio}, A., {Brook}, C.~B., {Macci{\`o}}, A.~V., {et~al.} 2014, \mnras,
  437, 415, \dodoi{10.1093/mnras/stt1891}

\bibitem[{{Di Teodoro} \& {Fraternali}(2015)}]{DiTeodoro15}
{Di Teodoro}, E.~M., \& {Fraternali}, F. 2015, \mnras, 451, 3021,
  \dodoi{10.1093/mnras/stv1213}

\bibitem[{{Einasto}(1965)}]{Einasto65}
{Einasto}, J. 1965, Trudy Astrofizicheskogo Instituta Alma-Ata, 5, 87

\bibitem[{{Elbert} {et~al.}(2015){Elbert}, {Bullock}, {Garrison-Kimmel},
  {Rocha}, {O{\~n}orbe}, \& {Peter}}]{Elbert15}
{Elbert}, O.~D., {Bullock}, J.~S., {Garrison-Kimmel}, S., {et~al.} 2015,
  \mnras, 453, 29, \dodoi{10.1093/mnras/stv1470}

\bibitem[{{Evans} {et~al.}(2009){Evans}, {An}, \& {Walker}}]{Evans09}
{Evans}, N.~W., {An}, J., \& {Walker}, M.~G. 2009, \mnras, 393, L50,
  \dodoi{10.1111/j.1745-3933.2008.00596.x}

\bibitem[{{Fitts} {et~al.}(2017){Fitts}, {Boylan-Kolchin}, {Elbert}, {Bullock},
  {Hopkins}, {O{\~n}orbe}, {Wetzel}, {Wheeler}, {Faucher-Gigu{\`e}re},
  {Kere{\v{s}}}, {Skillman}, \& {Weisz}}]{Fitts17}
{Fitts}, A., {Boylan-Kolchin}, M., {Elbert}, O.~D., {et~al.} 2017, \mnras, 471,
  3547, \dodoi{10.1093/mnras/stx1757}

\bibitem[{{Ghigna} {et~al.}(2000){Ghigna}, {Moore}, {Governato}, {Lake},
  {Quinn}, \& {Stadel}}]{Ghigna00}
{Ghigna}, S., {Moore}, B., {Governato}, F., {et~al.} 2000, \apj, 544, 616,
  \dodoi{10.1086/317221}

\bibitem[{{Governato} {et~al.}(2010){Governato}, {Brook}, {Mayer}, {Brooks},
  {Rhee}, {Wadsley}, {Jonsson}, {Willman}, {Stinson}, {Quinn}, \&
  {Madau}}]{Governato10}
{Governato}, F., {Brook}, C., {Mayer}, L., {et~al.} 2010, \nat, 463, 203,
  \dodoi{10.1038/nature08640}

\bibitem[{{Harvey} {et~al.}(2015){Harvey}, {Massey}, {Kitching}, {Taylor}, \&
  {Tittley}}]{Harvey15}
{Harvey}, D., {Massey}, R., {Kitching}, T., {Taylor}, A., \& {Tittley}, E.
  2015, Science, 347, 1462, \dodoi{10.1126/science.1261381}

\bibitem[{{Hayashi} {et~al.}(2020){Hayashi}, {Chiba}, \&
  {Ishiyama}}]{Hayashi20}
{Hayashi}, K., {Chiba}, M., \& {Ishiyama}, T. 2020, arXiv e-prints,
  arXiv:2007.13780.
\newblock \doarXiv{2007.13780}

\bibitem[{{Haynes} {et~al.}(2018){Haynes}, {Giovanelli}, {Kent}, {Adams},
  {Balonek}, {Craig}, {Fertig}, {Finn}, {Giovanardi}, {Hallenbeck}, {Hess},
  {Hoffman}, {Huang}, {Jones}, {Koopmann}, {Kornreich}, {Leisman}, {Miller},
  {Moorman}, {O'Connor}, {O'Donoghue}, {Papastergis}, {Troischt}, {Stark}, \&
  {Xiao}}]{Haynes18}
{Haynes}, M.~P., {Giovanelli}, R., {Kent}, B.~R., {et~al.} 2018, \apj, 861, 49,
  \dodoi{10.3847/1538-4357/aac956}

\bibitem[{{Hogan} \& {Dalcanton}(2000)}]{Hogan00}
{Hogan}, C.~J., \& {Dalcanton}, J.~J. 2000, \prd, 62, 063511,
  \dodoi{10.1103/PhysRevD.62.063511}

\bibitem[{{Hu} {et~al.}(2000){Hu}, {Barkana}, \& {Gruzinov}}]{Hu00}
{Hu}, W., {Barkana}, R., \& {Gruzinov}, A. 2000, \prl, 85, 1158,
  \dodoi{10.1103/PhysRevLett.85.1158}

\bibitem[{{Iorio} {et~al.}(2017){Iorio}, {Fraternali}, {Nipoti}, {Di Teodoro},
  {Read}, \& {Battaglia}}]{Iorio17}
{Iorio}, G., {Fraternali}, F., {Nipoti}, C., {et~al.} 2017, \mnras, 466, 4159,
  \dodoi{10.1093/mnras/stw3285}

\bibitem[{{Jardel} {et~al.}(2013){Jardel}, {Gebhardt}, {Fabricius}, {Drory}, \&
  {Williams}}]{Jardel13}
{Jardel}, J.~R., {Gebhardt}, K., {Fabricius}, M.~H., {Drory}, N., \&
  {Williams}, M.~J. 2013, \apj, 763, 91, \dodoi{10.1088/0004-637X/763/2/91}

\bibitem[{{Jarrett} {et~al.}(2013){Jarrett}, {Masci}, {Tsai}, {Petty},
  {Cluver}, {Assef}, {Benford}, {Blain}, {Bridge}, {Donoso}, {Eisenhardt},
  {Koribalski}, {Lake}, {Neill}, {Seibert}, {Sheth}, {Stanford}, \&
  {Wright}}]{Jarrett13}
{Jarrett}, T.~H., {Masci}, F., {Tsai}, C.~W., {et~al.} 2013, \aj, 145, 6,
  \dodoi{10.1088/0004-6256/145/1/6}

\bibitem[{{Jester} {et~al.}(2005){Jester}, {Schneider}, {Richards}, {Green},
  {Schmidt}, {Hall}, {Strauss}, {Vand en Berk}, {Stoughton}, {Gunn},
  {Brinkmann}, {Kent}, {Smith}, {Tucker}, \& {Yanny}}]{Jester05}
{Jester}, S., {Schneider}, D.~P., {Richards}, G.~T., {et~al.} 2005, \aj, 130,
  873, \dodoi{10.1086/432466}

\bibitem[{{Jiang} {et~al.}(2019){Jiang}, {Dekel}, {Freundlich}, {Romanowsky},
  {Dutton}, {Macci{\`o}}, \& {Di Cintio}}]{Jiang19}
{Jiang}, F., {Dekel}, A., {Freundlich}, J., {et~al.} 2019, \mnras, 487, 5272,
  \dodoi{10.1093/mnras/stz1499}

\bibitem[{{Jing} \& {Suto}(2002)}]{Jing02}
{Jing}, Y.~P., \& {Suto}, Y. 2002, \apj, 574, 538, \dodoi{10.1086/341065}

\bibitem[{{Jobin} \& {Carignan}(1990)}]{Marc90}
{Jobin}, M., \& {Carignan}, C. 1990, \aj, 100, 648, \dodoi{10.1086/115548}

\bibitem[{{Kahlhoefer} {et~al.}(2015){Kahlhoefer}, {Schmidt-Hoberg}, {Kummer},
  \& {Sarkar}}]{Kahlhoefer15}
{Kahlhoefer}, F., {Schmidt-Hoberg}, K., {Kummer}, J., \& {Sarkar}, S. 2015,
  \mnras, 452, L54, \dodoi{10.1093/mnrasl/slv088}

\bibitem[{{Kaplinghat} {et~al.}(2016){Kaplinghat}, {Tulin}, \&
  {Yu}}]{Kaplinghat16}
{Kaplinghat}, M., {Tulin}, S., \& {Yu}, H.-B. 2016, \prl, 116, 041302,
  \dodoi{10.1103/PhysRevLett.116.041302}

\bibitem[{{Kelly}(2007)}]{Kelly07}
{Kelly}, B.~C. 2007, \apj, 665, 1489, \dodoi{10.1086/519947}

\bibitem[{{Knebe} \& {Wie{\ss}ner}(2006)}]{Knebe06}
{Knebe}, A., \& {Wie{\ss}ner}, V. 2006, \pasa, 23, 125, \dodoi{10.1071/AS06013}

\bibitem[{{Kuzio de Naray} \& {Kaufmann}(2011)}]{KuziodeNaray11}
{Kuzio de Naray}, R., \& {Kaufmann}, T. 2011, \mnras, 414, 3617,
  \dodoi{10.1111/j.1365-2966.2011.18656.x}

\bibitem[{{Kuzio de Naray} {et~al.}(2008){Kuzio de Naray}, {McGaugh}, \& {de
  Blok}}]{KuziodeNaray08}
{Kuzio de Naray}, R., {McGaugh}, S.~S., \& {de Blok}, W.~J.~G. 2008, \apj, 676,
  920, \dodoi{10.1086/527543}

\bibitem[{{Kuzio de Naray} {et~al.}(2009){Kuzio de Naray}, {McGaugh}, \&
  {Mihos}}]{KuziodeNaray09}
{Kuzio de Naray}, R., {McGaugh}, S.~S., \& {Mihos}, J.~C. 2009, \apj, 692,
  1321, \dodoi{10.1088/0004-637X/692/2/1321}

\bibitem[{{Lake} {et~al.}(1990){Lake}, {Schommer}, \& {van Gorkom}}]{Lake90}
{Lake}, G., {Schommer}, R.~A., \& {van Gorkom}, J.~H. 1990, \aj, 99, 547,
  \dodoi{10.1086/115349}

\bibitem[{{Leisman} {et~al.}(2017){Leisman}, {Haynes}, {Janowiecki},
  {Hallenbeck}, {J{\'o}zsa}, {Giovanelli}, {Adams}, {Bernal Neira}, {Cannon},
  {Janesh}, {Rhode}, \& {Salzer}}]{Leisman17}
{Leisman}, L., {Haynes}, M.~P., {Janowiecki}, S., {et~al.} 2017, \apj, 842,
  133, \dodoi{10.3847/1538-4357/aa7575}

\bibitem[{{Lelli} {et~al.}(2017){Lelli}, {McGaugh}, {Schombert}, \&
  {Pawlowski}}]{Lelli17}
{Lelli}, F., {McGaugh}, S.~S., {Schombert}, J.~M., \& {Pawlowski}, M.~S. 2017,
  \apj, 836, 152, \dodoi{10.3847/1538-4357/836/2/152}

\bibitem[{{Leroy} {et~al.}(2008){Leroy}, {Walter}, {Brinks}, {Bigiel}, {de
  Blok}, {Madore}, \& {Thornley}}]{Leroy08}
{Leroy}, A.~K., {Walter}, F., {Brinks}, E., {et~al.} 2008, \aj, 136, 2782,
  \dodoi{10.1088/0004-6256/136/6/2782}

\bibitem[{{Lovell} {et~al.}(2014){Lovell}, {Frenk}, {Eke}, {Jenkins}, {Gao}, \&
  {Theuns}}]{Lovell14}
{Lovell}, M.~R., {Frenk}, C.~S., {Eke}, V.~R., {et~al.} 2014, \mnras, 439, 300,
  \dodoi{10.1093/mnras/stt2431}

\bibitem[{{Lu} {et~al.}(2006){Lu}, {Mo}, {Katz}, \& {Weinberg}}]{Lu06}
{Lu}, Y., {Mo}, H.~J., {Katz}, N., \& {Weinberg}, M.~D. 2006, \mnras, 368,
  1931, \dodoi{10.1111/j.1365-2966.2006.10270.x}

\bibitem[{{Macci{\`o}} {et~al.}(2007){Macci{\`o}}, {Dutton}, {van den Bosch},
  {Moore}, {Potter}, \& {Stadel}}]{Maccio07}
{Macci{\`o}}, A.~V., {Dutton}, A.~A., {van den Bosch}, F.~C., {et~al.} 2007,
  \mnras, 378, 55, \dodoi{10.1111/j.1365-2966.2007.11720.x}

\bibitem[{{Macci{\`o}} {et~al.}(2017){Macci{\`o}}, {Frings}, {Buck}, {Penzo},
  {Dutton}, {Blank}, \& {Obreja}}]{Maccio17}
{Macci{\`o}}, A.~V., {Frings}, J., {Buck}, T., {et~al.} 2017, \mnras, 472,
  2356, \dodoi{10.1093/mnras/stx2048}

\bibitem[{{Macci{\`o}} {et~al.}(2012){Macci{\`o}}, {Paduroiu}, {Anderhalden},
  {Schneider}, \& {Moore}}]{Maccio12}
{Macci{\`o}}, A.~V., {Paduroiu}, S., {Anderhalden}, D., {Schneider}, A., \&
  {Moore}, B. 2012, \mnras, 424, 1105, \dodoi{10.1111/j.1365-2966.2012.21284.x}

\bibitem[{{Mancera Pi{\~n}a} {et~al.}(2020){Mancera Pi{\~n}a}, {Fraternali},
  {Oman}, {Adams}, {Bacchini}, {Marasco}, {Oosterloo}, {Pezzulli}, {Posti},
  {Leisman}, {Cannon}, {di Teodoro}, {Gault}, {Haynes}, {Reiter}, {Rhode},
  {Salzer}, \& {Smith}}]{ManceraPina20}
{Mancera Pi{\~n}a}, P.~E., {Fraternali}, F., {Oman}, K.~A., {et~al.} 2020,
  \mnras, 495, 3636, \dodoi{10.1093/mnras/staa1256}

\bibitem[{{McMullin} {et~al.}(2007){McMullin}, {Waters}, {Schiebel}, {Young},
  \& {Golap}}]{McMullin07}
{McMullin}, J.~P., {Waters}, B., {Schiebel}, D., {Young}, W., \& {Golap}, K.
  2007, in Astronomical Society of the Pacific Conference Series, Vol. 376,
  Astronomical Data Analysis Software and Systems XVI, ed. R.~A. {Shaw},
  F.~{Hill}, \& D.~J. {Bell}, 127

\bibitem[{{Meidt} {et~al.}(2014){Meidt}, {Schinnerer}, {van de Ven},
  {Zaritsky}, {Peletier}, {Knapen}, {Sheth}, {Regan}, {Querejeta},
  {Mu{\~n}oz-Mateos}, {Kim}, {Hinz}, {Gil de Paz}, {Athanassoula}, {Bosma},
  {Buta}, {Cisternas}, {Ho}, {Holwerda}, {Skibba}, {Laurikainen}, {Salo},
  {Gadotti}, {Laine}, {Erroz-Ferrer}, {Comer{\'o}n}, {Men{\'e}ndez-Delmestre},
  {Seibert}, \& {Mizusawa}}]{Meidt14}
{Meidt}, S.~E., {Schinnerer}, E., {van de Ven}, G., {et~al.} 2014, \apj, 788,
  144, \dodoi{10.1088/0004-637X/788/2/144}

\bibitem[{{Milgrom}(1983)}]{Milgrom83}
{Milgrom}, M. 1983, \apj, 270, 365, \dodoi{10.1086/161130}

\bibitem[{{Moore}(1994)}]{Moore94}
{Moore}, B. 1994, \nat, 370, 629, \dodoi{10.1038/370629a0}

\bibitem[{{Moore} {et~al.}(1998){Moore}, {Governato}, {Quinn}, {Stadel}, \&
  {Lake}}]{Moore98}
{Moore}, B., {Governato}, F., {Quinn}, T., {Stadel}, J., \& {Lake}, G. 1998,
  \apjl, 499, L5, \dodoi{10.1086/311333}

\bibitem[{{Mould} {et~al.}(2000){Mould}, {Huchra}, {Freedman}, {Kennicutt},
  {Ferrarese}, {Ford}, {Gibson}, {Graham}, {Hughes}, {Illingworth}, {Kelson},
  {Macri}, {Madore}, {Sakai}, {Sebo}, {Silbermann}, \& {Stetson}}]{Mould00}
{Mould}, J.~R., {Huchra}, J.~P., {Freedman}, W.~L., {et~al.} 2000, \apj, 529,
  786, \dodoi{10.1086/308304}

\bibitem[{{Navarro} {et~al.}(1996){Navarro}, {Eke}, \& {Frenk}}]{Navarro96}
{Navarro}, J.~F., {Eke}, V.~R., \& {Frenk}, C.~S. 1996, \mnras, 283, L72,
  \dodoi{10.1093/mnras/283.3.L72}

\bibitem[{{Navarro} {et~al.}(1997){Navarro}, {Frenk}, \& {White}}]{Navarro97}
{Navarro}, J.~F., {Frenk}, C.~S., \& {White}, S. D.~M. 1997, \apj, 490, 493,
  \dodoi{10.1086/304888}

\bibitem[{{Nipoti} \& {Binney}(2015)}]{Nipoti15}
{Nipoti}, C., \& {Binney}, J. 2015, \mnras, 446, 1820,
  \dodoi{10.1093/mnras/stu2217}

\bibitem[{{Oh} {et~al.}(2011){Oh}, {de Blok}, {Brinks}, {Walter}, \&
  {Kennicutt}}]{Oh11}
{Oh}, S.-H., {de Blok}, W.~J.~G., {Brinks}, E., {Walter}, F., \& {Kennicutt},
  Robert~C., J. 2011, \aj, 141, 193, \dodoi{10.1088/0004-6256/141/6/193}

\bibitem[{{Oh} {et~al.}(2015){Oh}, {Hunter}, {Brinks}, {Elmegreen}, {Schruba},
  {Walter}, {Rupen}, {Young}, {Simpson}, {Johnson}, {Herrmann}, {Ficut-Vicas},
  {Cigan}, {Heesen}, {Ashley}, \& {Zhang}}]{Oh15}
{Oh}, S.-H., {Hunter}, D.~A., {Brinks}, E., {et~al.} 2015, \aj, 149, 180,
  \dodoi{10.1088/0004-6256/149/6/180}

\bibitem[{{Oman} {et~al.}(2019){Oman}, {Marasco}, {Navarro}, {Frenk}, {Schaye},
  \& {Ben{\'\i}tez-Llambay}}]{Oman19}
{Oman}, K.~A., {Marasco}, A., {Navarro}, J.~F., {et~al.} 2019, \mnras, 482,
  821, \dodoi{10.1093/mnras/sty2687}

\bibitem[{{Osterbrock} \& {Ferland}(2006)}]{Osterbrock06}
{Osterbrock}, D.~E., \& {Ferland}, G.~J. 2006, {Astrophysics of gaseous nebulae
  and active galactic nuclei}

\bibitem[{{Pineda} {et~al.}(2017){Pineda}, {Hayward}, {Springel}, \& {Mendes de
  Oliveira}}]{Pineda17}
{Pineda}, J. C.~B., {Hayward}, C.~C., {Springel}, V., \& {Mendes de Oliveira},
  C. 2017, \mnras, 466, 63, \dodoi{10.1093/mnras/stw3004}

\bibitem[{{Randall} {et~al.}(2008){Randall}, {Markevitch}, {Clowe}, {Gonzalez},
  \& {Brada{\v{c}}}}]{Randall08}
{Randall}, S.~W., {Markevitch}, M., {Clowe}, D., {Gonzalez}, A.~H., \&
  {Brada{\v{c}}}, M. 2008, \apj, 679, 1173, \dodoi{10.1086/587859}

\bibitem[{{Read} {et~al.}(2016{\natexlab{a}}){Read}, {Agertz}, \&
  {Collins}}]{Read16a}
{Read}, J.~I., {Agertz}, O., \& {Collins}, M.~L.~M. 2016{\natexlab{a}}, \mnras,
  459, 2573, \dodoi{10.1093/mnras/stw713}

\bibitem[{{Read} {et~al.}(2016{\natexlab{b}}){Read}, {Iorio}, {Agertz}, \&
  {Fraternali}}]{Read16b}
{Read}, J.~I., {Iorio}, G., {Agertz}, O., \& {Fraternali}, F.
  2016{\natexlab{b}}, \mnras, 462, 3628, \dodoi{10.1093/mnras/stw1876}

\bibitem[{{Read} {et~al.}(2019){Read}, {Walker}, \& {Steger}}]{Read19}
{Read}, J.~I., {Walker}, M.~G., \& {Steger}, P. 2019, \mnras, 484, 1401,
  \dodoi{10.1093/mnras/sty3404}

\bibitem[{{Riess} {et~al.}(2016){Riess}, {Macri}, {Hoffmann}, {Scolnic},
  {Casertano}, {Filippenko}, {Tucker}, {Reid}, {Jones}, {Silverman},
  {Chornock}, {Challis}, {Yuan}, {Brown}, \& {Foley}}]{Riess16}
{Riess}, A.~G., {Macri}, L.~M., {Hoffmann}, S.~L., {et~al.} 2016, \apj, 826,
  56, \dodoi{10.3847/0004-637X/826/1/56}

\bibitem[{{Rocha} {et~al.}(2013){Rocha}, {Peter}, {Bullock}, {Kaplinghat},
  {Garrison-Kimmel}, {O{\~n}orbe}, \& {Moustakas}}]{Rocha13}
{Rocha}, M., {Peter}, A. H.~G., {Bullock}, J.~S., {et~al.} 2013, \mnras, 430,
  81, \dodoi{10.1093/mnras/sts514}

\bibitem[{{Rong} {et~al.}(2017){Rong}, {Guo}, {Gao}, {Liao}, {Xie}, {Puzia},
  {Sun}, \& {Pan}}]{Rong17}
{Rong}, Y., {Guo}, Q., {Gao}, L., {et~al.} 2017, \mnras, 470, 4231,
  \dodoi{10.1093/mnras/stx1440}

\bibitem[{{Salvatier} {et~al.}(2016){Salvatier}, {Wiecki{\^a}}, \&
  {Fonnesbeck}}]{Salvatier16}
{Salvatier}, J., {Wiecki{\^a}}, T.~V., \& {Fonnesbeck}, C. 2016, {PyMC3: Python
  probabilistic programming framework}.
\newblock \doeprint{1610.016}

\bibitem[{{Santos-Santos} {et~al.}(2016){Santos-Santos}, {Brook}, {Stinson},
  {Di Cintio}, {Wadsley}, {Dom{\'\i}nguez-Tenreiro}, {Gottl{\"o}ber}, \&
  {Yepes}}]{Santos-Santos16}
{Santos-Santos}, I.~M., {Brook}, C.~B., {Stinson}, G., {et~al.} 2016, \mnras,
  455, 476, \dodoi{10.1093/mnras/stv2335}

\bibitem[{{Schive} {et~al.}(2014){Schive}, {Chiueh}, \&
  {Broadhurst}}]{Schive14}
{Schive}, H.-Y., {Chiueh}, T., \& {Broadhurst}, T. 2014, Nature Physics, 10,
  496, \dodoi{10.1038/nphys2996}

\bibitem[{{Schoenmakers} {et~al.}(1997){Schoenmakers}, {Franx}, \& {de
  Zeeuw}}]{Schoenmakers97}
{Schoenmakers}, R.~H.~M., {Franx}, M., \& {de Zeeuw}, P.~T. 1997, \mnras, 292,
  349, \dodoi{10.1093/mnras/292.2.349}

\bibitem[{{Schombert} {et~al.}(2019){Schombert}, {McGaugh}, \&
  {Lelli}}]{Schombert19}
{Schombert}, J., {McGaugh}, S., \& {Lelli}, F. 2019, \mnras, 483, 1496,
  \dodoi{10.1093/mnras/sty3223}

\bibitem[{{Shi} {et~al.}(2011){Shi}, {Helou}, {Yan}, {Armus}, {Wu}, {Papovich},
  \& {Stierwalt}}]{Shi11}
{Shi}, Y., {Helou}, G., {Yan}, L., {et~al.} 2011, \apj, 733, 87,
  \dodoi{10.1088/0004-637X/733/2/87}

\bibitem[{{Shi} {et~al.}(2018){Shi}, {Yan}, {Armus}, {Gu}, {Helou}, {Qiu},
  {Gwyn}, {Stierwalt}, {Fang}, {Chen}, {Zhou}, {Wu}, {Zheng}, {Zhang}, {Gao},
  \& {Wang}}]{Shi18}
{Shi}, Y., {Yan}, L., {Armus}, L., {et~al.} 2018, \apj, 853, 149,
  \dodoi{10.3847/1538-4357/aaa3e6}

\bibitem[{{Spekkens} {et~al.}(2005){Spekkens}, {Giovanelli}, \&
  {Haynes}}]{Spekkens05}
{Spekkens}, K., {Giovanelli}, R., \& {Haynes}, M.~P. 2005, \aj, 129, 2119,
  \dodoi{10.1086/429592}

\bibitem[{{Spergel} \& {Steinhardt}(2000)}]{Spergel00}
{Spergel}, D.~N., \& {Steinhardt}, P.~J. 2000, \prl, 84, 3760,
  \dodoi{10.1103/PhysRevLett.84.3760}

\bibitem[{{Telford} {et~al.}(2020){Telford}, {Dalcanton}, {Williams}, {Bell},
  {Dolphin}, {Durbin}, \& {Choi}}]{Telford20}
{Telford}, O.~G., {Dalcanton}, J.~J., {Williams}, B.~F., {et~al.} 2020, \apj,
  891, 32, \dodoi{10.3847/1538-4357/ab701c}

\bibitem[{{Tollet} {et~al.}(2016){Tollet}, {Macci{\`o}}, {Dutton}, {Stinson},
  {Wang}, {Penzo}, {Gutcke}, {Buck}, {Kang}, {Brook}, {Di Cintio}, {Keller}, \&
  {Wadsley}}]{Tollet16}
{Tollet}, E., {Macci{\`o}}, A.~V., {Dutton}, A.~A., {et~al.} 2016, \mnras, 456,
  3542, \dodoi{10.1093/mnras/stv2856}

\bibitem[{{Trachternach} {et~al.}(2008){Trachternach}, {de Blok}, {Walter},
  {Brinks}, \& {Kennicutt}}]{Trachternach08}
{Trachternach}, C., {de Blok}, W.~J.~G., {Walter}, F., {Brinks}, E., \&
  {Kennicutt}, R.~C., J. 2008, \aj, 136, 2720,
  \dodoi{10.1088/0004-6256/136/6/2720}

\bibitem[{{Tulin} \& {Yu}(2018)}]{Tulin18}
{Tulin}, S., \& {Yu}, H.-B. 2018, \physrep, 730, 1,
  \dodoi{10.1016/j.physrep.2017.11.004}

\bibitem[{{van Dokkum} {et~al.}(2016){van Dokkum}, {Abraham}, {Brodie},
  {Conroy}, {Danieli}, {Merritt}, {Mowla}, {Romanowsky}, \&
  {Zhang}}]{vanDokkum16}
{van Dokkum}, P., {Abraham}, R., {Brodie}, J., {et~al.} 2016, \apjl, 828, L6,
  \dodoi{10.3847/2041-8205/828/1/L6}

\bibitem[{{Vehtari} {et~al.}(2015){Vehtari}, {Gelman}, \& {Gabry}}]{Vehtari15}
{Vehtari}, A., {Gelman}, A., \& {Gabry}, J. 2015, arXiv e-prints,
  arXiv:1507.04544.
\newblock \doarXiv{1507.04544}

\bibitem[{{Walker} {et~al.}(2015){Walker}, {Olszewski}, \& {Mateo}}]{Walker15}
{Walker}, M.~G., {Olszewski}, E.~W., \& {Mateo}, M. 2015, \mnras, 448, 2717,
  \dodoi{10.1093/mnras/stv099}

\bibitem[{{Wang} {et~al.}(2020){Wang}, {Bose}, {Frenk}, {Gao}, {Jenkins},
  {Springel}, \& {White}}]{Wang20}
{Wang}, J., {Bose}, S., {Frenk}, C.~S., {et~al.} 2020, \nat, 585, 39,
  \dodoi{10.1038/s41586-020-2642-9}

\bibitem[{{Weinberg} {et~al.}(2015){Weinberg}, {Bullock}, {Governato}, {Kuzio
  de Naray}, \& {Peter}}]{Weinberg15}
{Weinberg}, D.~H., {Bullock}, J.~S., {Governato}, F., {Kuzio de Naray}, R., \&
  {Peter}, A. H.~G. 2015, Proceedings of the National Academy of Science, 112,
  12249, \dodoi{10.1073/pnas.1308716112}

\bibitem[{{Wright} {et~al.}(2010){Wright}, {Eisenhardt}, {Mainzer}, {Ressler},
  {Cutri}, {Jarrett}, {Kirkpatrick}, {Padgett}, {McMillan}, {Skrutskie},
  {Stanford}, {Cohen}, {Walker}, {Mather}, {Leisawitz}, {Gautier}, {McLean},
  {Benford}, {Lonsdale}, {Blain}, {Mendez}, {Irace}, {Duval}, {Liu}, {Royer},
  {Heinrichsen}, {Howard}, {Shannon}, {Kendall}, {Walsh}, {Larsen}, {Cardon},
  {Schick}, {Schwalm}, {Abid}, {Fabinsky}, {Naes}, \& {Tsai}}]{Wright10}
{Wright}, E.~L., {Eisenhardt}, P. R.~M., {Mainzer}, A.~K., {et~al.} 2010, \aj,
  140, 1868, \dodoi{10.1088/0004-6256/140/6/1868}

\bibitem[{{Yozin} \& {Bekki}(2015)}]{Yozin15}
{Yozin}, C., \& {Bekki}, K. 2015, \mnras, 452, 937,
  \dodoi{10.1093/mnras/stv1073}

\bibitem[{{Yu} {et~al.}(2019){Yu}, {Shi}, {Chen}, {Law}, {Bizyaev}, {Bing},
  {Li}, {Zhou}, {Chen}, {Riffel}, {Riffel}, {Zhang}, {Chen}, \& {Pan}}]{Yu19}
{Yu}, X., {Shi}, Y., {Chen}, Y., {et~al.} 2019, \mnras, 486, 4463,
  \dodoi{10.1093/mnras/stz1146}

\bibitem[{{Zavala} {et~al.}(2013){Zavala}, {Vogelsberger}, \&
  {Walker}}]{Zavala13}
{Zavala}, J., {Vogelsberger}, M., \& {Walker}, M.~G. 2013, \mnras, 431, L20,
  \dodoi{10.1093/mnrasl/sls053}

\bibitem[{{Zhao} {et~al.}(2003){Zhao}, {Mo}, {Jing}, \& {B{\"o}rner}}]{Zhao03}
{Zhao}, D.~H., {Mo}, H.~J., {Jing}, Y.~P., \& {B{\"o}rner}, G. 2003, \mnras,
  339, 12, \dodoi{10.1046/j.1365-8711.2003.06135.x}

\bibitem[{{Zibetti} {et~al.}(2009){Zibetti}, {Charlot}, \& {Rix}}]{Zibetti09}
{Zibetti}, S., {Charlot}, S., \& {Rix}, H.-W. 2009, \mnras, 400, 1181,
  \dodoi{10.1111/j.1365-2966.2009.15528.x}

\end{thebibliography}
\bibliographystyle{aasjournal}

\end{document}